\documentclass[onecolumn,tighten,times]{aastex62}
\usepackage{txfonts}
\usepackage{newlfont}

\usepackage{amssymb}
\usepackage{wrapfig}
\usepackage{cancel}

\usepackage{relsize,exscale}
\makeatletter

\usepackage{array,multirow,graphicx}
\usepackage{dcolumn}
\usepackage{txfonts}
\usepackage{newlfont}
\usepackage{bm}
\usepackage[figtopcap]{subfigure}

\shortauthors{G.~G.~L.~Nashed}
\usepackage{scalerel}
\usepackage{tikz}
\usetikzlibrary{svg.path}
\definecolor{orcidlogocol}{HTML}{A6CE39}
\tikzset{
  orcidlogo/.pic={
    \fill[orcidlogocol] svg{M256,128c0,70.7-57.3,128-128,128C57.3,256,0,198.7,0,128C0,57.3,57.3,0,128,0C198.7,0,256,57.3,256,128z};
    \fill[white] svg{M86.3,186.2H70.9V79.1h15.4v48.4V186.2z}
                 svg{M108.9,79.1h41.6c39.6,0,57,28.3,57,53.6c0,27.5-21.5,53.6-56.8,53.6h-41.8V79.1z M124.3,172.4h24.5c34.9,0,42.9-26.5,42.9-39.7c0-21.5-13.7-39.7-43.7-39.7h-23.7V172.4z}
                 svg{M88.7,56.8c0,5.5-4.5,10.1-10.1,10.1c-5.6,0-10.1-4.6-10.1-10.1c0-5.6,4.5-10.1,10.1-10.1C84.2,46.7,88.7,51.3,88.7,56.8z};}}
\newcommand\orcid[1]{\href{https://orcid.org/#1}{\mbox{\scalerel*{
\begin{tikzpicture}[yscale=-1,transform shape]
\pic{orcidlogo};
\end{tikzpicture}
}{|}}}}
\begin{document}

\title{\uppercase{The effect of $f(R,T)$ modified gravity  on mass and radius of pulsar ${\textit HerX1}$}}

\author{G.~G.~L.~Nashed \orcid{0000-0001-5544-1119}}
%\author[0000-0001-5544-1119]{G.~G.~L.~Nashed}
\affiliation{Centre for theoretical physics, the British University in Egypt, 11837 - P.O. Box 43, Egypt\\
 Center for Space Research, North-West University, Potchefstroom 2520, South Africa}
\correspondingauthor{G.~G.~L.~Nashed}
\email{nashed@bue.edu.eg}



\begin{abstract}
Recent findings from the Neutron Star Interior Composition Explorer (NICER) have opened up opportunities to investigate the potential coupling between matter and geometry, along with its resulting physical implications. Millisecond pulsars serve as an ideal subject for conducting such tests and examining these phenomena. We apply the field equations of modified gravity, $f(R, T)=R+\alpha\, T$ to a spherically symmetric spacetime, where $R$ is the Ricci scalar, $\alpha$ is a dimensional parameter, and $T$ is the matter of the geometry. Five unknown functions are present in the output system of differential equations, which consists of three equations. To close the system, we make explicit assumptions about the anisotropy and the radial metric potential, $g_{rr}$. We then solve the output differential equations and derive the explicit forms of the components of the energy-momentum tensor, namely, density, radial, and tangential pressures.  We explore the potential for expressing all physical parameters within the star using the compactness parameters, represented by the symbol $C$ which is defined as {\textrm C=2GM/Rc$^2$} and  the dimensional parameter  $\alpha$. Our findings demonstrate that within the framework of $f(R,T)$ theory, the matter-geometry interaction leads to a reduced size allowed by Einstein's general relativity for a given mass. The accuracy of this hypothesis was confirmed through observations involving an additional set of 22 pulsars. To achieve a boundary density consistent with that of a neutron star core, the mass-radius relationship permits for high masses, reaching up to 3.35 times the mass of the Sun ($M_\odot$).  It is important to highlight that no equation of state assumption is made in our analysis. Nevertheless, the model exhibits a good fit with a linear trend. Through a comparison of the surface densities of the twenty pulsars, we have categorized them into three distinct groups. We show that these three groups are compatible with neutron cores.
\end{abstract}

\section{Introduction}\label{Sec:Introduction}

While Einstein's General Relativity (GR) has achieved remarkable success in accurately predicting various gravitational phenomena observed within the solar system \citep{Will:2014kxa}.  It is worth noting that the evaluated  masses of the massive pulsars exceed the observed values, suggesting the necessity for tighter constraints on the parameter $\alpha_1$. This may entail adopting a smaller value of $\alpha_1$, such as $\alpha_1=0.02$, to ensure better agreement with the observational data \citep{LIGOScientific:2016lio,LIGOScientific:2018dkp}, it has thus far been unable to unravel the enigma of dark energy and other perplexing mysteries. Furthermore, within modern cosmology, numerous unanswered questions persist, leading many scientists to question whether  GR is the sole gravitational theory \citep{CANTATA:2021ktz}. Moreover, studies have shown that General Relativity (GR) cannot be renormalized unless it is formulated as a quantum field theory its action incorporates curvature invariants of higher order \citep{Stelle:1976gc,Vilkovisky:1992pb}.   Moreover, GR  necessitates adjustments at scales of both time and length that are small, along with energies that are comparable to the Planck energy scales. In this frame, a compelling argument put forth demonstrating that the  late-time accelerated expansion and early-time inflation of the Universe can be accounted for by modifying Einstein's geometric theory \citep{Starobinsky:1980te,Capozziello:2002rd,Carroll:2003wy,Nojiri:2006ri}.

 One of the fundamental alterations to  GR  involves substituting the Ricci scalar $R$ in the conventional Einstein-Hilbert action using a function of $R$ that is not predetermined or fixed. This modification gives rise to the class of gravitational theories known as $f(R)$ theories of gravity  \citep[see for exmaple][]{Sotiriou:2008rp,DeFelice:2010aj}. Comprehensive reviews discussing the cosmological applications of these theories can be found in references such as \citep[for exmaple][]{Capozziello:2011et,Nojiri:2010wj,Clifton:2011jh,Nojiri:2017ncd}. However, these theories bring about fundamental modification of the Tolman-Oppenheimer-Volkoff (TOV) equations, in the frame of astrophysical,  resulting in alterations to the astrophysical properties of compact stars. This includes alterations in properties such as mass-radius relationships, maximum masses, and moments of inertia. For a thorough investigation of non-relativistic and relativistic  stars under the framework of amended gravitational   theories, including both metric-affine and metric techniques, a comprehensive overview can be found in the referenced publication. \citep{Olmo:2019flu}. The majority of published works exploring the internal property of compact stars, both in modified and GR gravitational theories, typically assume the presence of isotropic, perfect fluid composition within these stars. However, there are many justifications showing that the presence of anisotropy cannot be ignored when examining nuclear matter under extremely high densities and pressures. This is evident in the literature, as referenced. \citep{Herrera:1997plx,Isayev:2017rci,Ivanov:2017kyr,Maurya:2017uyk,Biswas:2019gkw,Pretel:2020xuo,Bordbar:2022qhl} and references therein. Studies have demonstrated that the existence of anisotropy can result in considerable changes to the essential properties of compact stars  \citep{Maurya:2017uyk,Biswas:2019gkw,Pretel:2020xuo,Horvat:2010xf,Rahmansyah:2020gar,Roupas:2020mvs,Das:2021qaq,Das:2021giz,Roupas:2020jyv,Das:2022kwq}.  It is also worth mentioning that non-rotating anisotropic compact stars have recently been studied by some authors in Refs. \citep{Shamir:2017yza, Folomeev:2018ioy,Mustafa:2020jln,Nashed:2021sji,Nashed:2021gkp,Deb:2018sgt,Maurya:2019sfm,Biswas:2020gzd,Maurya:2019iup,Rej:2021qpi,Biswas:2021wfn,Vernieri:2019vlh,
Pretel:2022qng,Mota:2022zbq,Ashraf:2020yyo,Tangphati:2021tcy,
Tangphati:2021wng,Nashed:2021pkc,Solanki:2022yna,Pretel:2022plg} in the frame of extended gravitational theories. Moreover, the investigation of slowly rotating anisotropic NSs has been conducted within the framework of the scalar-tensor theory of gravity \citep{Silva:2014fca}.

In their work, \citep{Harko:2011kv} introduced $f(R, T)$ gravity as an extension of $f(R)$ modified theories of gravity. This formulation establishes a connection between geometry and matter by incorporating a coupling term involving the trace of the energy-momentum tensor, denoted as $T$. Undoubtedly, the $f(R, T) = R+ 2\alpha T$ gravity represents the simplest and extensively studied model that incorporates a minimal coupling between matter and gravity \citep[see also][]{Shabani:2017kis,Debnath:2018wct,Bhattacharjee:2020eec,Bhattacharjee:2020jsf,Gamonal:2020itt}. In recent studies, researchers have directed their attention towards examining the cosmological implications of this model. Additionally, other scholars have investigated the astrophysical ramifications of the $2\alpha T$ term within the equilibrium framework of both  anisotropic and isotropic systems  \citep{Moraes:2015uxq,Das:2016mxq,Deb:2017rhd,Deb:2018gzt,Lobato:2020fxt,Pretel:2020oae,Bora:2022dnu} and anisotropic compact stars \citep{Maurya:2019sfm,Biswas:2020gzd,Maurya:2019iup,Rej:2021qpi,Biswas:2021wfn,Nashed:2011fg,Vernieri:2019vlh,Pretel:2022qng,Mota:2022zbq,Ashraf:2020yyo,Tangphati:2021tcy}. One notable feature of this model is that the curvature scalar $R$ is zero beyond the boundaries of a compact star. As a result, the vacuum solution can  be prescribed using the Schwarzschild  spacetime. Hence,  it is shown  that the influence of the $2\alpha T$ term becomes negligible for central densities that reach a sufficiently high level. Nevertheless, below a specific critical core density threshold, the radius of isotropic  star exhibits notable difference from the predictions of general relativity  \citep{Biswas:2021wfn,Vernieri:2019vlh}.

%arXiv:2302.0390
For several decades, researchers have been studying the equations of state (EoSs) of equilibrated charge-neutral matter and their connections with neutron star (NS) properties \citep{Oppenheimer:1939ne,Tolman:1939jz,Glendenning:1992dr}.  Exact knowledge of NS properties and heavy ion collision data may constrain the behavior of EoSs at supra-saturation densities \citep{Huth:2021bsp}. The radius and tidal compressibility of a population of neutron stars with masses ranging from (1-3 $M_\odot$) would be used to investigate the EoS at densities several times ($\approx $6-8) that of the saturation density encountered at the center of finite nuclei. The tidal compressibility parameter of NS, which shows information about the EoS, has been deduced for the first time from GW170817, a gravitational wave event observed by advanced LIGO \citep{LIGOScientific:2014pky} and advanced Virgo detectors \citep{VIRGO:2014yos} from a binary neutron star (BNS) merger with a total mass of $2.74_{-0.01}^{+0.04}M_\odot$  of the system \citep{LIGOScientific:2018hze,LIGOScientific:2018jsj}. A further successive event, GW190425, was noted \citep{LIGOScientific:2020aai}, which was most likely caused by the coalescence of BNSs. The BNS signals emitted by coalescing neutron stars are presumed to be observed more frequently in the upcoming LIGO-Virgo-KAGRA runs and prospective detectors, such as the Einstein Telescope \citep{Punturo:2010zz} and Cosmic Explorer \citep{Reitze:2019iox}. The unexpected limitations on the EoS promised by gravitational wave astronomy, as revealed by a thorough analysis of gravitational wave parameter estimation, have prompted numerous theoretical investigations of neutron star characteristics \citep{LIGOScientific:2020aai,LIGOScientific:2017vwq,Malik:2018zcf,De:2018uhw,Nashed:2016tbj,Liliani:2021jne,Forbes:2019xaz,Landry:2018prl,Piekarewicz:2018sgy,
Biswas:2020puz,Thi:2021jhz}.Â  In recent times, two groups of Neutron Star Interior Composition Explorer (NICER) X-ray telescopes simultaneously supplied neutron star mass and radius for PSR J0030+0451 with $R=13.02^{+1.24}_{-1.06}$km for mass $1.44^{+0.15}_{-0.14} M_\odot$  \citep{Miller:2019cac} and $R=12.71^{+1.14}_{-1.19}$km for mass $1.34^{+0.15}_{-0.16} M_\odot$ \citep{Riley:2019yda}, which are supplementary restrictions on the EoS.  $R=13.7^{+2.6}_{-1.5}$km with mass $2.08 \pm 0.07 M_\odot$  \citep{Miller:2021qha} and $R=12.39^{+1.30}_{-0.98}$km with mass $2.072^{+0.067}_{-0.066} M_\odot$ \citep{Riley:2021pdl} were noted for the heavier pulsar PSR J0740+6620. The methodology lower bound on the maximum NS mass for the black-widow pulsar PSR J0952-0607 \citep{Romani:2022jhd} surpasses any prior measurements, including $M_{\textrm max}= 2.35\pm 0.17 M_\odot$ for PSR J2215-5135 \citep{Linares:2018ppq,Patra:2023jvv}. If the observational bounds are credible, stiffer EoSs are needed to support the NS with a higher mass $2M_\odot$.

Remarkably, the observations of PSR J0740+6020 and PSR J0030+0451 conducted by NICER supply compelling proof contradicting the viability of more compressible models. Despite their nearly identical sizes, PSR J0740+6020 possesses significantly greater mass compared to PSR J0030+0451. Therefore, it is logical to propose mechanisms that can explain the non-squeezability of neutron stars (NS) as their mass increases. Additionally, the presence of pulsars with high masses like PSR J0740+6020  suggests that the inclusion of matter-geometry coupling in $f(R,T)=R+\alpha_1 T$ gravity allows for surpassing the maximum limit set by the conformal sound speed, ${c_s}^2 = c^2/3$. Even in situations with low-density, this poses an extra difficulty for theoretical models, as demonstrated by previous studies \citep{Bedaque:2014sqa}  \citep[see also][]{Cherman:2009tw,Landry:2020vaw}.  The investigation of PSR J0740+6020 conducted by \citep{Legred:2021hdx} determined that the conformal sound speed is significantly broken within the core of the neutron star. Specifically, it was found that $c_s^2=0.75 c^2$ at a density of approximately $3.60, \rho_{nuc}$, indicating a deviation from the expected conformal sound speed.

As NS accumulates mass, it is expected that gravity will intensify, leading to the collapse of the NS into a smaller size. This collapse results in a higher density within the NS. Nevertheless, the observations made by NICER on the pulsars PSR J0740+6020 and  PSR J0030+0451  contradict this particular mechanism. We propose that the existence of high-density regions in neutron stars (NS) at larger masses would give rise to anisotropy, where the radial and tangential pressures differ. This phenomenon, as demonstrated by the Tolman-Oppenheimer-Volkoff (TOV) equation  \citep{Bowers:1974tgi}, results in a repulsive anisotropic force that counteracts the attractive gravitational force when ${\textrm p_t>p_r}$.  The objective of this study is to create a compact stellar object and examine its physical implications within the framework of non-minimal matter theory, specifically using the $f(R,T)=R+\alpha T$, where $\alpha$ represents a dimensional parameter.

The organization of this study is outlined as follows: In Sec. \ref{Sec2}, we  give a short review of the modification of the general theory of relativity as proposed Harko et al. \citep{Harko:2011kv} . In Section \ref{Sec:Model}, we establish the fundamental assumptions of the current investigation. In Section \ref{Sec:Stability}, we utilize the data from the pulsar ${\textit HerX1}$ to impose constraints on the parameters of the model.  Furthermore, we analyze the physical characteristics and stability of the pulsar according to the findings obtained from our model in Section \ref{S8}. Additionally, we compare our model with data from other pulsars in Section \ref{S8}. In Section \ref{MR}, we determine the upper limit on compactness imposed by physical limitations, and demonstrate mass-radius relationships for various surface density choices, highlighting the maximum mass achievable for a configuration that remains stable in each scenario. Finally, in Section \ref{Sec:Conclusion}, we provide a summary of the findings obtained in this study.
%%%%%%%%%%%%%%%%%%%%%%%%%%%%%%%%%%%%%%%%%%%%%% Section 2 %%%%%%%%%%%%%%%%%%%%%%%%%%%%%%%%%%%%%%%%%%%%%%%%%%%%%%%%%%%%%%%

\section{Fundamentals of $f(R,T)$ gravitational theory}\label{Sec2}

The construction of $f(R)$-amended gravitational theoriesinvolves incorporating an explicit coupling between gravity and matter through an arbitrary function that depends on the Ricci scalar tensor and the trace part of the energy-momentum tensor. Hence, the amended Einstein  action for $f(R,T)$ gravity is \citep{Harko:2011kv}:
\begin{equation}\label{1}
 {\mathrm   S = \frac{1}{16\pi}\int  {\mathrm f(R,T)} \sqrt{-g}d^4x + \int\mathrm{L}_m \sqrt{-g}d^4x}\, ,
\end{equation}
where  $\mathrm{L}_m$ corresponds to the Lagrangian of the matter,  and $g$ represents the determinant of the metric tensor $g_{\mu\nu}$. The equation of motions governing ${\mathrm f(R,T)}$ modified gravity can be derived by varying Eq.~(\ref{1}) w.r.t.  the metric.
\begin{eqnarray}\label{2}
   {\mathit f_R(R,T) R_{\mu\nu}} &{\mathit - \frac{1}{2}f(R,T) g_{\mu\nu} + [g_{\mu\nu}\square - \nabla_\mu\nabla_\nu] f_R(R,T)= \kappa T_{\mu\nu} -(T_{\mu\nu} + \Theta_{\mu\nu})f_T(R,T)}\,.
\end{eqnarray}
In the given equations, ${\mathit R_{\mu\nu}}$ represents the Ricci tensor, ${\mathit f_R \equiv \frac{\partial f}{\partial R}}$ denotes the derivative of the function $f$ with respect to $R$, ${\mathit T_{\mu\nu}}$ is the energy-momentum tensor, ${\mathrm f_T \equiv \frac{\partial f}{\partial T}}$ represents the derivative of $f$ with respect to $T$, ${\mathit \square \equiv \nabla_\mu\nabla^\mu}$ represents the d'Alembertian operator, where ${\mathit \nabla_\mu}$ denotes the  derivative. Additionally, the tensor ${\mathrm \Theta_{\mu\nu}}$ is represented by the variation of ${\mathrm T_{\mu\nu}}$ w.r.t. the metric.
\begin{eqnarray}\label{3}
   {\mathrm \Theta_{\mu\nu}} &{\mathrm\equiv g^{\alpha\beta}\frac{\delta T_{\alpha\beta}}{\delta g^{\mu\nu}}=  -2T_{\mu\nu} + g_{\mu\nu}\mathrm{L}_m - 2g^{\alpha\beta} \frac{\partial^2\mathcal{L}_m}{\partial g^{\mu\nu} \partial g^{\alpha\beta}}}\, .
\end{eqnarray}

Similar to the case of ${\mathrm f(R)}$ gravity \citep{Sotiriou:2008rp,DeFelice:2010aj}, in ${\mathit f(R,T)}$ modified theories, the Ricci scalar can be considered as a dynamic quantity and can be fixed by   by taking the trace of Eq.~ (\ref{2}), which can be expressed as:
\begin{eqnarray}\label{4}
    {\mathit 3\square f_R(R,T)} &{\mathrm+ Rf_R(R,T) - 2f(R,T)= \kappa T - (T+\Theta)f_T(R,T)} .
\end{eqnarray}
Here, we define $\Theta= \Theta_\mu^{\ \mu}$. Moreover, taking the divergence of Eq. (\ref{2}) yields  \citep{BarrientosO:2014mys}
\begin{eqnarray}\label{5}
   {\mathit \nabla^\mu T_{\mu\nu} = \frac{f_T(R,T)}{\kappa - f_T(R,T)}\bigg[ (T_{\mu\nu} + \Theta_{\mu\nu})\nabla^\mu \ln f_T(R,T)   + \nabla^\mu\Theta_{\mu\nu} - \frac{1}{2}g_{\mu\nu}\nabla^\mu T \bigg]} \,,
\end{eqnarray}
where ${\mathit \kappa}$  is the Einstein coupling constant which is defined as ${\mathrm \kappa=\frac{8\pi G}{c^4}}$. Furthermore, ${\mathrm G}$ represents the gravitational constant in Newtonian physics, while ${\mathrm c}$ symbolizes the speed of light.

To acquire numerical solutions that depict compact stars, it is necessary to make an assumption about the specific form of $f(R,T)$ gravity. In this context, we consider the simplest form of $f(R,T)$ gravity, which involves a minimal matter-gravity coupling \citep{Harko:2011kv}. This form, denoted as $f(R,T)= R+ \alpha T$ gravity, has been extensively studied in the context of astrophysical and cosmological scales. In this study, the parameter $\alpha$ is a dimensional parameter with the same dimension as the gravitational constant $\kappa$. Therefore, we will express it in terms of another parameter to maintain consistency in dimensions \[\alpha=\kappa\alpha_1\,.\]

Consequently, Eqs.~(\ref{2}), (\ref{4}) and (\ref{5}) can be rewritten as:
\begin{eqnarray}
   {\mathrm G_{\mu\nu}} &{\mathrm= \kappa [T_{\mu\nu} + \alpha_1 Tg_{\mu\nu} - 2\alpha_1(T_{\mu\nu} + \Theta_{\mu\nu})]}\, ,   \label{6}   \\
    {\mathrm R} &{\mathrm= -\kappa [T + 2\alpha_1(T- \Theta) ]}\,,   \label{7}    \\
   {\mathrm  \nabla^\mu T_{\mu\nu}} &{\mathrm= \frac{2\alpha_1}{(1 - 2\alpha_1)} \left[ \nabla^\mu \Theta_{\mu\nu} - \frac{1}{2}g_{\mu\nu}\nabla^\mu T \right]}\, ,   \label{8}
\end{eqnarray}
with ${\mathrm G_{\mu\nu}}$ being the Einstein tensor. In this study, ${\mathrm \Theta _{\mu \nu }}$ is regarded as a scalar expansion and is defined as
${\mathrm \Theta _{\mu \nu }=-2 T_{\mu \nu }-P g_{\mu \nu }}$, where we assume  ${\mathrm {L}_{m}=- {P}}$, with ${\mathrm {P}=\frac{1}{3}(p_{1}+2 p_{2})}$ \citep{Harko:2011kv}.

%%%%%%%%%%%%%%%%%%%%%%%%%%%%%%%%%%%%%%%%%%%%%% Section 3 %%%%%%%%%%%%%%%%%%%%%%%%%%%%%%%%%%%%%%%%%%%%%%%%%%%%%%%%%%%%%%%
\section{The linear form of $f(R,T)$ Model}\label{Sec:Model}
By considering a spherically symmetric and static line element in a 4-dimensional spacetime, we can express it using spherical polar coordinates ($t,r,\theta,\phi$) as follows:
\begin{equation}\label{eq:metric}
   {\mathrm  ds^2=-e^{\mu(r)}c^2 dt^2 + e^{\nu(r)} dr^2+ r^2 (d\theta^2+\sin^2 \theta \, d\phi^2)}\,.
\end{equation}
Here ${\mathrm \mu(r)}$ and ${\mathrm \nu(r)}$ are the ansatz of the laps functions. Furthermore, we make the assumption that the energy-momentum tensor of the anisotropic fluid, which exhibits spherical symmetry, can be represented in the following manner:
\begin{equation}\label{Tmn-anisotropy}
    {\mathrm \mathfrak{T}{^\alpha}{_\beta}}= \mathrm {(p_{2}+\rho c^2)v{^\alpha} v{_\beta}+p_{2} \delta ^\alpha _\beta + (p_{1}-p_{2}) w{^\alpha} w{_\beta}}\,.
\end{equation}
Here ${\mathrm \rho=\rho(r)}$ is the fluid energy density, ${\mathrm p_{1}}={\mathrm p_{1}(r)}$ its radial pressure (in the direction of the time-like four-velocity ${\mathrm v_\alpha}$), ${\mathrm p_{2}}={\mathrm p_{2}(r)}$ its tangential pressure (perpendicular to ${\mathrm v_\alpha}$) and ${\mathrm w{^\alpha}}$ is the unit space-like vector in the radial direction. Then, the energy-momentum tensor takes the diagonal form ${\mathrm \mathfrak{T}{^\alpha}{_\beta}}=\mathrm{diag(-\rho c^2,\,p_{1},\,p_{2},\,p_{2})}$.
Applying the field equation  (\ref{6})  Eqs. (\ref{eq:metric})  and (\ref{Tmn-anisotropy}), we get the following non-vanishing components:
\begin{eqnarray}\label{sys}
\mathrm {\kappa \rho c^2}&=&\mathrm{\frac{1}{1+3\,\alpha_1
 }\left[\frac {\nu' r+{
e^\nu-1}}{{r}^{2}{e^\nu}}+\frac{5\kappa \alpha_1}{3} \left(p_1  +2\,p_2  \right)\right]}\,,\nonumber\\
\mathrm{\kappa p_1}&=&\mathrm{\frac{3}{3+11\,\alpha_1
 }\left[\frac {\mu'r -{
e^\nu+1}}{{r}^{2}{e^\nu}}+\frac{\kappa \alpha_1}{3} \left(3\rho c^2  -10p_2  \right)\right]}\,,\nonumber\\
\mathrm{\kappa p_2}&=&\mathrm{\frac{3}{3+16\,\alpha_1
 }\left[\frac {2\mu'' r+\mu'[\mu'-\nu']+2}{4{r}{e^\nu}}+\kappa \alpha_1\left(\rho c^2-\frac{5}{3} p_1  \right)\right]}\,,\nonumber\\
&&
\label{eq:Feqs}
\end{eqnarray}
Here, the prime represents the derivative with respect to the radial coordinate $r$. As a result, the anisotropy function, denoted as, $\mathrm{\Delta(r) = p_2-p_1}$, as:
\begin{equation}\label{eq:Delta1v}
\mathrm{\Delta(r) =\frac {2\mu'' {r}^{2}+\mu'^2{r}^{2}- \left[\nu'{r}-2 \right] \,r\mu' +4 e^\nu-2\nu'r-4}{4\kappa\left(2\alpha_1+1 \right){r}^{2}e^\nu}}.
\end{equation}
Interestingly, when considering a spherically symmetric spacetime configuration, the connection between matter and geometry that emerges from the trace ${\mathrm \mathfrak{T}}$ does not play a role in the anisotropy, as observed in \citep{Nashed:2022zyi}.  Hence, the presence of anisotropic effects does not undermine the deviations from General Relativity (GR) caused by matter-geometry coupling. In the case of $\alpha_1=0$, the differential equations (\ref{eq:Feqs}) align with the GR field equations for a spherically symmetric interior spacetime \citep[c.f.,][]{Roupas:2020mvs}.

 {The set of equations (\ref{eq:Feqs}) comprises three distinct nonlinear differential equations involving five unknowns: $\mathrm{\mu}$, $\mathrm{\nu}$, $\mathrm{\rho}$, $\mathrm{p_1}$, and $\mathrm{p_2}$. Consequently, to fully determine the system, two additional conditions must be imposed. To achieve this, we employ the ansatz proposed in \citep{Das:2019dkn} (also utilized in \citep{Nashed:2020kjh}) for the metric potential $\nu$, which takes the following form:
\begin{equation}\label{eq:Delta11v}
\nu=\ln\left(\frac{1}{\left(1-\frac{a_2{}^2r^2}{\textbf{L}^2}\right)^4}\right)\,,
\end{equation}
where $a_2$ is a dimensionless constant and $\textbf{L}$ is the boundary surface of the star. Equation (\ref{eq:Delta11v}) shows that the metric potential $\nu$ has a finite value at the center of the star and at the boundary surface.
Using Eq.~(\ref{eq:Delta11v}) in Eq.~(\ref{eq:Delta1v}) we get:
%\newpage
\begin{eqnarray}\label{eq:Delta1c}
&\Delta(r) ={\frac {{a_2}^{4}{r}^{2} \left( 6\,{\textbf{L}}^{4}-8\,{a_2}^{2}{r}^{2}{\textbf{L}}^{2}+3\,{r}^{4}{a_2}^{4} \right) }{{\kappa}^{2} \left( 2\alpha_1+1 \right) {\textbf{L}}^{8}}}\nonumber\\
&+\frac {1 }{4 \kappa\left(1+2\,\alpha_1 \right) r{\textbf{L}}^{8}}\bigg[\bigg\{ \mu'^2r{\textbf{L}}^{2}- \mu'^2{r}^{3}a_2{}^{2}-6\mu'a_2{}^{2}{r}^{2} -2\mu' {\textbf{L}}^{2}+2\mu''r{\textbf{L}}^{2}  -2\mu''{r}^{3}a_2{}^{2} \bigg\}\left(
\textbf{L}-a_2r \right) ^{3} \left( \textbf{L}+a_2r \right) ^{3}\bigg].
\end{eqnarray}
Now, we assume the anisotropy to have the form:
\begin{eqnarray}\label{eq:Delta11c}
&\Delta(r) ={\frac {{a_2}^{4}{r}^{2} \left( 6\,{\textbf{L}}^{4}-8\,{a_2}^{2}{r}^{2}{R}^{2}+3\,{r}^{4}{a_2}^{4} \right) }{{\kappa}^{2} \left( 2\alpha_1+1 \right) {\textbf{L}}^{8}}}\,.
\end{eqnarray} Then if we use  Eq.~(\ref{eq:Delta11c}) in Eq. (\ref{eq:Delta1c}) we get:
\begin{eqnarray}\label{const}
&\frac {1 }{4 \kappa\left(1+2\,\alpha_1 \right) r{\textbf{L}}^{8}}\bigg[\bigg\{ \mu'^2r{\textbf{L}}^{2}- \mu'^2{r}^{3}a_2{}^{2}-6\mu'a_2{}^{2}{r}^{2}-2\mu' {\textbf{L}}^{2}+2\mu''r{\textbf{L}}^{2}  -2\mu''{r}^{3}a_2{}^{2} \bigg\}\left(
\textbf{L}-a_2r \right) ^{3} \left( \textbf{L}+a_2r \right) ^{3}\bigg]=0\,.
\end{eqnarray}
The solution of Eq.~(\ref{const}) gives the metric potential $\mu$  in the form:
\begin{eqnarray}\label{eq:Delta111}
&\mu(r) =\ln\left(\frac{\left(c_1+2c_2a_2{}^2\textbf{L}^2-2c_2a_2{}^4 r^2\right)^2}{16a_2{}^2(\textbf{L}^2-a_2{}^2 r^2)^2}\right)\,.
\end{eqnarray}
Eq.~(\ref{eq:Delta111}) shows that $c_2$ is a constant and $c_1$ is a dimensionalful parameter which we assume here to be $c_1=a_0\textbf{L}^2$ and we also assume $c_2=a_1$.
The two Eqs. (\ref{eq:Delta11v}) and (\ref{eq:Delta111}) are the two extra conditions that we need to make  the system given by  Eq. (\ref{sys}) in a closed form . The use of the two Eqs. (\ref{eq:Delta11v}) and (\ref{eq:Delta111}) in  (\ref{sys}) yields   the explicate form of the components of the energy momentum tensor.}
Now we define the formula of the mass content in a radius $r$  as:
\begin{equation} \label{mass}
    \mathbb{M}(r)=4 \pi \int_0^r \rho({r}) \, {r}{^2} dr.
\end{equation}
%Using the form of density from Eq. (\ref{sys}) in Eq. (\ref{mass}) we get:
%\newpage
%\begin{eqnarray}\label{eq:Mass}
 %   &\mathbb{M}(r)=\frac{4\pi \,{a_2}^{2}}{3 {\kappa}^{2} \left\{ 2\,\alpha_1+1 \right\} \left( 1+8\,\alpha_1\right) {\textbf{L}}^{8}{c}^{2}} \left\{ \frac{2}{7{a_1}}\left\{{\textbf{L}}^{2}{a_2}^{2} \left( 15\, \alpha_1\,a_0+42\,a_1\,{a_2}^{2}+224\,a_1 \,{a_2}^{2}\alpha_1 \right) {r}^{7}\right\}-{a_2}^{6} \left( 3+16\,\alpha_1 \right) {r}^{9}\right.\nonumber\\
    %
  %  &\left.+{\frac { 3{\textbf{L}}^{4}\alpha_1 \left( {a_0}^{2}-4a_1{a_2}^ {2}a_0-\frac{6{a_1}^{2}{a_2}^{4}}{\alpha_1}-32{a_1}^{2}{a_2}^{4} \right) {r}^{5}}{{a_1}^{2} {a_2}^{2}}}+{\frac {{\textbf{L}}^{6}\alpha_1 \left( 5{a_0}^{3}-10a_1{a_2}^{2}{a_0}^{2}+20 {a_1}^{2}{a_2}^{4}a_0+\frac{24{a_1}^ {3}{a_2}^{6}}{\alpha_1}+128{a_1}^{3}{a_2}^{6} \right) {r}^{3}}{2{a_1}^{3}{a_2}^{6}}}\right.\nonumber\\
    %
   % &\left.+{\frac {15}{4}}\,{ \frac {{a_0}^{4}{\textbf{L}}^{8}\alpha_1\,r}{{a_1}^{4}{a_2 }^{10}}}-{\frac {15}{8}}\,{a_0}^{4}{\textbf{L}}^{9}\alpha_1\, \left( a_0+2\,a_1\,{a_2}^{2} \right) \sqrt {2}{\arctanh} \left( {\frac {a_1\,{a_2}^{2}r\sqrt {2}}{\textbf{L}\sqrt { \left( a_0+2\,a_1\,{a_2}^{2} \right) a_1}}} \right){\frac {1}{ { a_1}^{4}{a_2}^{12}\sqrt { \left( a_0+2\, a_1\,{a_2}^{2} \right) a_1}}} \right\}\,.
%\end{eqnarray}
%We note that the mass function (\ref{eq:Mass}) reduces to the GR version when $\alpha_1=0$ \citep{Nashed:2020kjh}.
%%%%%%%%%%%%%%%%%%%%%%%%%%%%%%%%%%%%%%%%%%%%%%%%%%%%
\subsection{Matching conditions}\label{Sec:Match}
Since the exterior solution of $f(R,T)$ is  the Schwarzschild one because the vacuum solutions of GR and $f(R,T)$ are equivalent \citep{Ganguly:2013taa,Sheykhi:2012zz} then we use the exterior spacetime in the form:
\begin{equation}
 {\mathrm   ds^2=-\left(1-\frac{2GM}{c^2r}\right) c^2 dt^2+\frac{dr^2}{\left(1-\frac{2GM}{c^2 r}\right)}+r^2 (d\theta^2+\sin^2 \theta d\phi^2)}\,.
\end{equation}
{\bf Using the interior solution given by Eq.~ (\ref{eq:metric}) to match it with Schwarzschild  solution which is given by\footnote{{\bf In this study and due to the specific form of $f(R,T)=R+\alpha T$ we will use the exterior solution of Schwarzschild spacetime because it is the only solution for the specific forms of  $f(R,T)= R$.}}
\begin{equation}\label{eq:bo}
{\mathbf {\mathrm   \mu(r=\textbf{L})=\ln(1-C)\,, \qquad \qquad  \mu(r=\textbf{L})=-\ln(1-C)}}\,,
\end{equation}}
where  ${\textrm C}$  is the compactness parameter defined as:
\begin{eqnarray}\label{comp11}
C=\frac{2GM}{c^2 \textbf{L}}\,.
\end{eqnarray}
Additionally, we assume the radial pressure to be  vanishing  on the surface of the star, i.e.
\begin{equation}
   \mathrm{{p}_1}(r=\textbf{L})=0\,.
\end{equation}
{ Using the  ansatz given by Eqs.~ (\ref{eq:Delta11v}) and (\ref{eq:Delta111}) and the radial pressure given by Eq.~(\ref{sys}) in addition to the aforementioned boundary conditions, we determine the shape of the dimensional constant .  $a_0$,  $a_1$,  and $a_2$,  in terms of the compactness. Due to the lengthy of the forms of these constants we  display them in the Equation (\ref{df1}) of Appendix~\ref{A}.}

{ The most interesting thing is the fact that all the physical quantities of the above solution, Eq. (\ref{sys}) after using Eq. (\ref{eq:Delta11v}) and (\ref{eq:Delta111}), and  for the pulsar, $0\leq \frac{r}{\textbf{L}}=x\leq 1$, can be expressed in dimensionless forms using the dimensionless parameter $\alpha_1$ and the compactness parameters, specifically $\rho(\alpha_1, C)$, ${p}_1(\alpha_1, C)$, and ${p}_2(\alpha_1, C)$.} In practical terms,  the compactness  is fixed using observational data, which provides a direct limit for estimating the non-minimal coupling between geometry and  matter. Moreover, this approach offers a means to determine an upper bound for the allowable compactness of the anisotropic neutron star and. This aspect explored in subsequent sections.

%%%%%%%%%%%%%%%%%%%%%%%%%%%%%%%%%%%%%%%%%%%%%% Section 4 %%%%%%%%%%%%%%%%%%%%%%%%%%%%%%%%%%%%%%%%%%%%%%%%%%%%%%%%%%%%%%%
\section{Observational constrains and the stability conditions}\label{Sec:Stability}
Now we are going to use  the pulsar ${\textit HerX1}$   whose  mass and radius are given as $M=0.85\pm 0.15 M_\odot$ and $\textbf{L}=8.1\pm0.41$ km \citep{Legred:2021hdx} where the mass is given by $M_\odot=1.9891\times 10^{30}$ kg. In order for a star to exhibit proper physical behavior, it must meet the following conditions, enumerated from (\textbf{1}) to (\textbf{12}) as outlined below:
%%%%%%%%%%%%%%%%%%%%%%%%%%%%
\subsection{Matter sector}\label{Sec:matt}
\begin{figure*}
\centering
\subfigure[~The energy-density]{\label{fig:density}\includegraphics[scale=0.25]{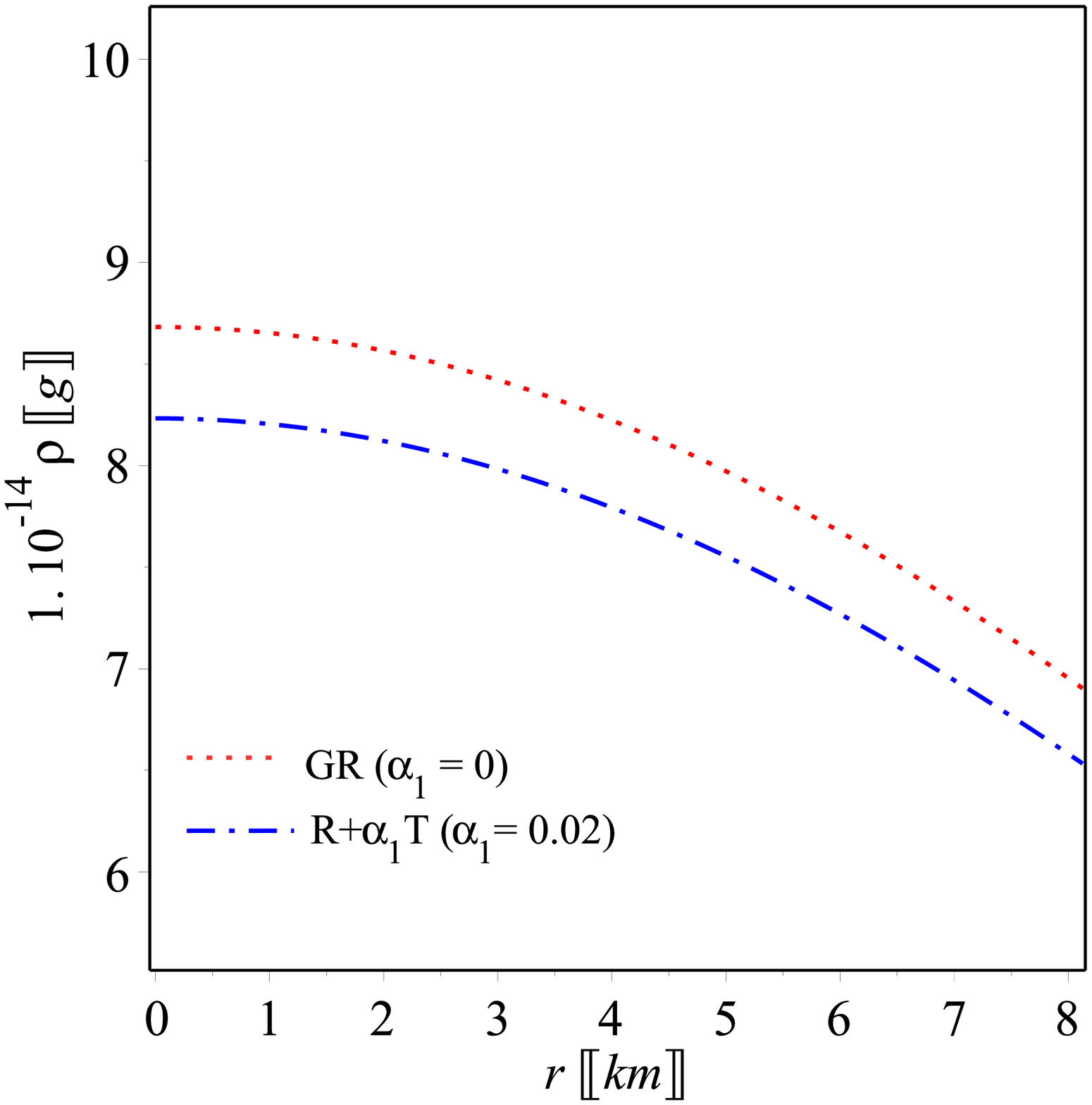}}
\subfigure[~The radial pressure]{\label{fig:radpressure}\includegraphics[scale=0.25]{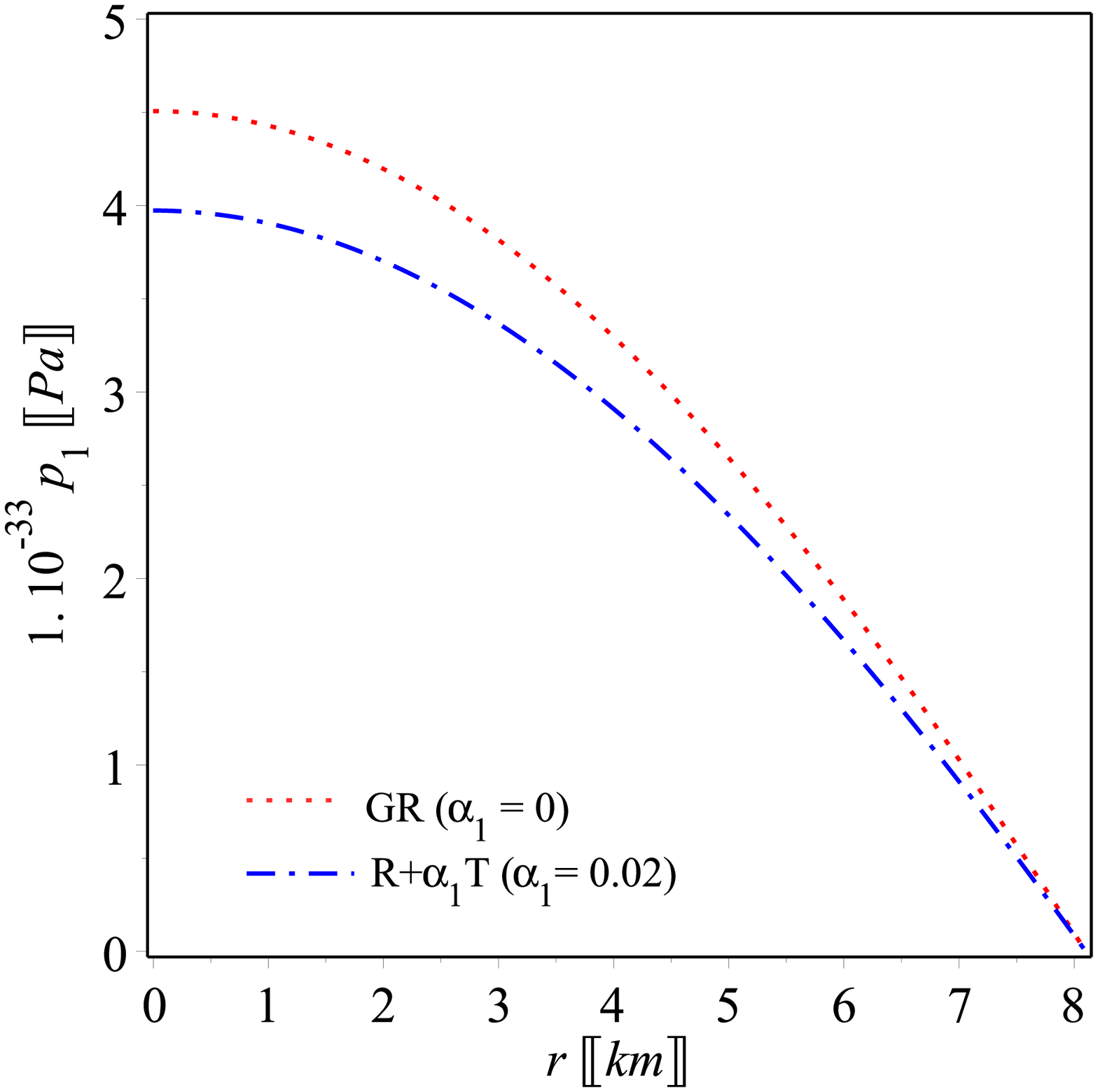}}
\subfigure[~The tangential pressure]{\label{fig:tangpressure}\includegraphics[scale=0.25]{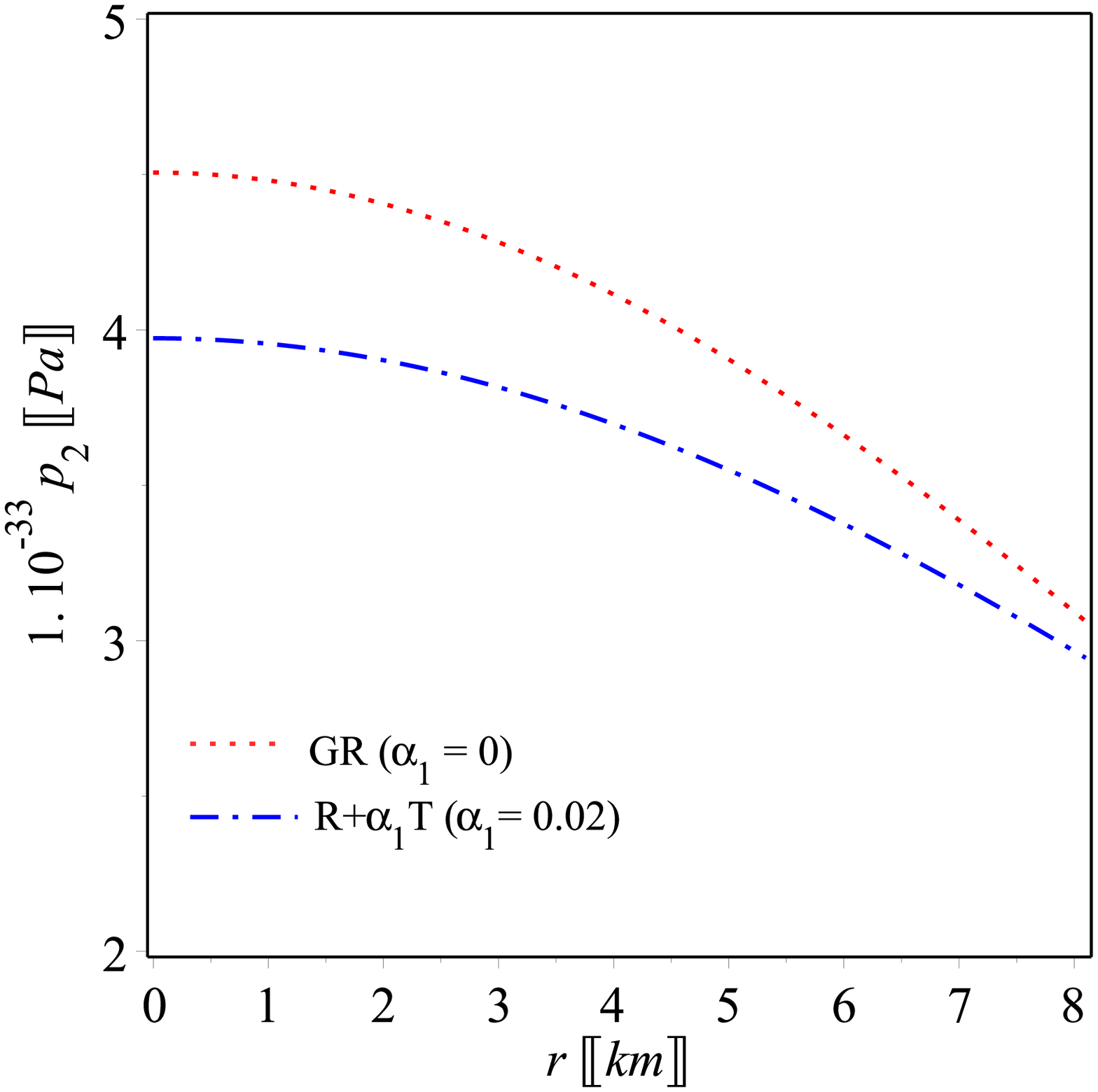}}\\
\subfigure[~The gradients (GR)]{\label{fig:GRgrad}\includegraphics[scale=0.25]{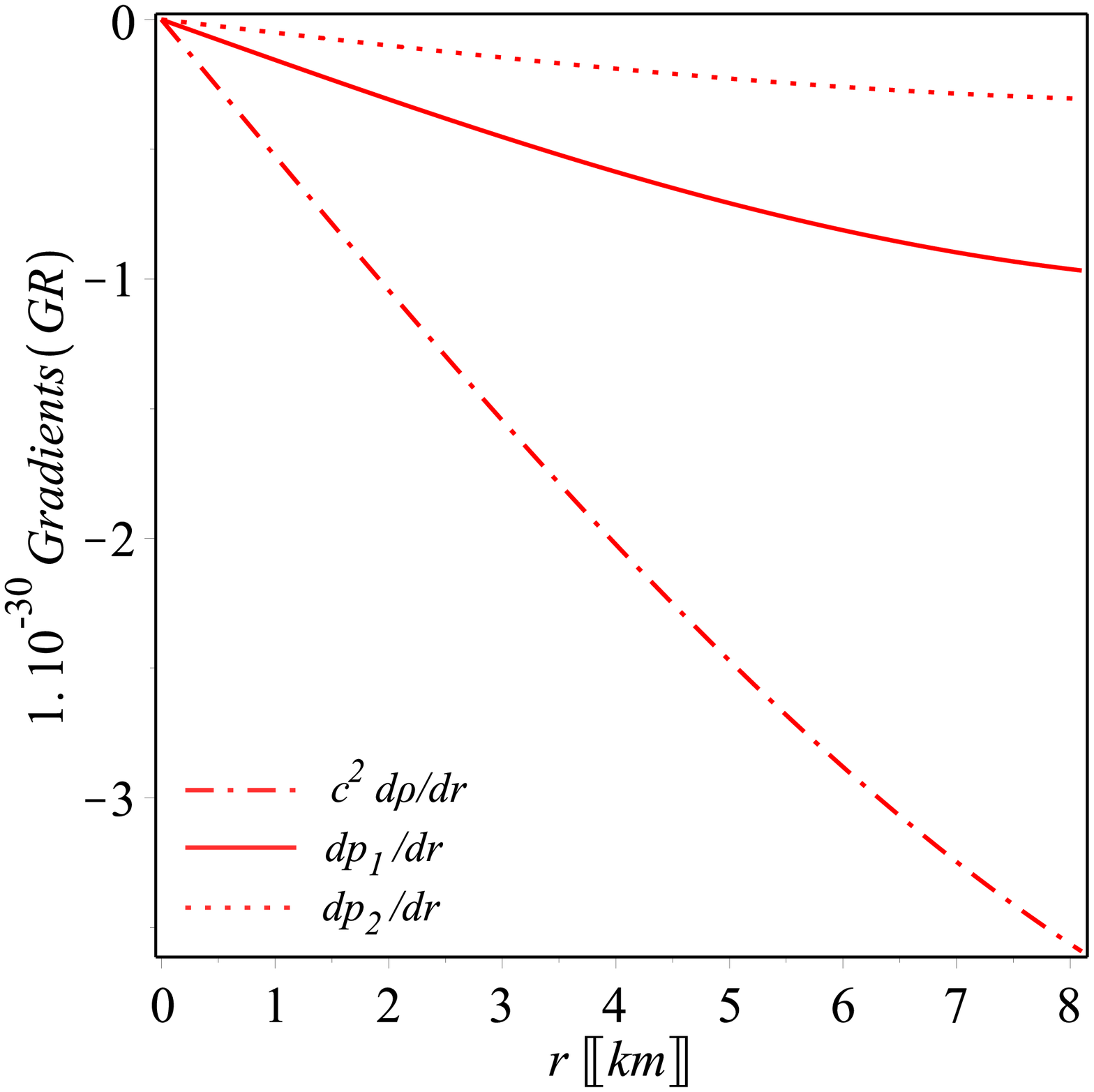}}
\subfigure[~The gradients f(R,T)]{\label{fig:RTgrad}\includegraphics[scale=0.25]{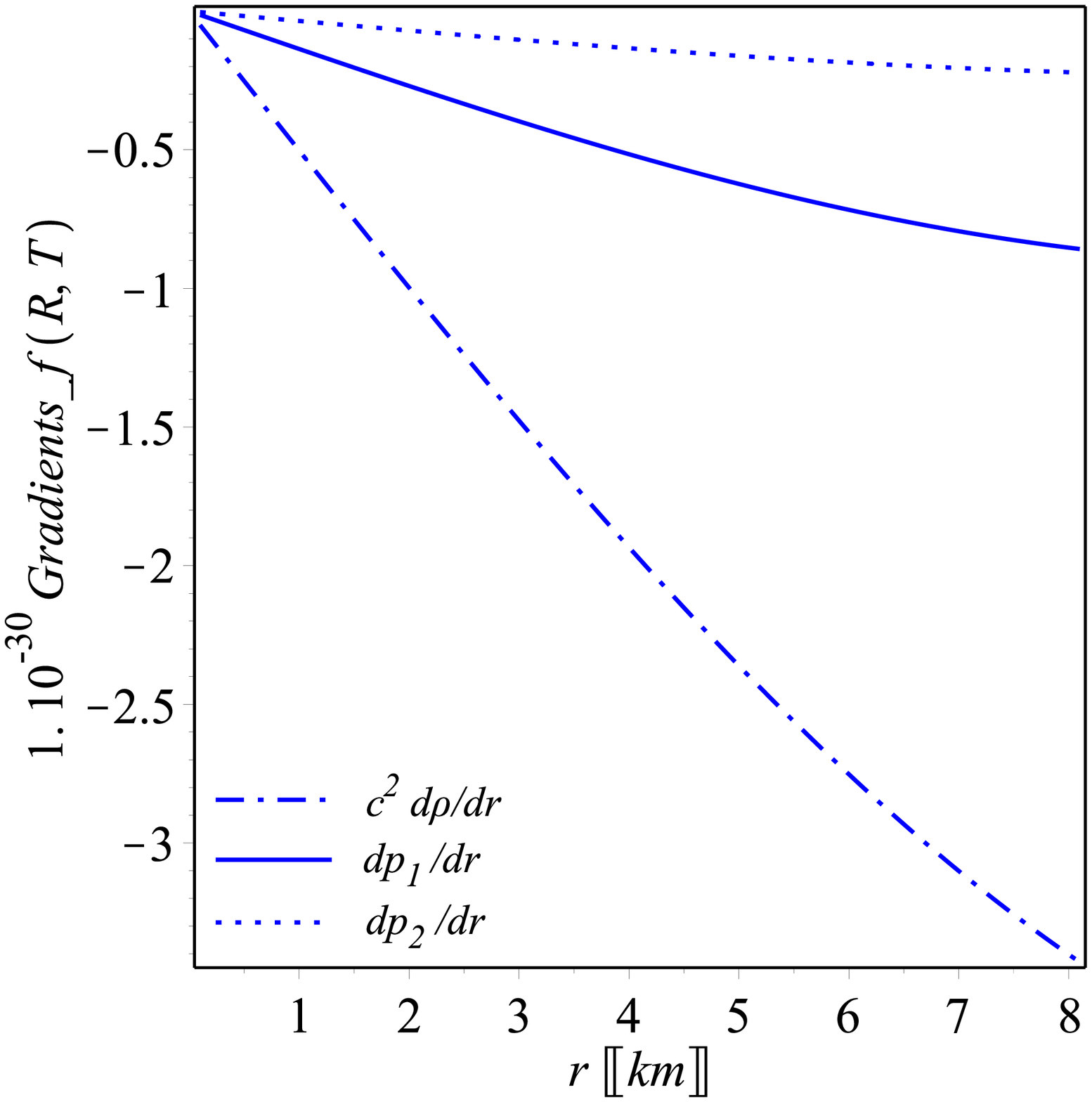}}
\subfigure[~The anisotropy (GR)]{\label{fig:anisotg}\includegraphics[scale=0.25]{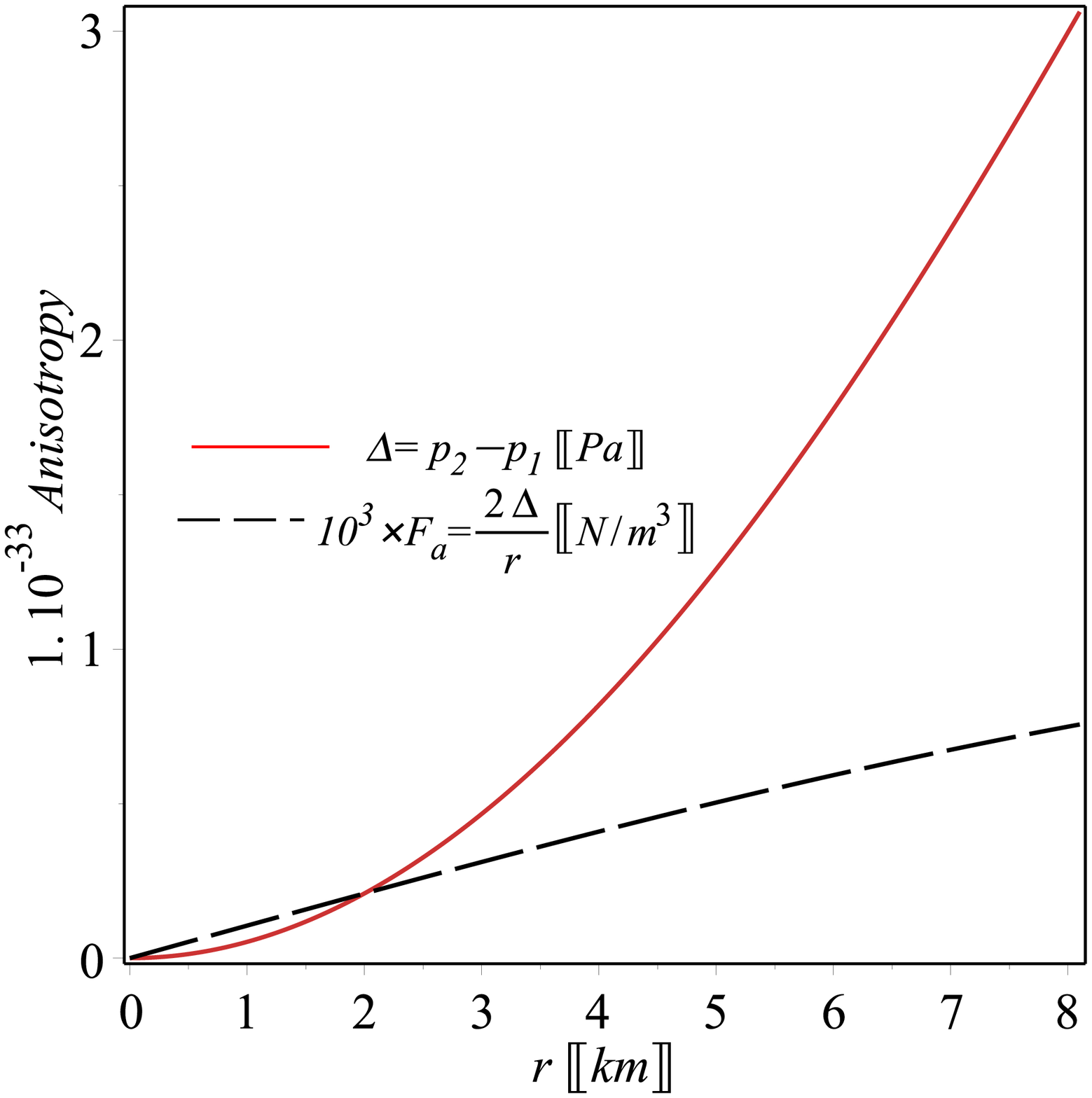}}
\subfigure[~The anisotropy f(R,T)]{\label{fig:anisotf}\includegraphics[scale=0.25]{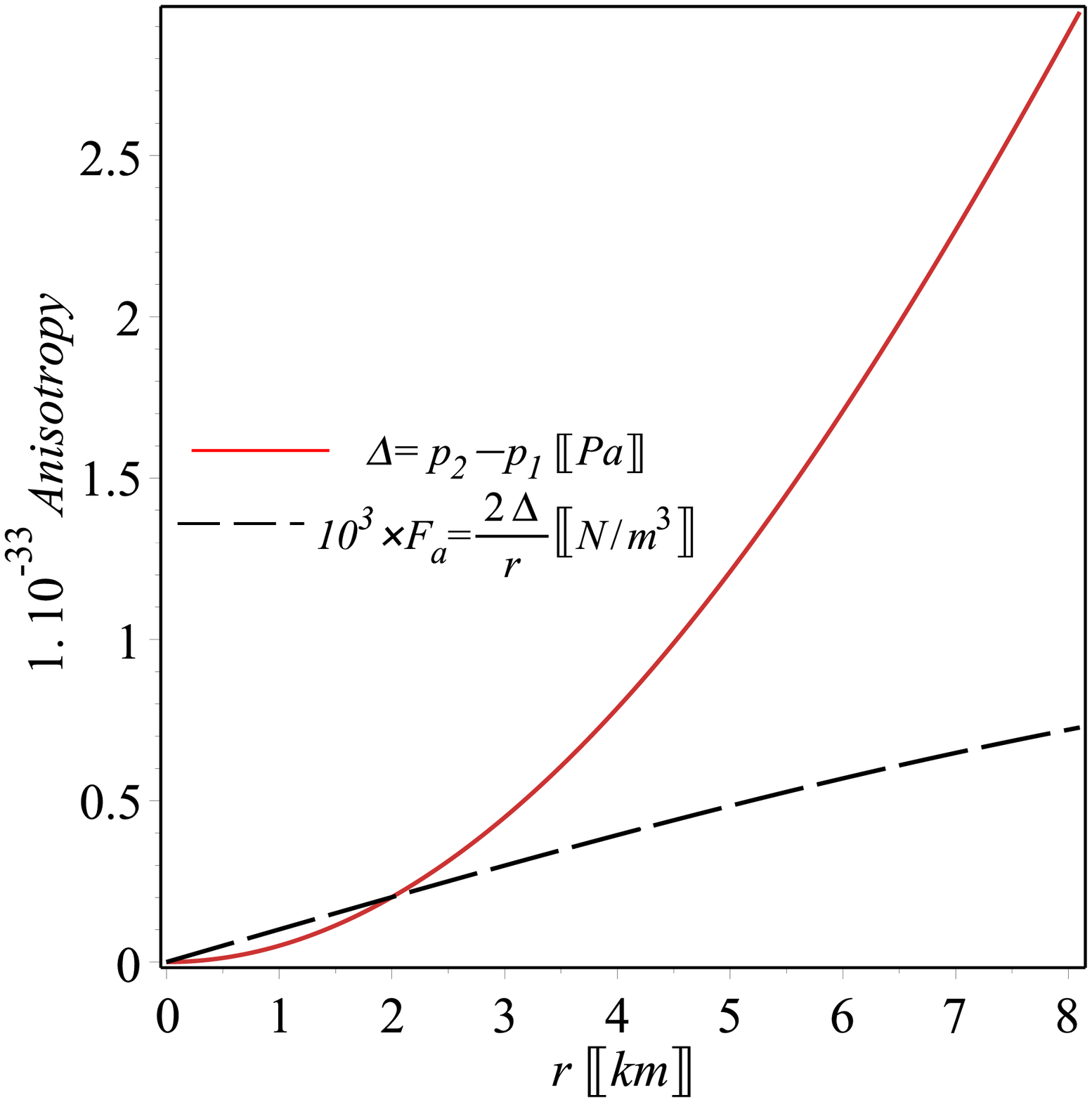}}
\caption{The behaviors of the components of the energy-tensor of Eqs. (\ref{sys}), after using Eq. (\ref{eq:Delta11v}) and (\ref{eq:Delta111}), of the pulsar HerX1 are depicted in  Figs. \subref{fig:density}--\subref{fig:tangpressure}. The  derivatives of the energy-momentum  components are depicted in Figs. \subref{fig:GRgrad} and \subref{fig:RTgrad} for  $f(R,T)$ and GR. Figures \subref{fig:anisotg} and \subref{fig:anisotf} represent  the behavior of the { anisotropy function} $\Delta$ for GR and $f(R,T)$ cases which are given analytically by  Eq. (\ref{eq:Delta11c}). All the plots displayed in Fig. \ref{Fig:dens_press} ensure that conditions listed from  (\textbf{1}) to (\textbf{4}) are verified.}
\label{Fig:dens_press}
\end{figure*}
\noindent Condition (\textbf{1}):  It is well known that any interior solution must have  a  regular behavior, thus the components of the energy-momentum  of the fluid under consideration must be regular at the center of the star and behave regularly   everywhere inside the stellar. { Moreover, such  physical quantities should have maximum values at the center of the star and behave in a  decreasing  way towards the boundary of the stellar as shown in Fig.\ref{Fig:dens_press} \subref{fig:density}--\subref{fig:tangpressure}.}
%\begin{itemize}
%\item[i-] ${\mathrm \rho(r=0)>0}$, \quad ${\mathrm\rho'(r=0)=0}$, \quad ${\mathrm \rho''(r=0)<0}$,\quad and \quad ${\mathrm \rho'(0< r \leq \textbf{L})< 0}$\,,
 % \item[ii-] ${\mathrm p_1(r=0)>0}$, \quad ${\mathrm p'_1(r=0)=0}$, \quad ${\mathrm p''_1(r=0)<0}$, \quad and \quad ${\mathrm p'_1(0< r \leq \textbf{L})< 0}$\,,
  %\item[iii-] ${\mathrm p_2(r=0)>0}$, \quad ${\mathrm p'_2(r=0)=0}$, \quad ${\mathrm p''_2(r=0)<0}$, \quad and \quad ${\mathrm p'_2(0< r \leq \textbf{L})< 0}$\,.
%\end{itemize}

\noindent Condition (\textbf{2}): The components of the energy momentum tensor in the star ($0 < r < \textbf{L}$), must  be non-negative, namely  ${\mathrm \rho(0 < r < \textbf{L})>0}$, ${\mathrm  p_1(0 < r < \textbf{L})>}0$ and ${\mathrm  p_2(0 < r < \textbf{L})>0}$.

\noindent Condition (\textbf{3}): The  pressure in the radial direction of the fluid   must have zero value at the boundary   of the star, i.e. ${\mathrm  p_1(r=\textbf{L})=0}$. On the contrary,  the  pressure in the tangential direction is not necessarily to be zero at the boundary of the star.

 Conditions from (\textbf{1})--(\textbf{3}) \textit{are verified for the pulsar ${\textit HerX1}$ as indicated  in Fig. \ref{Fig:dens_press}\subref{fig:density}--\subref{fig:RTgrad}}.

As shown in Fig. \ref{Fig:dens_press} \subref{fig:density}  that the model being considered provides an estimation of the core density of NS as ${\mathrm \rho_{core}\approx 8.2\times 10^{14}>}$ g/cm$^{3} \approx 3.1\rho_{nuc}$ for the pulsar ${\textit HerX1}$. This implies that the model being studied does not rule out the likelihood of the pulsar's core being composed of neutrons.  Moreover,  this value of the density (at the core) makes the assumption of the anisotropy form of the  fluid  to be a logic one.

\noindent Condition (\textbf{4}): The { anisotropy function } $\Delta$  should have a zero value  at the core namely ${\mathrm p_1(r=0)=p_2(r=0)}$, and growing in the direction of the surface, i.e. $\Delta'(0 \leq r\leq \textbf{L})>0$. Therefore, the anisotropic force ${\mathit F_a=\frac{2\Delta}{r}}$ must have a zero value at the center. As Eq. (\ref{eq:Delta11c}) shows that the limit $r\to 0$  yields $\Delta\to 0$.

 As shown in Fig. \ref{Fig:dens_press} \subref{fig:anisotg} and \ref{Fig:dens_press}\subref{fig:anisotf} that condition  (\textbf{4}) \textit{is satisfied  for the pulsar ${\textit HerX1}$}.

Moreover, one can show that the {\bf anisotropy function} ${\mathrm \Delta(r>0) >} 0$ which means that  $\mathrm {p_2>p_1}$.   This condition is essential to ensure that the anisotropy force acts as a repulsive force, thereby enabling a larger size for the neutron star in comparison to the situation of an isotropic perfect fluid. Nevertheless, when  $\alpha_1>0$, the level of anisotropy in $f(R,T)=R+ \alpha_1T$ is  lower compared to the case of GR.
%%%%%%%%%%%%%%%%%%%%%%%%%%%
\subsection{Zeldovich condition}
\noindent Condition (\textbf{5}): According to  the study presented in \citep{1971reas.book.....Z}, at the center of the star the radial  pressure   is required to be either  less than or equal to the central energy  density, namely
\begin{equation}\label{eq:Zel}
  \mathrm  {\frac{{p}_1(0)}{{\rho}(0)}\leq 1.}
\end{equation}
We obtain the components of the energy momentum tensor at the center as:
\begin{eqnarray}
 &\mathrm {{\rho}(r=0) =4{\frac { \left( 3a_0+6a_1{a_2}^{2}+32a_1{a_2}^{2}\alpha_1+21\alpha_1a_0 \right) { a_2}^{2}}{{\textbf{L}}^{2}{\kappa}^{2} \left( 1+2\alpha_1 \right) \left( 8\alpha_1+1 \right)  \left( a_0+2a_1{a_2}^{2} \right) {c}^{2}}}}\,, \nonumber \\
 & \mathrm {p_1(r=0) ={\frac {12a_2^{2}\alpha_1a_0-8a_1a_2^{4}}{{\textbf{L}}^{2}{\kappa}^{2} \left( 10\alpha_1+1+16\alpha_1{}^{2} \right)  \left( a_0+2a_1a_2^{2} \right) }}= {p}_2(r=0)}.\quad
&\end{eqnarray}
{ For the pulsar ${\textit HerX1}$ we calculate the compactness  $C =0.3100165206+\pm 0.0547087977$. Therefore, by applying the Zeldovich condition (\ref{eq:Zel}), we can establish a valid range for the parameter $\alpha_1$ as $0\leq \alpha_1 \leq 0.02$, with the understanding that as $\alpha_1$ approaches zero, it is expected to converge to the GR version.}
%%%%%%%%%%%%%%%%%%%%%%%%%%%%%%%%%%%%%%%%%%
\subsection{The radius and mass observational limits using pulsar  ${\textit HerX1}$ }
%$M=0.85\pm 0.15 M_\odot$ and $R=8.1\pm0.41$
\begin{figure}
\centering
\subfigure[~The Mass function]{\label{Fig:Mass}\includegraphics[scale=0.25]{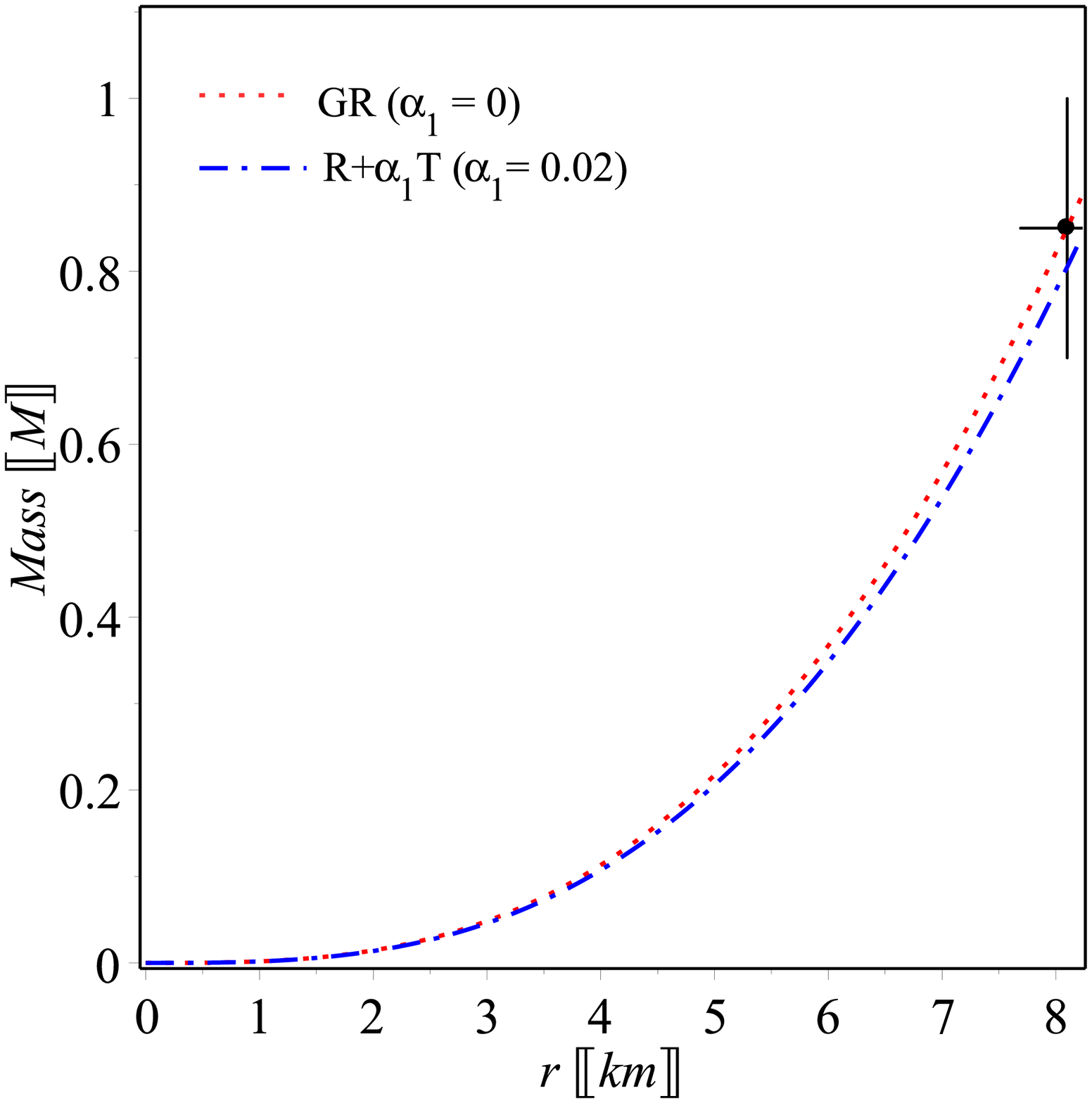}}
\subfigure[~The Compactness parameter]{\label{Fig:Comp}\includegraphics[scale=0.25]{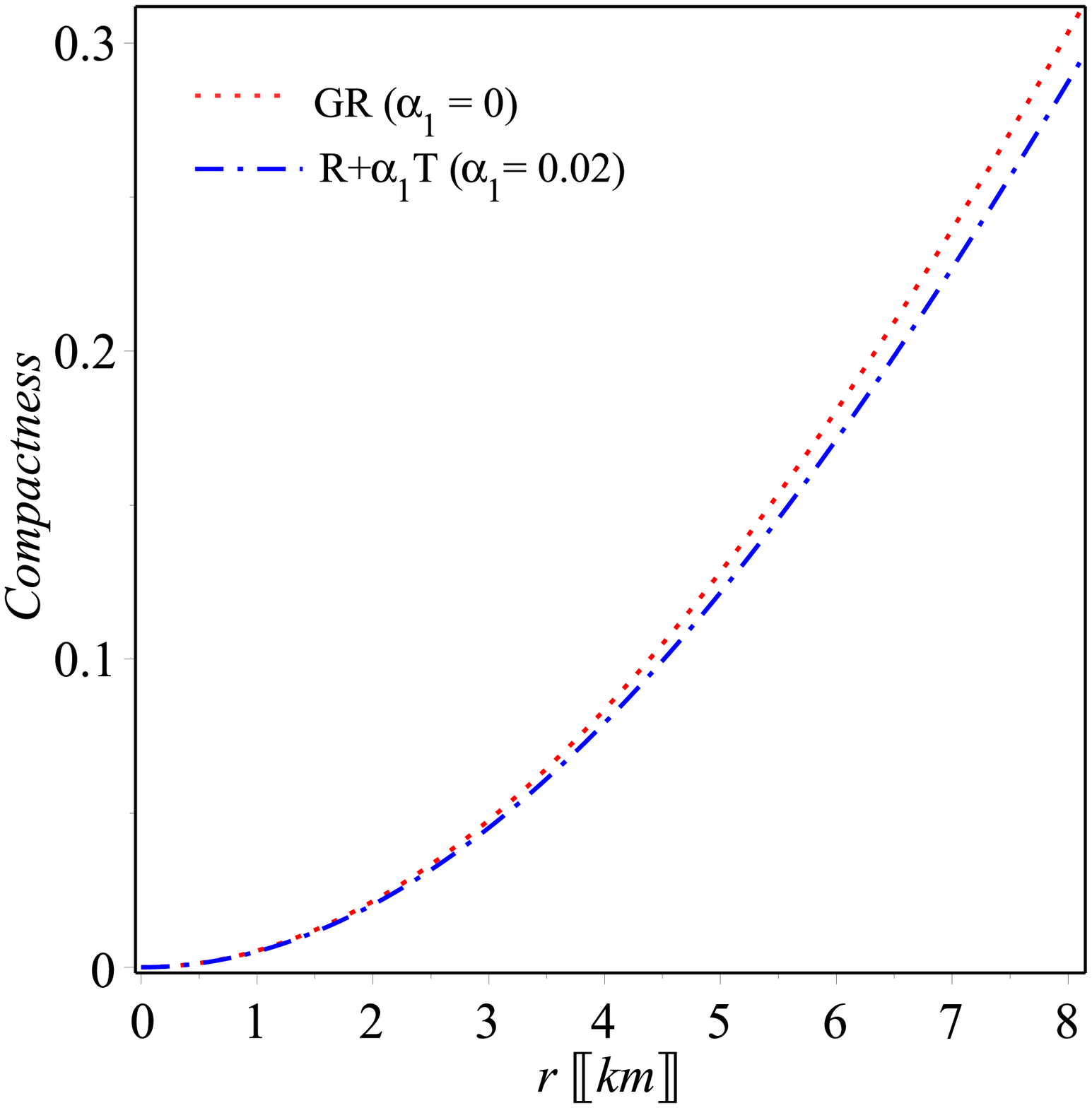}}
\caption{The figures of the mass function  (\ref{mass}) and compactness parameter  (\ref{comp11}) for the pulsar  ${\textit HerX1}$. The curves are in good consistent  with the observational data ($M=0.85\pm 0.15 M_\odot$ and $\textbf{L}=8.1\pm0.41$ km) \citep{Legred:2021hdx}. For the case  $f(R,T)=R+\alpha_1\,T$   we employ \{$\alpha_1=0.02$, $\kappa=2.302\times 10^{-43}\,N^{-1}$, $C=0.3100165206$, $a_0 =0.2966665917$, $a_1 =-0.175664532$, $a_2 =0.2976552341$\} and for the case of GR  we put  ($\alpha_1=0$) and their corresponding values are \{ $a_0 =0.3100165210$, $a_1 =  -0.258327738$\}.}
\label{Fig:Mass1}
\end{figure}
Taking into account the limitations on the mass and radius of the pulsar ${\textit HerX1}$,   the mass function (\ref{mass}) approximates   $M=0.85 M_\odot$ at $\textbf{L} =8.1$ km with a compactness $C=0.3100165206$ in consistent  with the observed value, ($M=0.85 \pm 0.15 M_\odot$ and $\textbf{L}=8.1\pm0.41$ km) \citep{Legred:2021hdx}, when we choose $\alpha_1$ parameter $\alpha_1=0.02$. This fixes  the set of constants presented in Eq.~(\ref{df1}) of Appendix~\ref{A} \{ $a_0 =0.2966665917$, $a_1 =-0.175664532$, $a_2 =0.2976552341$,$\kappa=2.302\times 10^{-43}\,N^{-1}$\}. These numerical values clearly verify the Zeldovich condition (\ref{eq:Zel}). We depict the patterns of the mass function and compactness parameter in Fig. \ref{Fig:Mass1} \subref{Fig:Mass} and \subref{Fig:Comp}  to illustrate the concurrence between the anticipated mass-radius relationship of the pulsar ${\textit HerX1}$ and the observed measurements. Comparison with  GR prediction, $f(R,T)=R+\alpha_1\,T$ with positive value of the dimensional parameter $\alpha_1$ predicts   same mass like the GR however,  within a star of greater size (or reduced mass at equivalent size). Hence, the model $f(R,T)=R+\alpha_1,T$ predicts a lower compactness value compared to GR for a given mass. This implies that the function $f(R,T)=R+\alpha_1,T$ can handle greater masses or, in other words, higher levels of compactness while still maintaining stability requirements.
%We are going to discuss this issue  in more details in Sec.  \ref{MR}. Remarkably, for $\alpha_1 < 0$, the matter-geometry coupling turns the star to be slightly has  small   sizes in comparison with GR for the same mass \citep{Nashed:2022zyi}; and thus we regard away this choice in this study.
%%%%%%%%%%%%%%%%%%%%%%%%%%%
\subsection{Geometric sector}\label{Sec:geom}
\noindent Condition (\textbf{6}):In the interior region of the stellar object, ranging from $0$ to $\textbf{L}$, it is essential that the geometric sector, specifically the metric potentials ${\textit g_{tt}}$ and ${\textit g_{rr}}$, do not exhibit any coordinate or physical singularities. Clearly, the metric (\ref{eq:metric}) verifies these conditions since at the core, ${\mathrm g_{tt}(r=0)=-\frac { \left( a_0+2\,a_1\,{a_2}^{2} \right) ^
{2}}{16{a_2}^{4}}}$ and ${\mathrm g_{rr}(r=0)=1}$, and both are well defined in the interior of the stellar $0 \leq r/\textbf{L} \leq 1$.

\noindent Condition (\textbf{7}): The  interior  and the exterior solution of the metric potentials must match smoothly at the surface of the star.
\begin{figure}
\centering
{\includegraphics[scale=0.3]{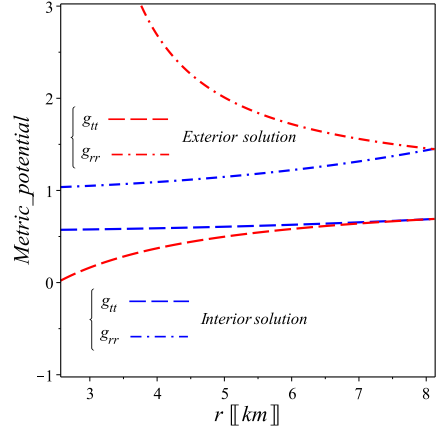}}
\caption{Inside the pulsar ${\textit HerX1}$, the metric potentials $g_{00}$ and $g_{rr}$ have finite values, and they seamlessly align with the Schwarzschild exterior vacuum solution. The graph guarantees the fulfillment of conditions (\textbf{6}) and (\textbf{7}).}
\label{Fig:Matching}
\end{figure}

\textit{Clearly conditions} (\textbf{6}) and (\textbf{7}) \textit{are verified for the pulsar ${\textit HerX1}$ as indicated in Fig. \ref{Fig:Matching}}.

\noindent Condition (\textbf{8}): The red-shift of the metric potential (\ref{eq:metric}) is defined as:
\begin{equation}\label{eq:redshift}
    Z=\frac{1}{\sqrt{-g_{tt}}}-1=\frac{1}{\sqrt{\frac { \left( a_0\,{\textbf{L}}^{2}+2\,a_1\,{a_2}^{2}{
\textbf{L}}^{2}-2\,a_1\,{a_2}^{4}{r}^{2} \right) ^{2}}{16  \left( \textbf{L}^2-a_2{}^2\,r^2 \right) ^{2}{a_2}^{4}}
}}-1.
\end{equation}
It is known that the gravitational red-shift should be  positive and has a finite value everywhere within the interior of the stellar and decreases monotonically in the direction of  the surface, namely,  $Z'<0$ and $Z>0$.
\begin{figure}
\centering
{\includegraphics[scale=0.3]{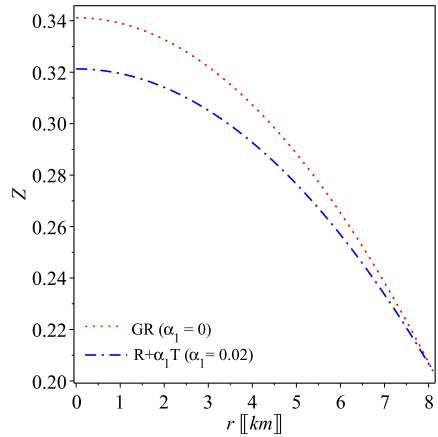}}
\caption{The red-shift function (\ref{eq:redshift}) of the pulsar ${\textit HerX1}$. Fig. \ref{Fig:Redshift} ensures that condition (\textbf{8}) is verified.}
\label{Fig:Redshift}
\end{figure}

\textit{Clearly condition} (\textbf{8}) \textit{is verified  as displayed in Fig. \ref{Fig:Redshift}}.

At the center of the star, the value of $Z(0)$ is approximately $0.3326085846$, which is lower than the corresponding value in GR ($Z(0)\approx 0.3390521449$). However, at the boundary, $Z_\textbf{L}$ is approximately $0.2038729440$, which is similar to the GR value. Notably, this value of $Z_\textbf{L}$ remains below the upper red-shift limit of $Z_\textbf{L}=2$, as derived by Buchdahl (1959) \citep{Buchdahl:1959zz}. For further exploration of the anisotropic nature, one can refer to studies by Ivanov and Sibgatullin (2002) \citep{Ivanov:2002xf} and Barraco et al. (2003) \citep{Barraco:2003jq}, as well as the study by Boehmer et al. (2006) that includes a cosmological constant \citep{Boehmer:2006ye}. We observe that the maximum red-shift limit is not a highly restrictive condition, and therefore it does not provide a suitable means to determine a maximum limit on the compactness, as previously discussed in works such as those by  \citep[c.f.,][]{Ivanov:2002xf,Barraco:2003jq,Boehmer:2006ye}. Nevertheless, the scenario shifts when the energy conditions related to the matter configuration  are taken into account, imposing more stringent limitations. This was demonstrated in the investigation conducted by Roupas (2020) \citep{Roupas:2020mvs}.
%%%%%%%%%%%%%%%%%%%%%%%%%%%%%%%%%%%%%%
\subsection{The energy conditions}\label{Sec:Energy-conditions}
The focusing theorem within the framework of GR states that in the Raychaudhuri equation, the trace of the tidal tensor, given by ${\mathrm{R}{\alpha\beta} u^{\alpha} u^{\beta} \geq 0}$ and ${\mathrm {R}{\alpha\beta} \ell^{\alpha} \ell^{\beta} \geq 0}$, is always positive. Here, ${\mathrm u^{\alpha}}$ represents any arbitrary timelike vector, and ${\mathrm \ell^{\alpha}}$ represents null vector.  This gives rise to four energy conditions that impose constraints on   ${\mathrm \mathbb{T}^{\alpha\beta}}$, which are applicable not only in the context of General Relativity but can also be extended to modified gravitational theories. Specifically, in the case of $f(R,T)=R+\alpha_1T$ gravity, these energy conditions can be reformulated in terms of the energy-momentum tensor. ${\mathrm {{T}}{^\alpha}{_\beta}}=\mathrm{diag(-\rho c^2,p_1, p_2, p_2)}$, since ${\mathrm \mathbb{R}_{\alpha\beta}=\kappa\left({{T}}_{\alpha\beta}-\frac{1}{2} g_{\alpha\beta} {{T}}\right)}$.\\
\begin{figure*}[t!]
\centering
\subfigure[WEC \& NEW (radial)]{\label{fig:Cond1}\includegraphics[scale=0.29]{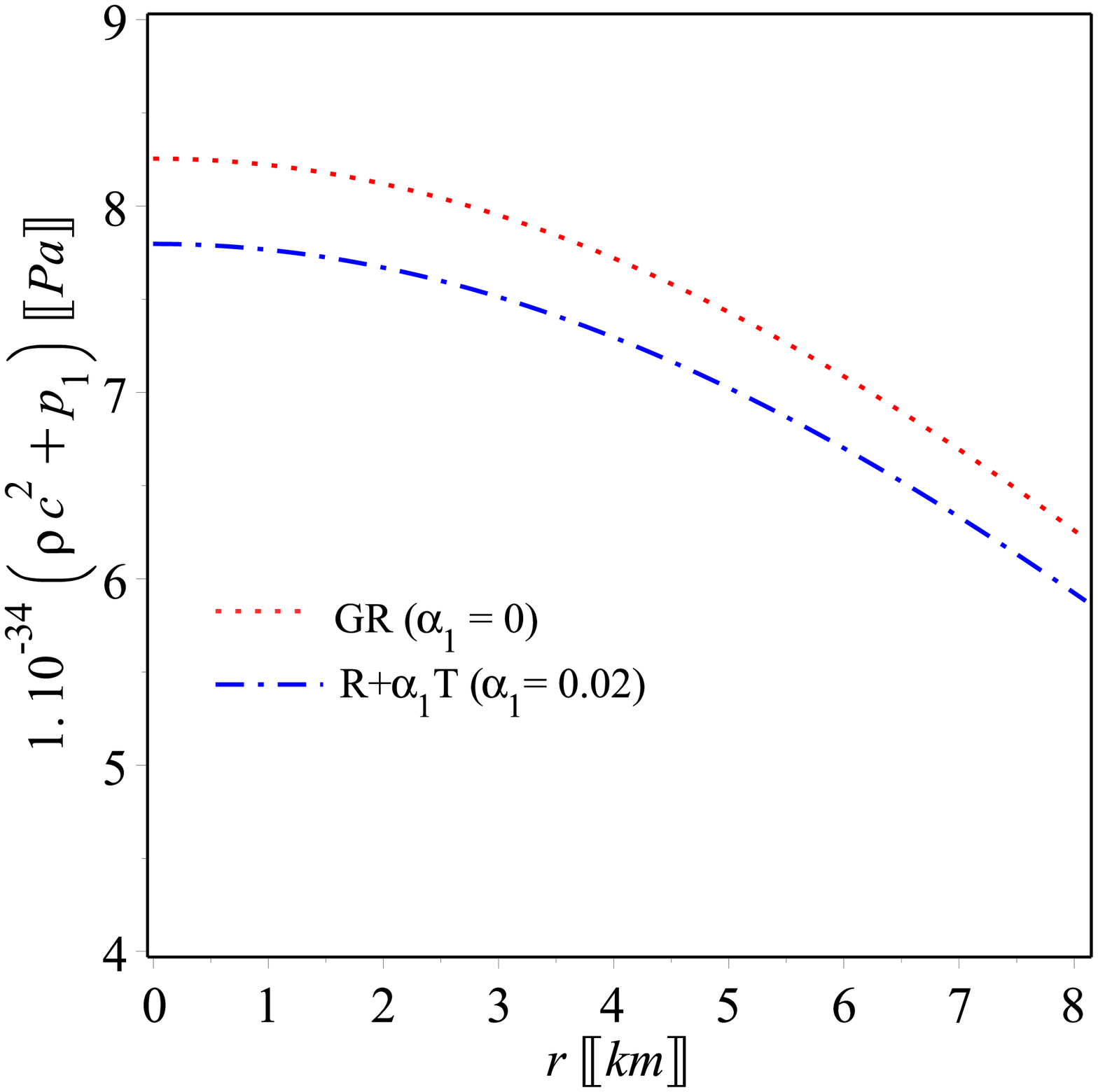}}\hspace{0.1cm}
\subfigure[WEC \&NEC (tangential)]{\label{fig:Cond2}\includegraphics[scale=0.29]{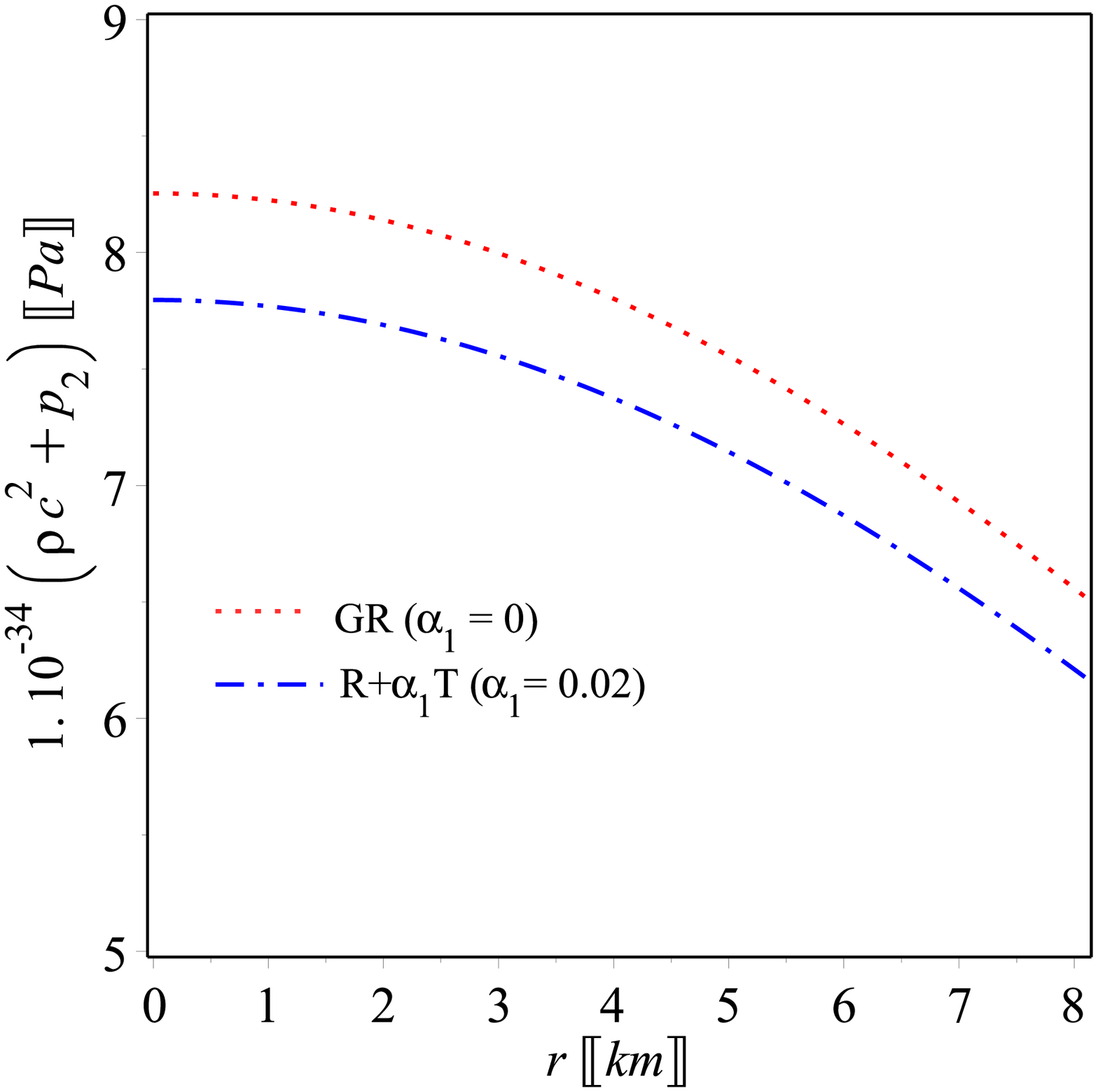}}\hspace{0.1cm}
\subfigure[SEC]{\label{fig:Cond3}\includegraphics[scale=.29]{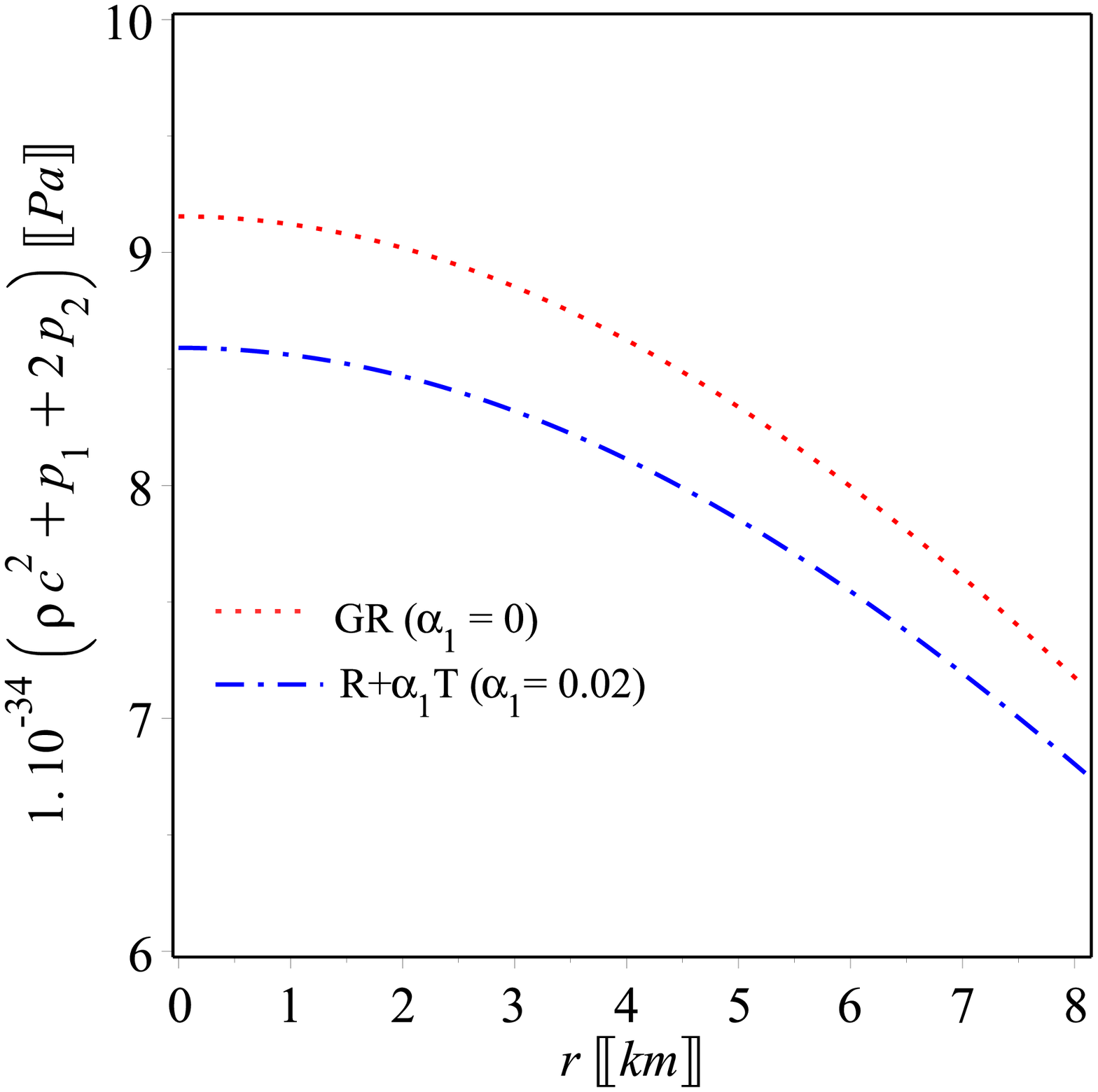}}
\subfigure[~DEC (radial)]{\label{fig:DEC}\includegraphics[scale=.23]{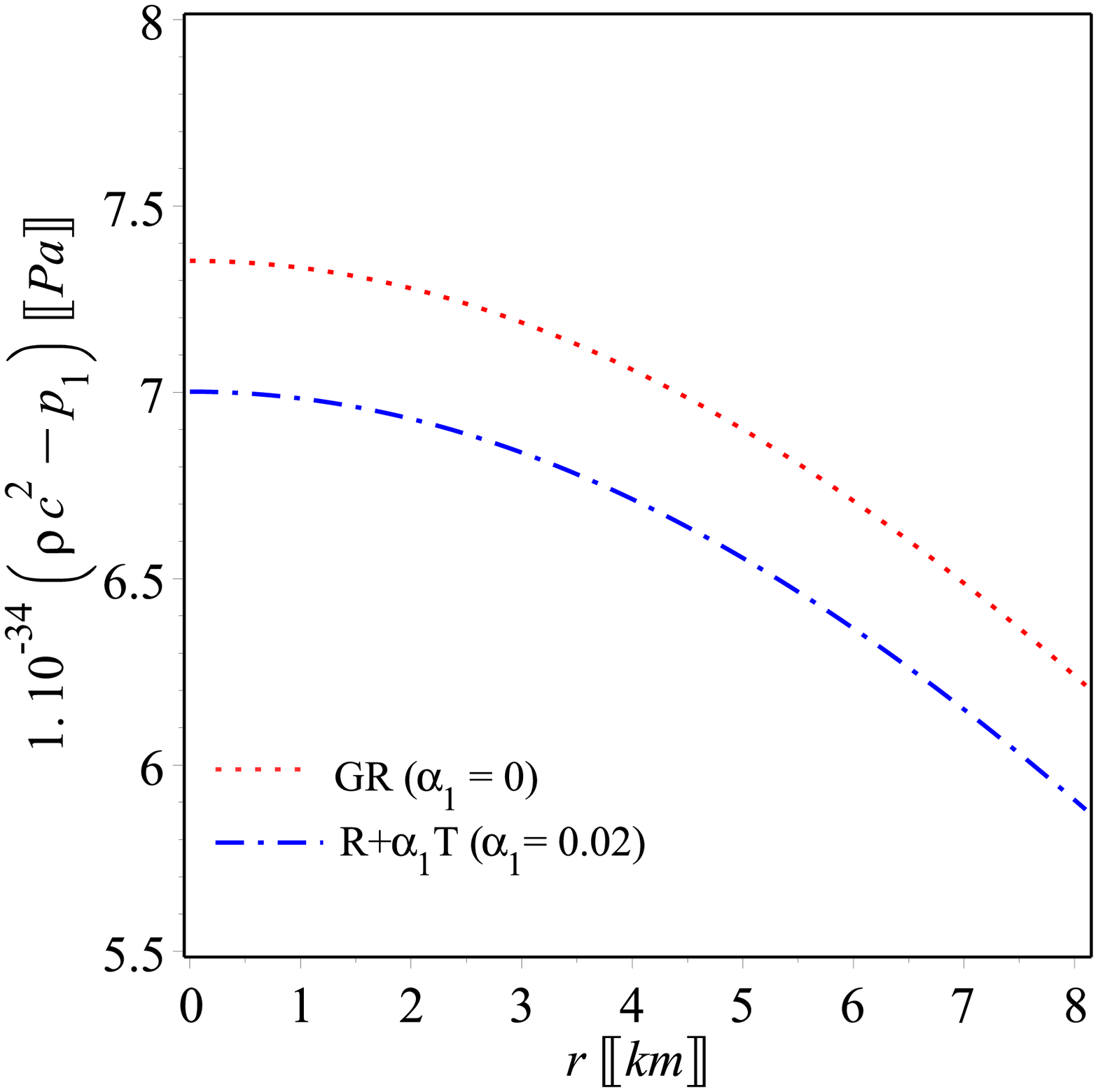}}\hspace{2cm}
\subfigure[~DEC (tangential)]{\label{fig:DEC2}\includegraphics[scale=.23]{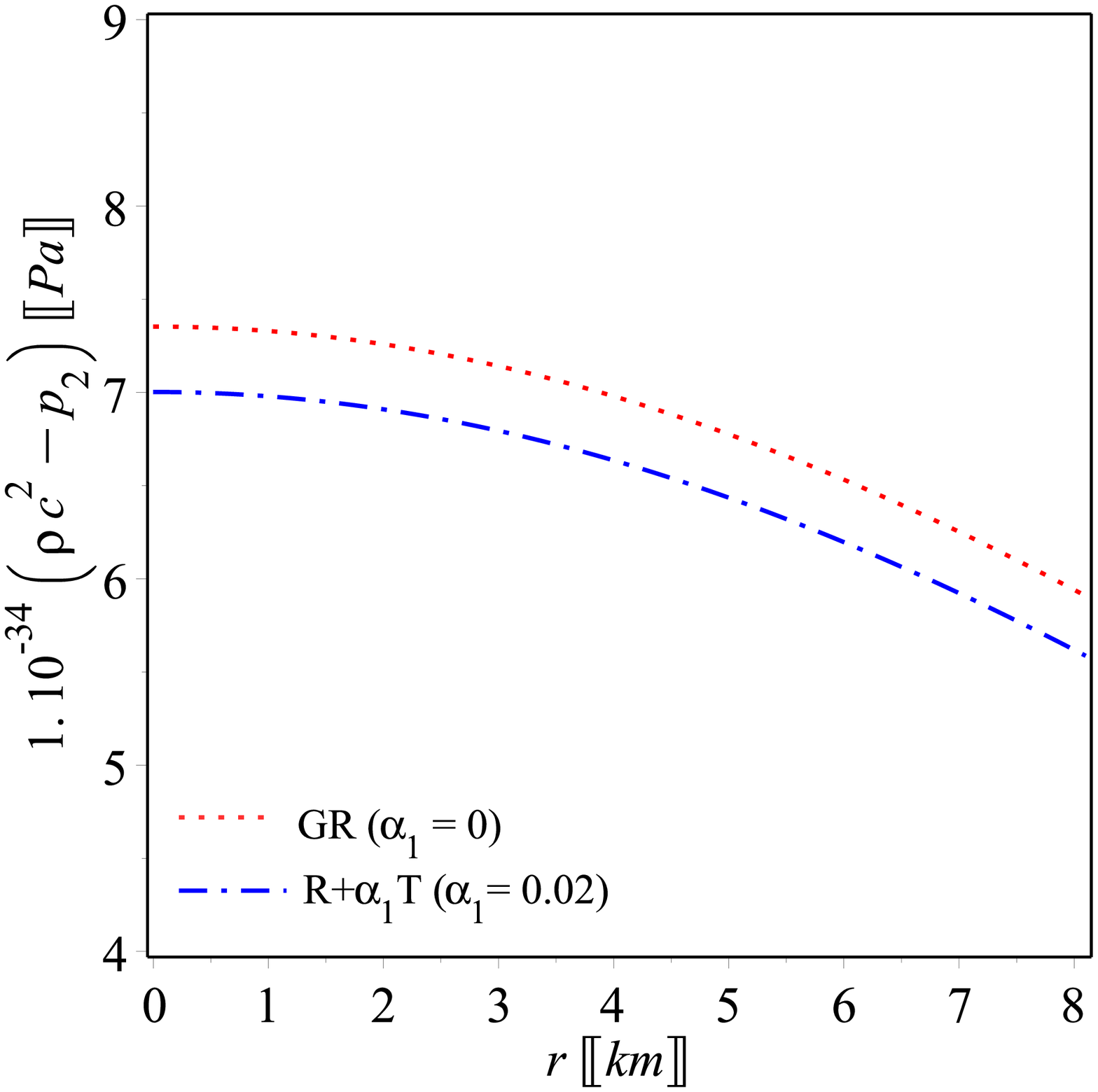}}
\caption{The characteristics of energy conditions when considering   Eq.~(\ref{sys}) of the pulsar ${\textit HerX1}$. The plots of Fig. \ref{Fig:EC} ensures that all the energy conditions (\textbf{9})  are satisfied.}
\label{Fig:EC}
\end{figure*}
\noindent Condition (\textbf{9}): Any  physically acceptable stellar model   must be compatible with the energy conditions:
\begin{itemize}
   \item[{\textrm i-}] The null energy condition (NEC)  which should satisfy the following identities: ${\mathrm \rho c^2+ p_1 \geq 0}$, \, ${\mathrm \rho c^2+ p_2 \geq 0}$\,,
    \item[{\textrm ii-}] The weak energy condition (WEC) which should satisfy the following identities: ${\mathrm \rho\geq 0}$, \,${\mathrm \rho c^2+ p_1 > 0}$, \,${\mathrm\rho c^2+p_2 > 0}$\,,
  \item[{\textrm iii}-] The dominant energy conditions (DEC) which should satisfy the following identities: ${\mathrm \rho \geq 0}$, \,${\mathrm\rho c^2-p_1 \geq 0}$, and ${\mathrm\rho c^2-p_2 \geq 0}$\,,
      \item[{\textrm iv-}] The strong energy condition (SEC) which should satisfy the following identities: ${\mathrm \rho c^2+p_1+2p_2 \geq 0}$, \,${\mathrm\rho c^2+p_1 \geq 0}$, \,${\mathrm\rho c^2+p_2 \geq 0}$\,.\\
\end{itemize}

Now let us rewrite the field equations (\ref{eq:Feqs}) as:
\begin{eqnarray}\label{eq:Eff-dens-press}
\nonumber {\mathrm \rho c^2} &=&{\mathrm  \rho_{GR} c^2 +\frac{5\alpha_1}{3(1+3\,\alpha_1)
 }\left( p_1  +2\,p_2 \right)},\\
\nonumber {\mathrm p_1} &=& {\mathrm {p_1}_{GR} +\frac{\alpha_1}{3+11\alpha_1} \left(3\rho c^2  -10p_2  \right)},\\
{\mathrm p_2} &=&{\mathrm { p_2}_{GR} + \frac{3\alpha_1}{3+16\,\alpha_1
 }\left(\rho c^2-\frac{5}{3} p_1  \right)}\,.
\end{eqnarray}
When $\alpha_1=0$ one can derive the GR case. Now let us discuss condition (\textbf{9}) analytically. {\bf Equation (\ref{eq:Eff-dens-press}) shows  clearly that the effect of the coupling constant which is a small effect as shown in Fig. \ref{Fig:EC}\subref{fig:Cond1}--\subref{fig:DEC2}

 %It is straightforward to prove that once the NEC is fulfilled in GR, it is also fulfilled in $f(R,T)$ whereas $\tilde{\rho} c^2 + \tilde{p}_r=\rho c^2 + p_r$ and $\tilde{\rho} c^2 + \tilde{p}_t=\rho c^2 + p_t$.
 Using conditions  (\textbf{1}) and  (\textbf{2}), it can be demonstrated that both the density and pressures within the star are consistently positive  which means that the NEC is verified. Moreover, we can show that:}

\begin{eqnarray}\label{eq:Ras_EC}
\nonumber& {\mathrm \rho c^2}+{\mathrm p_1} = ({\mathrm \rho_{GR} c^2} +{\mathrm {p_1}_{GR}}) + {\mathrm \alpha_1}\frac { \left( 27\,{\alpha_1}+9\, \right) {\mathrm \rho c^2}+ \left( 55\,{\alpha_1}+15\, \right){\mathrm p_1} +20\,{\mathrm {\alpha_1}}{\mathrm p_2}  }{9+99\,{\alpha}^
{2}+60\,\alpha},\\
&\nonumber  {\mathrm \rho c^2}+{\mathrm p_2} = ({\mathrm \rho_{GR} c^2} +{\mathrm {p_2}_{GR}})- \frac{\alpha_1\, \left(10{\mathrm p_2}[3
-16\alpha_1]- 125\,{\mathrm p_1} \alpha_1+9\,
{\mathrm \rho} c^2[1+3 \alpha_1] \right) }{3
 \left( 1+3\,\alpha_1 \right)  \left(3-16\,\alpha\right) }
,\\
\nonumber& {\mathrm\rho} c^2+{\mathrm p_1}+2{\mathrm p_2} = ({\mathrm\rho_{GR}} c^2 +{\mathrm{p_1}{_GR}}+2 {\mathrm {p_2}_{GR}}) \\
&+{\frac { \alpha_1\left( 162\,{\alpha_1}^{2}+297{\alpha_1}+81
 \right) {\mathrm \rho c^2}- \alpha_1\left( 1870{\alpha_1}^{2}+675{
\alpha_1}+45 \right){\mathrm p_1} +20{\mathrm p_2} {\alpha_1}^{2} \left( 3-16\alpha_1 \right) }{27-1584
\,{\alpha_1}^{3}-663\,{\alpha_1}^{2}+36\,\alpha_1}},\\
\nonumber &{\mathrm {\rho} c^2}- {\mathrm {p}_1} = ({\mathrm \rho_{GR} c^2} - {\mathrm {p_1}_{GR}}) +\frac {\alpha_1\, \left( 15\,{\mathrm p_1} +55{\mathrm p_1}\alpha_1+60\,p_2 +200\,{\mathrm p_2}  \alpha_1-9\,{\mathrm \rho} c^2[1+3 \alpha_1] \right) }{3 \left( 1+3\,\alpha_1 \right)  \left( 3+11\,
\alpha_1 \right) }
,\\
\nonumber &{\mathrm {\rho} c^2}- {\mathrm {p}_2} = ({\mathrm \rho_{GR} c^2} - {\mathrm {p_2}_{GR}}) +{\frac {\alpha_1 \left( {\mathrm p_1}[30 -35
  \alpha_1]+{\mathrm p_2}[30  -160{\mathrm p_2}
  \alpha_1]-9{\mathrm \rho c^2}[1+\alpha_1] \right) }{9-144{\alpha_1}^{2}-21\,\alpha_1}}
,\\
& {\mathrm \rho c^2}-{\mathrm p_1}-2{\mathrm p_2} = ({\mathrm \rho_{GR} c^2} -{\mathrm {p_1}{_GR}}-2 {\mathrm {p_2}_{GR}}) \\
&+\frac {\alpha_1 \left({\mathrm p_1}[ 135  +525  \alpha_1+110 {\alpha_1
}^{2}]+{\mathrm p_2}[180  -360
\alpha_1-3200\,{\alpha_1}^{2}] -3\,{\mathrm \rho c^2}[27+99\alpha_1+54 {\alpha}^{2}] \right) }{3\left( 1+3\,\alpha_1 \right)  \left( 3+
11\,\alpha_1 \right)  \left( 3-16\,\alpha_1 \right) }
.\qquad
\end{eqnarray}
Since  $\alpha_1>0$  (in this study $\alpha_1=0.02$) in addition to the fact that ${\mathrm p_r\geq 0}$,  ${\mathrm p_t\geq 0}$, and ${\mathrm \rho\geq 0}$; it is still necessary to demonstrate that \[ {\mathrm p_1}[30 -35
  \alpha_1]+{\mathrm p_2}[30  -160p_2
  \alpha_1]>9\,{\mathrm \rho c^2}[1+\alpha_1]>0\,,\] \[\left( 162\,{\alpha_1}^{2}+297{\alpha_1}+81
 \right) {\mathrm \rho c^2} +20{\mathrm p_2} {\alpha_1} \left( 3-16\alpha_1 \right)>\left( 1870{\alpha_1}^{2}+675{
\alpha_1}+45 \right)p_1>0\,,\] and \[10{\mathrm p_2}[3
-16\alpha_1]+9{\mathrm \rho c^2}[1+3 \alpha_1]>125\,{\mathrm p_1} \alpha_1>0\,,\] to ensure that the energy conditions are satisfied ({\textrm 1}--{\textrm 4}).

\textit{Clearly the energy conditions} (\textbf{9}) \textit{are verified for the pulsar  ${\textit HerX1}$ as shown in Fig. \ref{Fig:EC}\subref{fig:Cond1}--\subref{fig:DEC2}}.

Lastly, we observe  that the density prevails over the  radial and tangential pressures is ensured in $f(R,T)=R+\alpha_1 T$ as far as it is satisfied in GR, because \[{\mathrm {\rho} c^2} - \mathrm{p_1}- 2\mathrm{p_2} =\frac {\alpha_1 \left({\mathrm p_1}[ 135  +525  \alpha_1+110 {\alpha_1
}^{2}]+p_2[180  -360
\alpha_1-3200\,{\alpha_1}^{2}] -3\,{\mathrm \rho c^2}[27+99\alpha_1+54 {\alpha}^{2}] \right) }{3\left( 1+3\,\alpha_1 \right)  \left( 3+
11\,\alpha_1 \right)  \left( 3-16\,\alpha_1 \right) }.\]

The above condition is commonly mentioned as the strong energy condition by certain researchers \citep[c.f.][]{1988CQGra...5.1329K,Ivanov:2017kyr,2019EPJC...79..853D,Roupas:2020mvs}.

Based on the aforementioned discussion, we can demonstrate a strong connection between the energy conditions and the matter fluid in $f(R,T)$ gravity.
%However, the DEC  ${\mathrm \rho c^2} -{\mathrm p_1} - 2{\mathrm p_2}\geq 0$ is the essential  condition that allows for other conditions to be satisfied in general where  the matter density and total  pressures are positive.
 %Similar limitations on $f(R,T)=R+\alpha_1 T$ parameter motivated by thermodynamics aspects (positivity of the horizon entropy) have been obtained by \citep{Moradpour:2016fur}. Moreover, in cosmological applications, it has been shown that the second law of thermodynamics is fulfilled in $f(R,T)$, if the WEC is satisfied for the matter sector \citep{2016PhLB..757..187M}.

%%%%%%%%%%%%%%%%%%%%%%%%%%%%%%%%%%%%%%%%%%%%%%%%%%%%%
\subsection{ Stability conditions and causality}
\begin{figure*}
\centering
\subfigure[~Radial speed of sound]{\label{fig:vr}\includegraphics[scale=0.28]{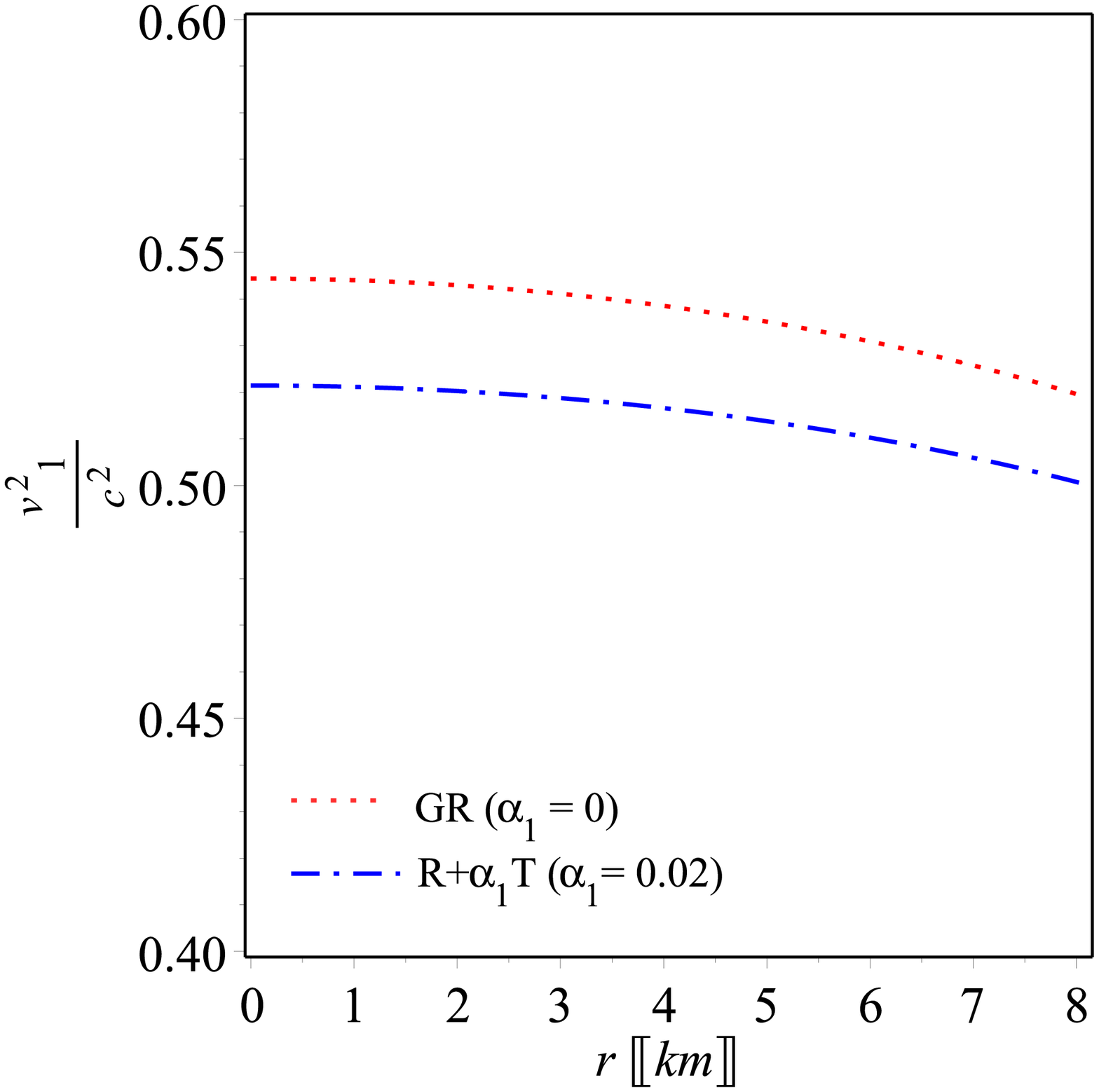}}
\subfigure[~Tangential speed of sound]{\label{fig:vt}\includegraphics[scale=.28]{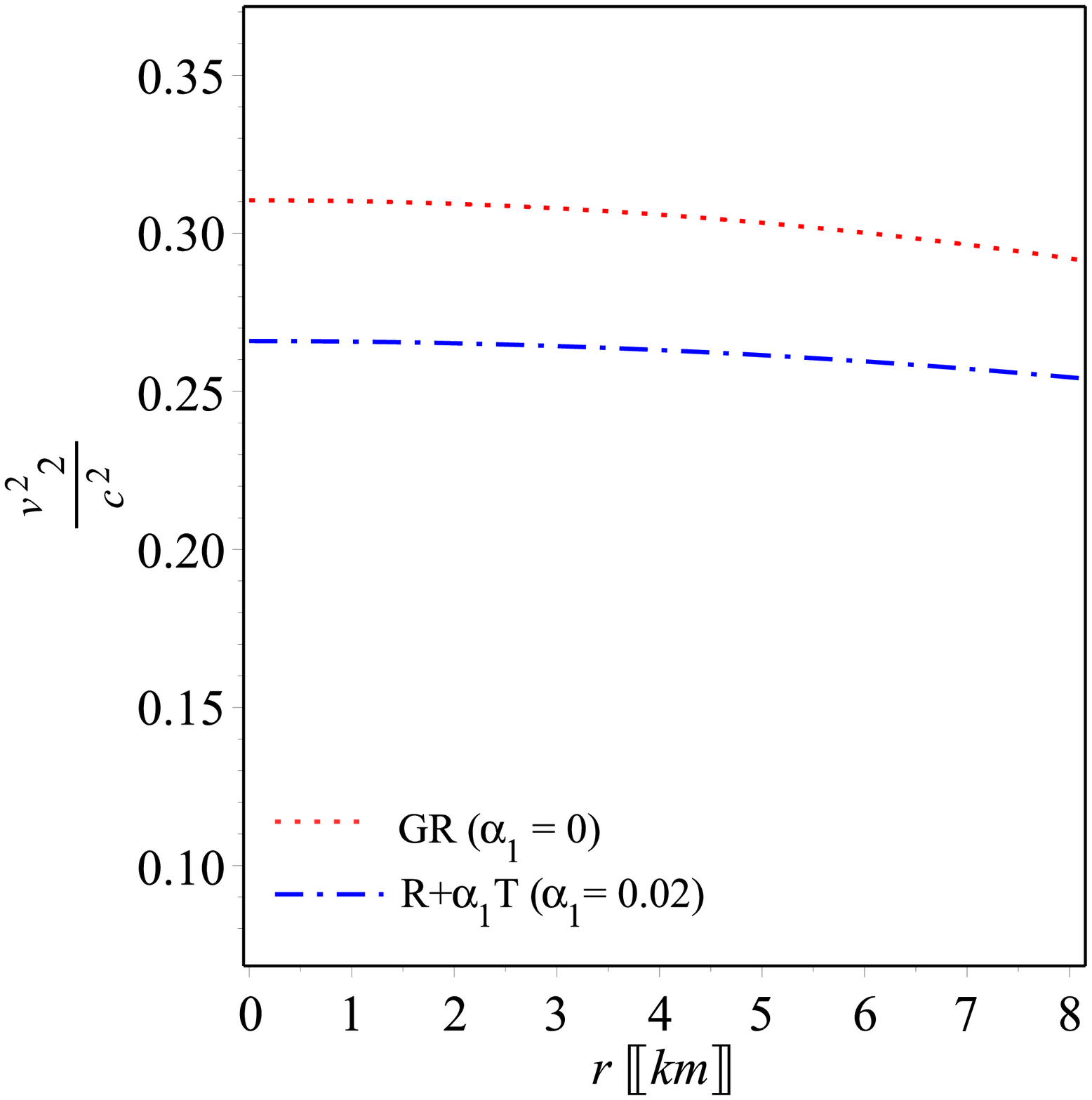}}
\subfigure[~difference between radial and tangential  speed of sounds]{\label{fig:vt-vr}\includegraphics[scale=.28]{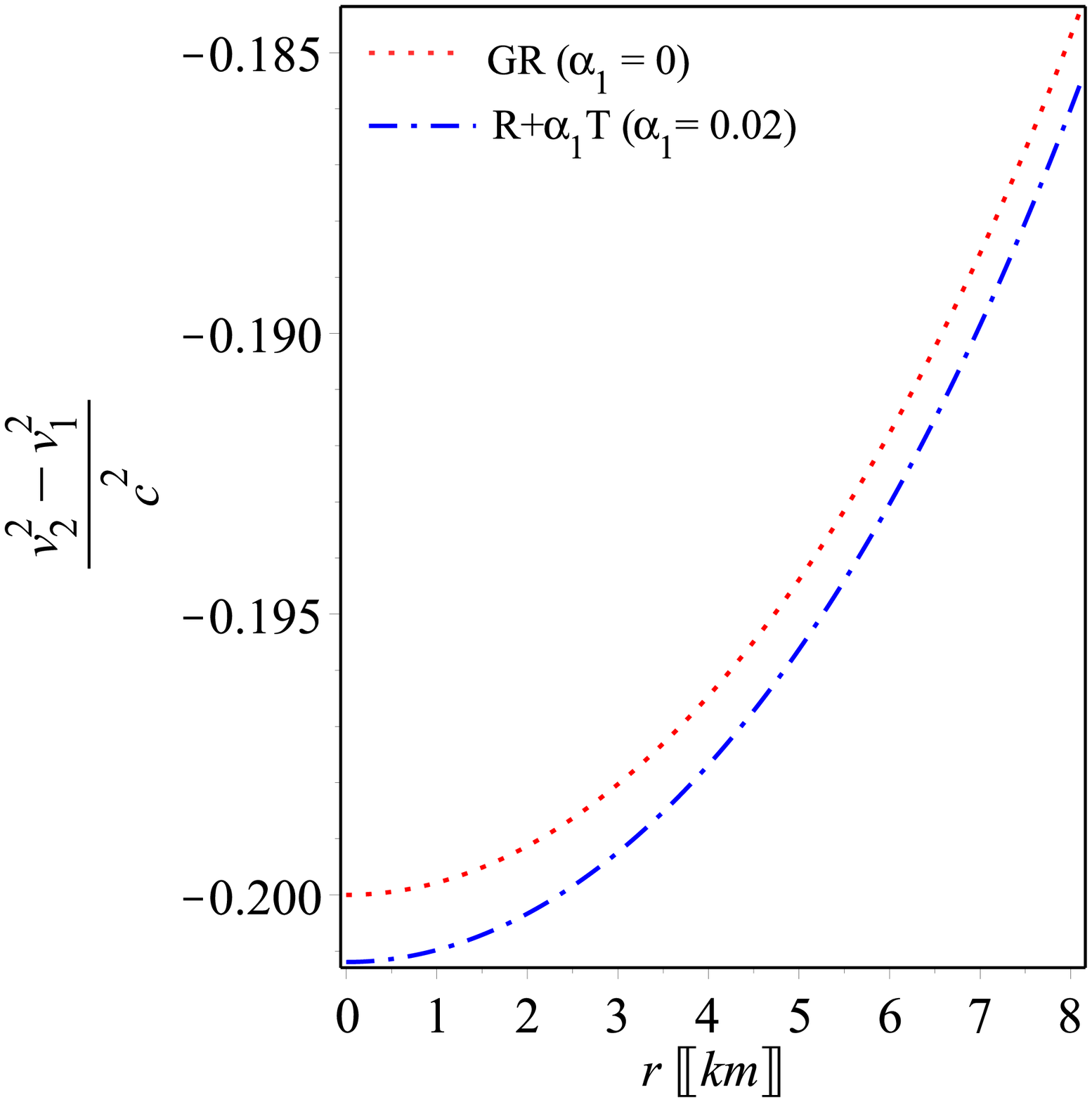}}
\caption{Tangential and radial sound speeds (\ref{eq:sound_speed}) of  ${\textit HerX1}$ ensure that the causality and stability conditions  (\textbf{10}) and (\textbf{11}) are met.}
\label{Fig:Stability}
\end{figure*}
The radial and tangential speeds of sound are defined as \citep{Herrera:1992lwz,Abreu:2007ew}:
\begin{equation}\label{eq:sound_speed}
  {\mathrm v_1^2} =  \frac{ d{\mathrm p}_1}{d {\mathrm \rho}}=  \frac{{\mathrm  p'}_1}{{\mathrm \rho'}}\,, \quad  {\mathrm  v_2^2} = \frac{d{\mathrm  p}_2}{d{\mathrm   \rho}}= \frac{{\mathrm  p'}_2}{{\mathrm \rho'}}\,,
\end{equation}
where the energy density and the derivative of the pressures  are given by Eqs. (\ref{eq:dens_grad})--(\ref{eq:pt_grad}).\\

\noindent Condition (\textbf{10}): The star's structure must adhere to causality, meaning it must meet the condition that sound speeds are positive and less than unity within the star ($0\leq \frac{{\mathrm v_1}}{c}\leq 1$, $0\leq \frac{{\mathrm v_2}}{c}\leq 1$), and they should decrease monotonically towards the surface ($({\mathrm v'_1})^2<0$, $({\mathrm v'_2})^2<0$).

\noindent Condition (\textbf{11}): For the star to be stable, its structure must fulfill the stability condition $-1< \frac{{\mathrm v_t^2}-{\mathrm v_r^2}}{c^2} < 0$ throughout the entire star \citep{Herrera:1992lwz}.

\textit{The causality and stability conditions, as stated in equations (10) and (11), are clearly satisfied for the pulsar ${\textit HerX1}$, as depicted in Fig. \ref{Fig:Stability}\subref{fig:vr}--\subref{fig:vt-vr}}.

%Although the sound speed fulfills the causality and stability conditions for the pulsar PSR J0740+6620, the GR prediction of the sound speed within the region $r \lesssim 10.14$ km from the center exceeds the conjectured conformal upper bound on the sound speed $v_1^2=c_s^2\leq c^2/3$. This violation is strongly suggested for NS with radii less than $\sim 11.8$ km \citep{Bedaque:2014sqa}. Also, the violation of the conformal upper limit of the sound speed has been obtained in other models when hadronic EoS is assumed \citep{Cherman:2009tw,Landry:2020vaw} or when a non-parametric EoS approach based on Gaussian processes is applied to the pulsar PSR J0740+6620 using the X-ray NICER+XMM observations \citep{Legred:2021hdx}. On the contrary, as seen in Fig. \ref{Fig:Stability}\subref{fig:vr}, the conformal bound on the sound speed for the pulsar PSR J0740+6620 is not violated in RT from the core to the surface.
%%%%%%%%%%%%%%%%%%%%%%%%%%%%%%%%%%%%%%%%%%%%%%%%%%%%%%%%%%%%%%%%%%
\subsection{The adiabatic and hydrodynamic equilibrium conditions}
In order to validate the reliability of the considered model, we conduct two additional tests to assess the stability of the obtained model in $f(R,T)=R+\alpha_1 T$ gravity. Initially, we examine the  adiabatic indices   of a spacetime having spherical symmetric. These indices, that determine the ratio of specific heats, are analyzed to assess the stability of the model \citep{1964ApJ...140..417C, chan1993dynamical} as:
\begin{equation}\label{eq:adiabatic}
{\mathrm \gamma}=\frac{4}{3}\left(1+\frac{{\mathrm \Delta}}{|{\mathrm  p}'_1|}\right)_{max}\,,\qquad \quad
{\mathrm  \Gamma_1}=\frac{{\mathrm  \rho c^2}+{\mathrm  p_1}}{{\mathrm  p_1}} {\mathrm  v_1^2}\,, \qquad \quad
{\mathrm  \Gamma_2}=\frac{{\mathrm  \rho c^2}+{\mathrm  p_2}}{{\mathrm  p_2}} {\mathrm v_2^2.}
\end{equation}
In the case of an anisotropic fluid, the stability of the sphere is guaranteed when ${\mathrm \Gamma}={\mathrm \gamma}$ (or when ${\mathrm \Gamma}>{\mathrm \gamma}$), as discussed in the work by Chan and Santos \citep[see]{chan1993dynamical}. Clearly, in the case where ${\mathrm \gamma}=4/3$ and ${\mathrm \Gamma_1}={\mathrm \Gamma_2}$, the adiabatic  lead to an isotropic sphere, as indicated by the research conducted by Harrison \citep{1975A&A....38...51H}.

Next, we consider the truth of the TOV equation \citep{Hansraj:2018jzb, Oppenheimer:1939ne} by considering the assumption of hydrostatic equilibrium in the star. The TOV equation has been modified to incorporate the newly introduced force of   $f(R,T)=R+\alpha_1 T$ term, and  is denoted as $F_T$. It can be rewritten  as follows:
\begin{equation}\label{eq:RS_TOV}
{\mathit F_a}+{\mathit F_g}+{\mathit F_h}+{\mathit F_T=0}\,.
\end{equation}
Here  ${\mathrm  F_h}$ and ${\mathrm  F_g}$  are  the hydrostatic and the gravitational forces. We defined the different forces as \citep{Nashed:2022zyi,ElHanafy:2022kjl}:
\begin{eqnarray}\label{eq:Forces}
  {\mathrm F_a} =&\frac{ 2{\mathrm  \Delta}}{\mathrm r} ,\qquad
  {\mathrm F_g} = -\frac{{\mathrm  M_g}}{r}({\mathrm  \rho c^2}+{\mathrm p_1})e^{{\mathrm  \gamma/2}} ,\qquad\nonumber\\
  {\mathrm  F_h} =&-{\mathrm  p'_1} ,\qquad
  {\mathrm F_T} = \frac{\alpha_1}{3(1-\alpha_1)}({\mathrm  c^2 \rho}'-{\mathrm p}'_1-2{\mathrm  p}'_2),
\end{eqnarray}
where ${\mathrm \gamma}\equiv {\mathrm  \gamma}(r)=\mu-\nu$ and the mass (energy) ${\mathrm M_g}$ of an isolated   systems   is given by  \citep{1930PhRv...35..896T}
\begin{eqnarray}\label{eq:grav_mass}
{\mathrm M_g(r)}&=&{\int_{\mathrm V}}\Big(\mathbb{{\mathrm T}}{^r}{_r}+\mathbb{{\mathrm T}}{^\theta}{_\theta}+\mathrm{T}{^\phi}{_\phi}-\mathrm{T}{^t}{_t}\Big)\sqrt{-g}\,dV\nonumber\\
&=&\frac{(e^{\mu/2})'}{e^{\mu}} e^{\nu/2} r =\frac{\mu'}{2} r e^{-\gamma/2}\,.
\end{eqnarray}
 Here, ${\mathrm F_g}$ represents the gravitational force, which can be defined as $-\frac{a_0 r}{\textbf{L}^2}({\mathrm \rho c^2}+{\mathrm p_1})$. We now incorporate the stability conditions associated with the relativistic modified TOV equation and adiabatic indices.
\begin{figure}
\centering
\subfigure[~Radial adiabatic index]{\label{fig:gamar}\includegraphics[scale=0.30]{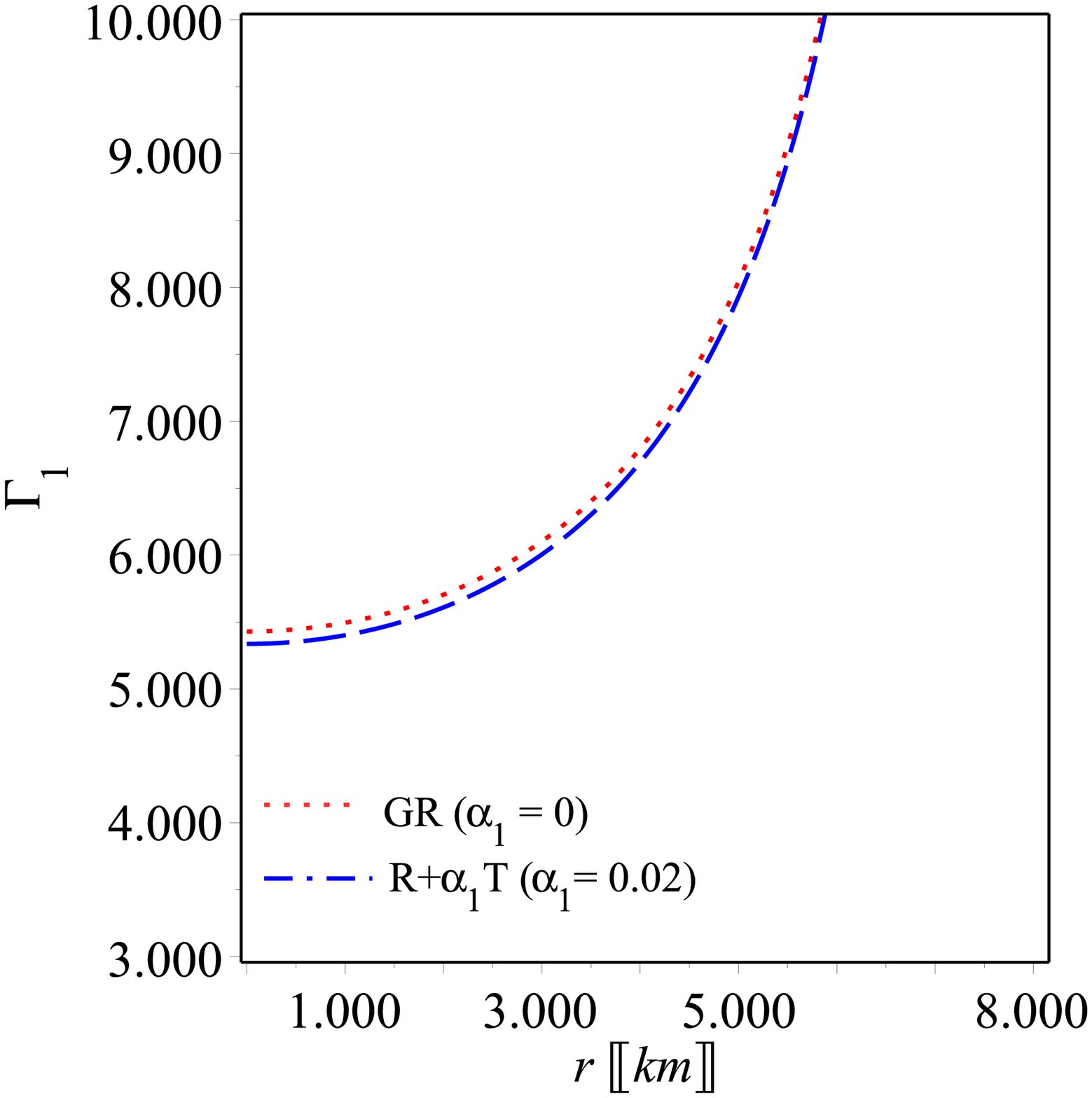}}
\subfigure[~Tangential adiabatic index]{\label{fig:gamat}\includegraphics[scale=0.30]{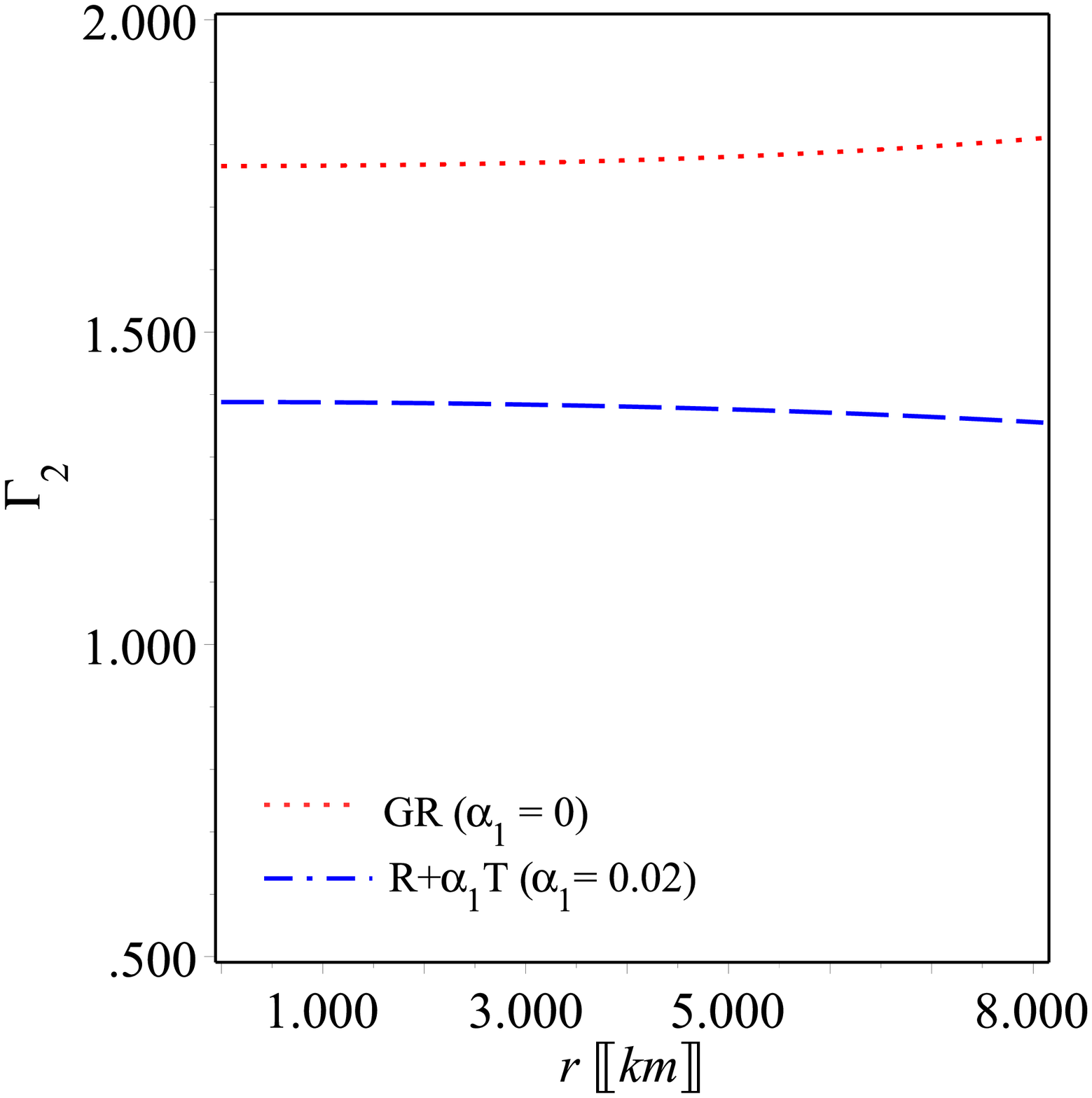}}
\subfigure[~The adiabatic index]{\label{fig:gamar1}\includegraphics[scale=0.28]{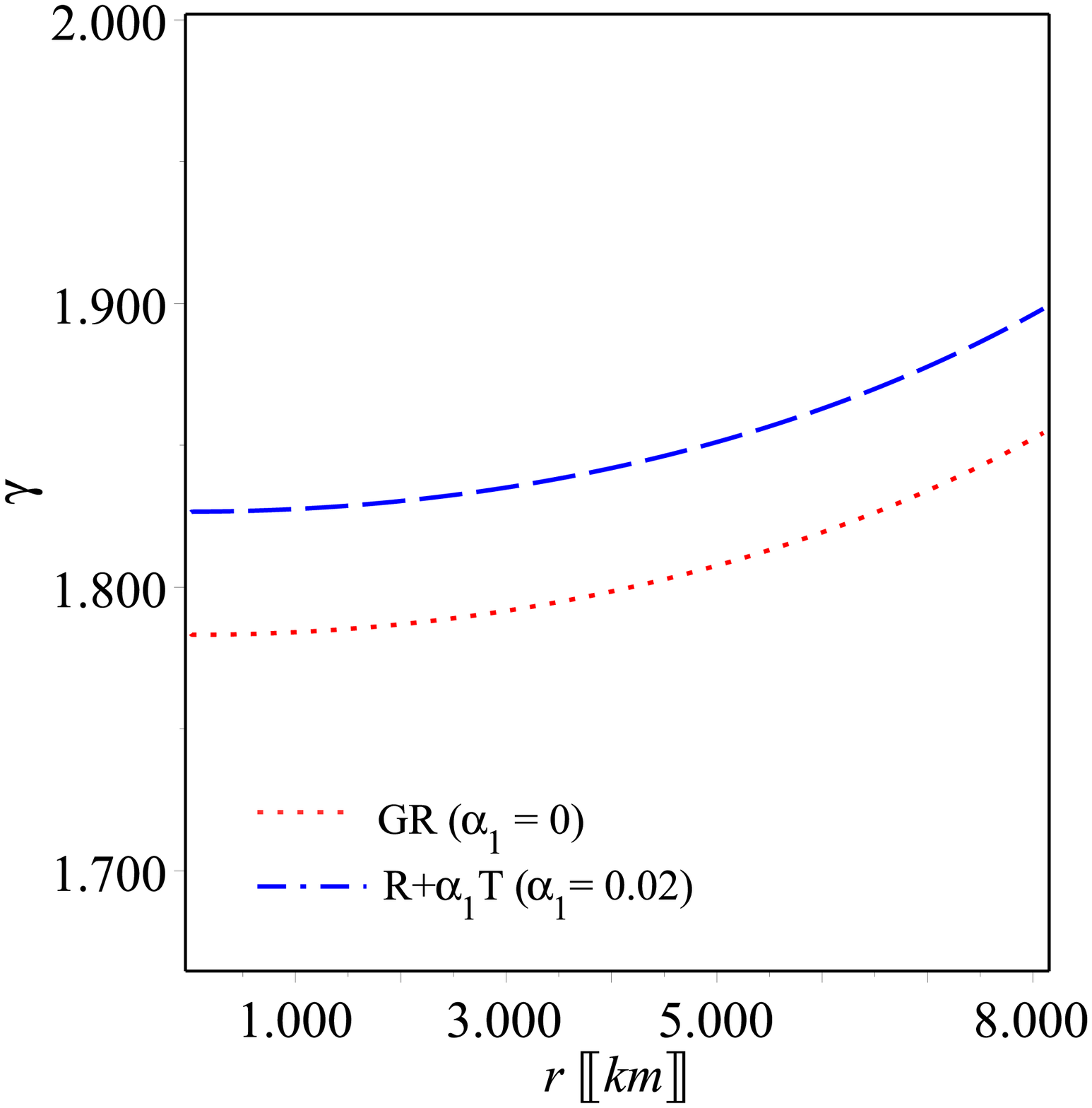}}
\caption{The adiabatic indices given by Eq.~(\ref{eq:adiabatic}) of the pulsar ${\textit HerX1}$. The graphs provide confirmation that the stability conditions (\textbf{12}) hold true throughout the pulsar.}
\label{Fig:Adiab}
\end{figure}
\begin{figure}
\centering
%\label{fig:GRTOV}\includegraphics[scale=0.28]{Fig8.pdf}
\subfigure[~GR]{\label{fig:GRTOV}\includegraphics[scale=0.25]{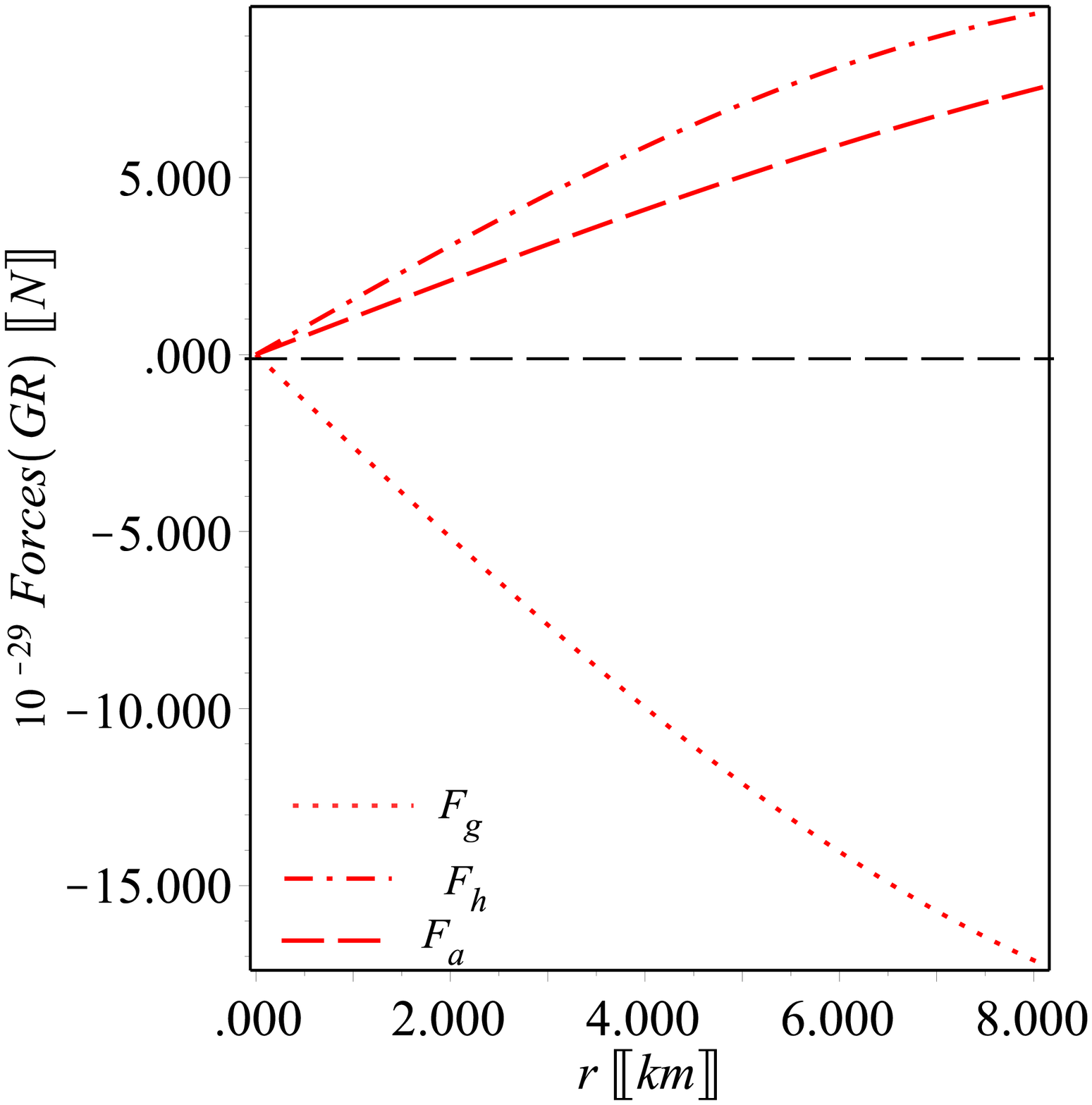}}
\subfigure[~$f(R,T)=R+\alpha_1\,T$]{\label{fig:RTTOV}\includegraphics[scale=0.25]{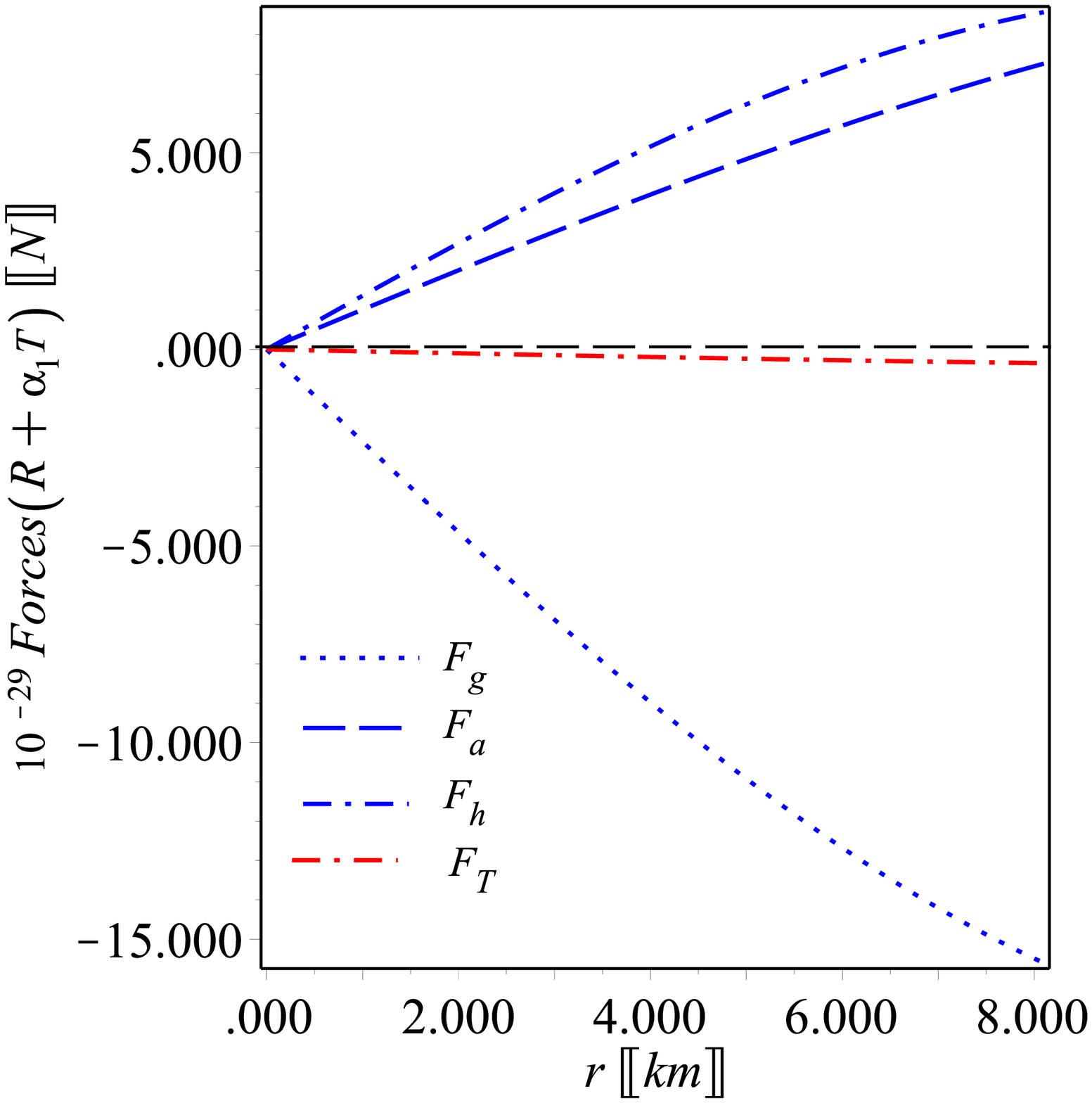}}
\caption{The different forces (\ref{eq:Forces}) of the amended TOV equation (\ref{eq:RS_TOV}) of the pulsar ${\textit HerX1}$.  The plots ensure the stability condition (\textbf{13}) which is verified for ${\textit HerX1}$ .}
\label{Fig:TOV}
\end{figure}
\noindent Condition (\textbf{12}): The anisotropic star model maintains a stable behavior since the adiabatic indices verify  ${\mathrm \Gamma_t}> {\mathrm \gamma}$ and ${\mathrm \Gamma_r}> {\mathrm \gamma}$  throughout the entire pulsar. By employing the  pressures and density  described in Eqs. (\ref{sys}), along with their derivatives (\ref{eq:dens_grad})--(\ref{eq:pt_grad}), we illustrate the adiabatic  (\ref{eq:adiabatic}) of the pulsar ${\textit HerX1}$ for both the   $f(R,T)$ and GR   as shown in Figure \ref{Fig:Adiab}.

\textit{The stability conditions and adiabatic indices of condition (\textbf{12}) are evidently satisfied for the pulsar ${\textit HerX1}$, as illustrated in Figure \ref{Fig:Adiab}.}

\noindent Condition (\textbf{13}): The anisotropic star achieves a state of hydrodynamic equilibrium, as the forces verify the  TOV equation (\ref{eq:RS_TOV}). Utilizing Eqs. (\ref{sys}) and (\ref{eq:dens_grad})--(\ref{eq:pt_grad}), we calculate the forces (\ref{eq:Forces}), which are illustrated in Figure \ref{Fig:TOV} for both the general relativity (GR) and $f(R,T)$ cases. The figures clearly demonstrate that the negative   force counterbalances the positive ones, resulting in hydrodynamic equilibrium and supporting the requirement for a stable configuration.

\textit{The condition of hydrodynamic equilibrium (\textbf{13}) is evidently satisfied for the pulsar ${\textit HerX1}$, as depicted in Figure \ref{Fig:TOV}.}.
%%%%%%%%%%%%%%%%%%%%%%%%%%%%%%%%%%%%%%%%%%%%%%%%%
\subsection{EoS}

By utilizing the observational data from the pulsar ${\textit HerX1}$, we were able to determine a reasonable value for the parameter $\alpha_1=0.02$. Consequently, we compute the surface density, which has been determined to be approximately $\rho_R \approx 6.5\times 10^{14}$ g/cm$^{3}$, while at the core, the density increases to approximately $\rho_{core} \approx 8.2\times 10^{14}$ g/cm$^3$.
%which is only $\sim 2.1$ times the nuclear saturation density $\rho_{nuc}=2.7\times 10^{14}$ g/cm$^{3}$.
It is worth mentioning that the central density in $f(R,T)$ gravity with $\alpha_1>0$ is lower than the corresponding value in general relativity (GR) ($\alpha_1=0$). This is due to the fact that within a given radius, the estimated mass in $f(R,T)$ gravity is lower than that in Einstein gravity, as illustrated in Figure \ref{Fig:Mass1} \subref{Fig:Mass}. The value of the central density implies that the core of ${\textit HerX1}$ is composed of neutrons.
\begin{figure}[th!]
\centering
{\subfigure[~Radial EoS]{\label{fig:REoS}\includegraphics[scale=0.25]{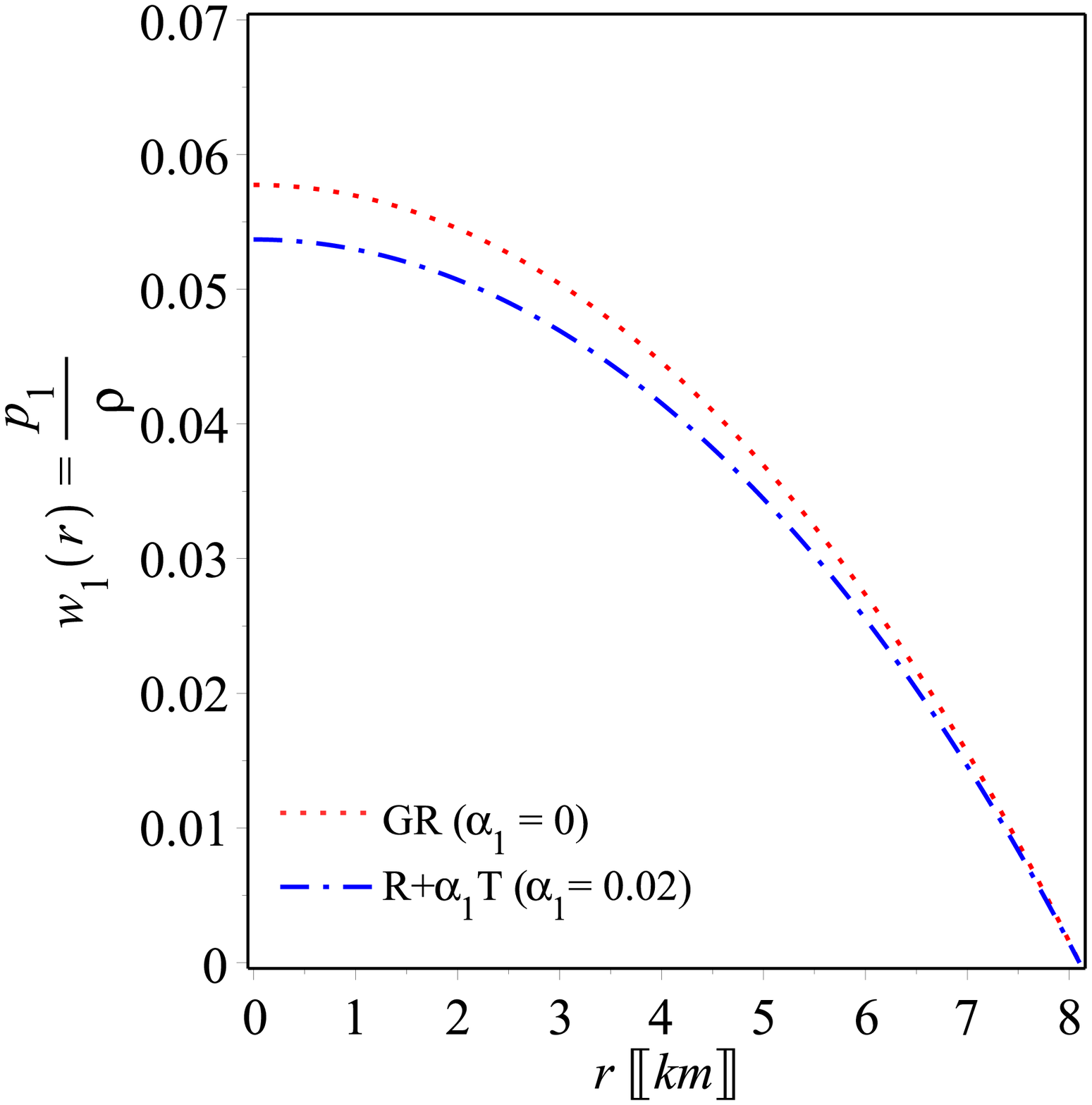}}
\subfigure[~Linear radial EoS]{\label{fig:OREoS}\includegraphics[scale=0.25]{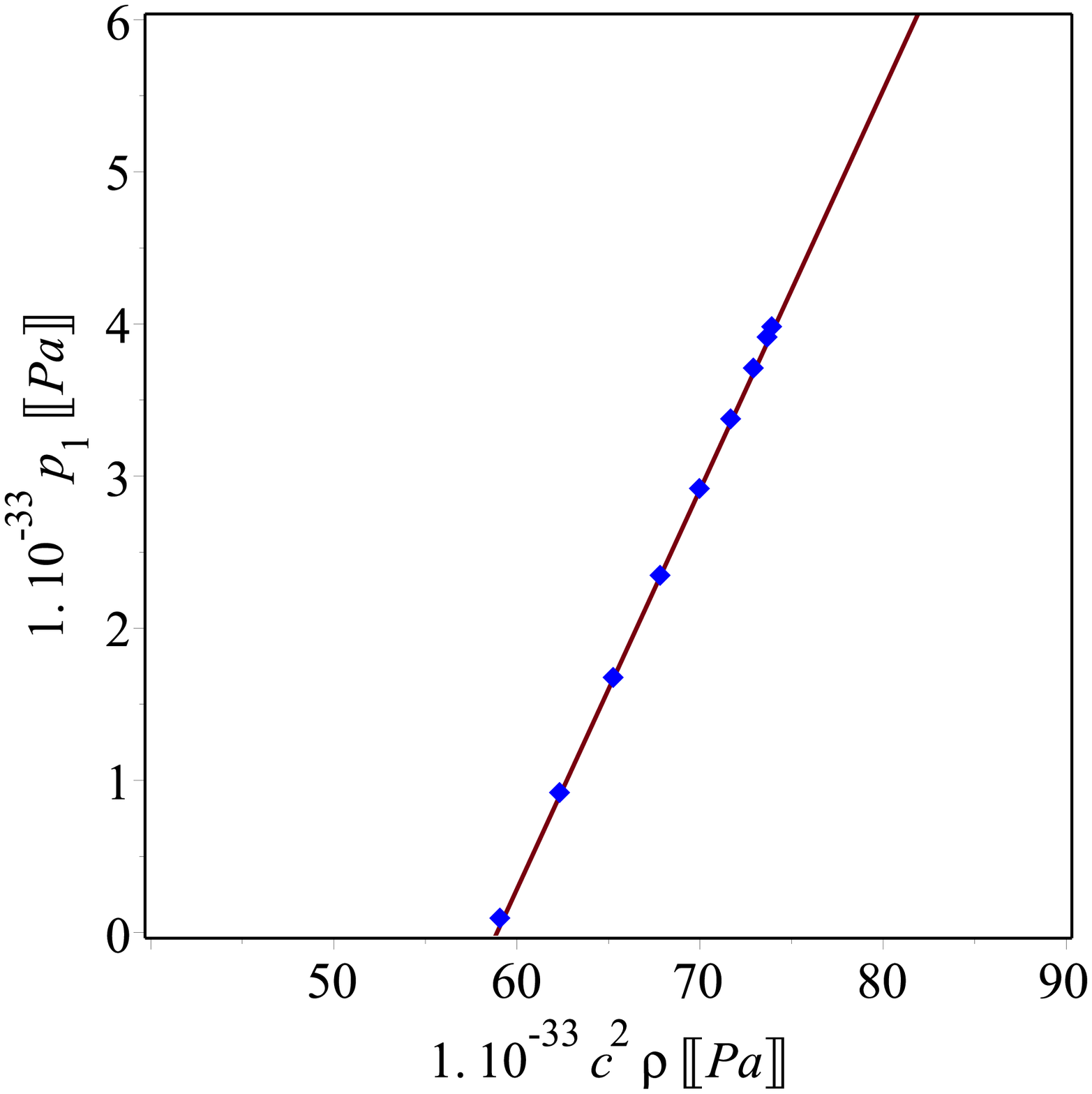}}
\subfigure[~Tangential EoS]{\label{fig:TEoS}\includegraphics[scale=.25]{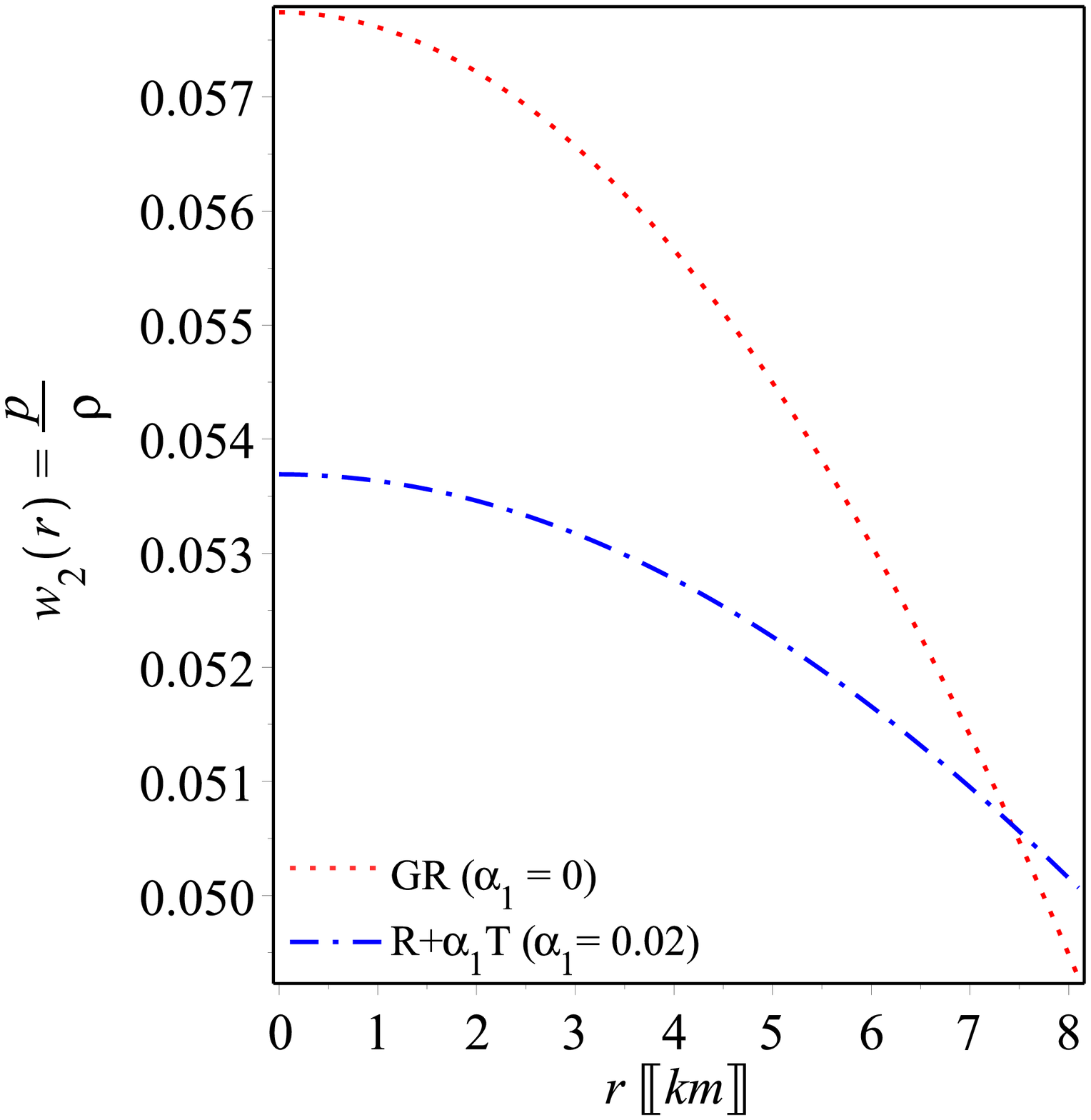}}
\subfigure[~Linear tangential EoS]{\label{fig:OTEoS}\includegraphics[scale=0.25]{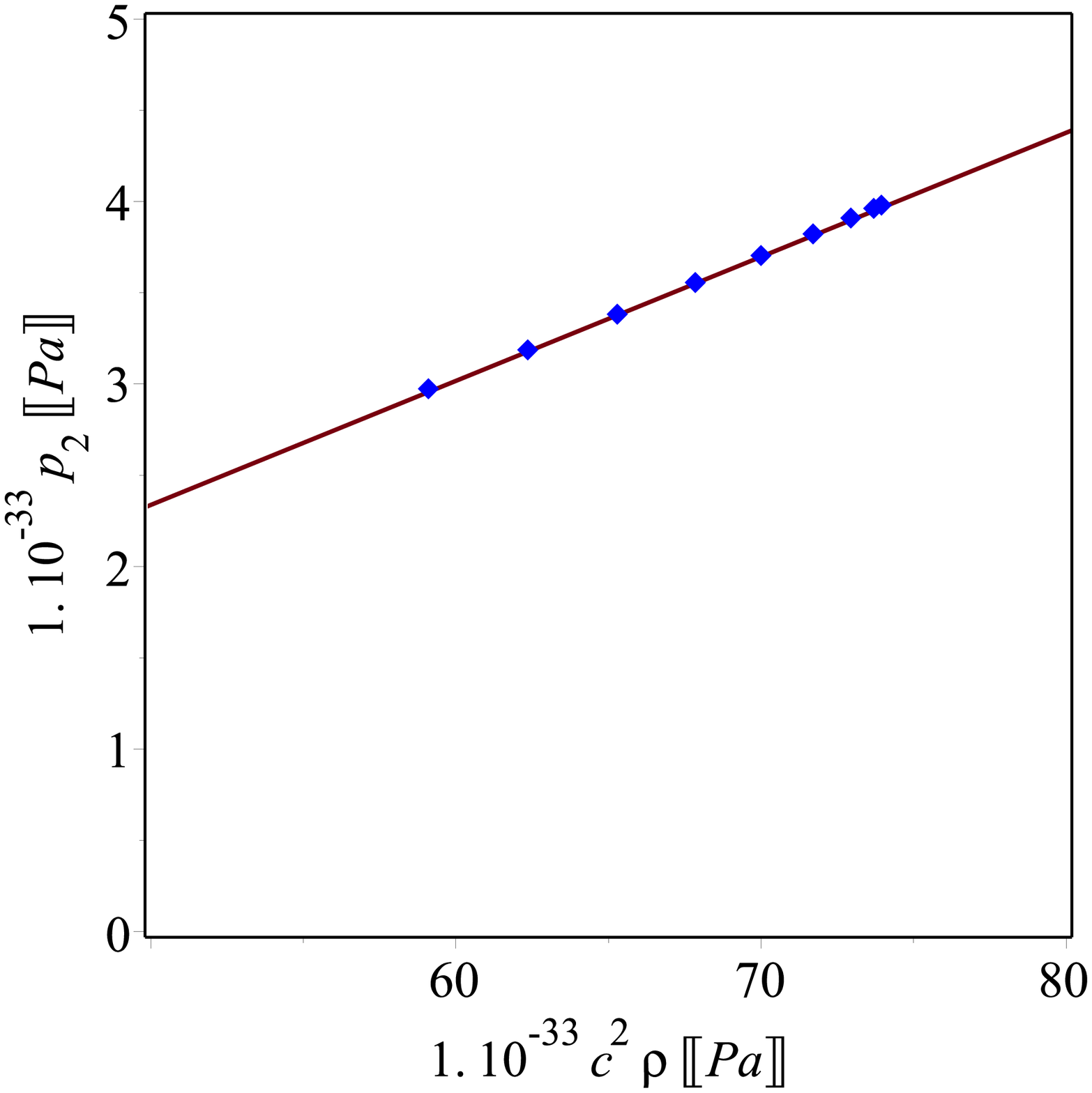}}}
\caption{The relationship between pressures and density in the pulsar ${\textit HerX1}$ aligns well with a linear equation of state (EoS) incorporating a bag constant. The slopes $dp_1/d\rho=0.27 c^2$ and $dp_2/d\rho=0.068 c^2$ provide clear evidence that the matter within the neutron star adheres to the proposed conformal bound on the sound speed, $c_s^2\leq c^2/3$, throughout the star's interior.}
\label{Fig:EoS}
\end{figure}
 demonstrate  the equation of state (EoS) parameters that exhibits  slightly variation from their maximum values at the center, gradually decreasing in a monotonic manner towards the surface. At the boundary, it is expected that the parameter $w_1$ approaches zero, aligning with the anticipated behavior.
{ In Figs. \ref{Fig:EoS}\subref{fig:REoS} and \subref{fig:TEoS}, we plot  EoS, i.e. we plot  $\mathrm {w_1=p_1/\rho}$ and $\mathrm{w_2=p_2/\rho}$,  when the dimensionless parameter $\alpha_1=0$ which is the GR case, and  $\alpha_1\neq 0$. The plots, Figs. \ref{Fig:EoS}\subref{fig:REoS} and \subref{fig:TEoS}, demonstrate  the equation of state (EoS) parameters that exhibits  slightly variation from their maximum values at the center, gradually decreasing in a monotonic manner towards the surface. At the boundary, it is expected that the parameter $w_1$ approaches zero, aligning with the anticipated behavior. It is important to acknowledge that the changes in the equation of state (EoS) parameters are more constrained in the context of $R+\alpha_1T$ compared to the case of  GR.  Moreover, we stress that we have  not assume any formula of the EoS in the model under consideration, \ref{Fig:EoS}\subref{fig:REoS} and \subref{fig:TEoS}. Nevertheless, we demonstrate that EoS $p(\rho)$ of the radial and tangential directions can be accurately approximated by linear relationships. For the pulsar ${\textit HerX1}$, the best-fit equations are found to be $p_1\approx 0.262,\rho-15.47$ and $p_2=0.06799, \rho-1.0623$.  More interestingly, the slopes of the fitted lines $dp_1/d\rho=0.262=v_1^2$ and $dp_2/d\rho=0.06799=v_2^2$ are in agreement  with the numerical values listed in Table \ref{Table2} for the model of the present study.
%Moreover, the present anisotropic NS model derives an EoS which is not very stiff as usually shown for NS models, but in agreement with the EoS as expected from gravitational wave signals where no clear evidence on a tidal deformation in the observed GW patterns.\\
}
%%%%%%%%%%%%%%%%%%%%%%%%%%%%%%%%%%%%%%%%%%%%%%%%%%%%%%%%%%%%%%%%%%%%%%%%%%%%%%%%%%%%%%%%
%%%%%%%%%%%%%%%%%%%%%%% Section 8 %%%%%%%%%%%%%%%%%%%%%%%%%%%%%%%%%
\section{Additional Observational Constraints}\label{S8}
%%%%%%%%%%%%%%%%%%%%%%%%%%%%%%%%%%%%%%%%%%%%%%%%%%%%%%%%%%%%%%%%%%%
Next, we will compare the model being considered with data from other pulsars in order to assess its validity across a broad range of astrophysical observations. Additionally, we generate a mass-radius profile by considering different boundary densities that align with the nuclear saturation density. This analysis reveals the model's consistency in predicting masses within the lower mass range of 3.5 to 5.3 times the mass of the Sun ($M_\odot$), often referred to as the ``mass gap."  Furthermore, we compare the compactness  of the present study with the Buchdahl one bound.
%%%%%%%%%%%%%%%%%%%%%%%%%%%%%%%%%%%%%%
\subsection{Stars data}
%%%%%%%%%%%%%%%%%%%%%%%%%%%%%%%%%%%%%%
In addition to the analysis conducted on the pulsar ${\textit Her X-1}$, the same methodology is applied to twenty-two other stars, spanning a range from 0.8 times the mass of the Sun ($M_\odot$) to heavy pulsars with a mass of 2.01 times $M_\odot$.  Table \ref{Table1}, presents the observed radii and masses of each pulsar, along with the respective model parameters {$a_0$, $a_1$, $a_2$}, assuming a value of $\alpha_1=0.2$ for the $f(R,T)$ parameter. The results demonstrate that the model proposed in this study successfully predicts masses for these pulsars that are consistent to the observed ones. Table \ref{Table2}, displays the evaluated values of various important physical quantities. As indicated in Table 2, the density values obtained are in accordance with the expected nuclear density. It is important to emphasize that this study does not rely on any specific form of equations of state (EoSs). The obtained results are found to align well with a linear behavior, with the slopes $dp_r/d\rho$ and $dp_t/d\rho$ of the best-fit lines being in agreement.  Furthermore, the values presented in Table \ref{Table2} ensure  stablity and causality, satisfying all the necessary physical conditions discussed in Section \ref{Sec:matt}.In addition, we provide the predicted boundary redshift numerical; values for the twenty-two pulsars based on the current model. It is noteworthy that all of these values adhere to the upper bound limit of $Z_{\textbf{L}}\leq 2$ proposed by Buchdahl \citep{Buchdahl:1959zz}. This consistency also holds true for anisotropic spheres \citep{Ivanov:2002xf,Barraco:2003jq}.
\begin{table*}
\caption{The mass-radius of twenty two stars with their corresponding parameters ($\alpha_1=0.02$).}
\label{Table1}
\begin{tabular*}{\textwidth}{@{\extracolsep{\fill}}llcccccc@{}}
\hline
%Equil. & \multicolumn{1}{c}{$x$} & \multicolumn{1}{c}{$y$} & \multicolumn{1}{c}{$z$} & \multicolumn{1}{c}{$C$} & S \\
%Points \\
{{Pulsar}}   &Ref.& observed mass ($M_{\odot}$) &  obs. radius [{km}]& estimated mass ($M_{\odot}$) &  {$a_0$}    & {$a_1$}     & {$a_2$}   \\
\hline
Her X-1 &\citep{Abubekerov:2008inw}         &  $0.85\pm 0.15$    &  $8.1\pm 0.41$   &$0.905$&  $0.297$    & $-0.176$    & $0.298$     \\
M13 &\citep{Webb:2007tc}            &  $1.38\pm 0.2$     &  $9.95\pm 0.27$  &$1.459$&  $0.394$    & $-0.285$    & $0.351$     \\
RX J185635-3754 &\citep{Pons:2001px}     &  $0.9\pm 0.2$      &  $6$             &$0.949$&  $0.427$    & $-0.324$    & $0.369$     \\
GW170817-2  &\citep{LIGOScientific:2018cki}     &  $1.27\pm 0.09$    &  $11.9\pm 1.4$   &$1.351$&  $0.302$    & $-0.181$    & $0.301$     \\
EXO 1785-248 &\citep{Ozel:2008kb}    &  $1.3\pm 0.2$      &  $8.849\pm 0.4$  &$1.372$&  $0.418$    & $-0.313$    & $0.364$     \\
PSR J0740+6620 &\citep{Raaijmakers:2019qny}  &  $1.34\pm 0.16$    &  $12.71\pm 1.19$ &$1.426$&  $0.298$    & $-0.177$    & $0.298$     \\
LIGO    &\citep{LIGOScientific:2020zkf}         &  $1.4$             &  $12.9\pm 0.8$   &$1.489$&  $0.307$    & $-0.187$    & $0.303$     \\
X7  &\citep{Rybicki:2005id}             &  $1.4$             &  $14.5\pm 1.8$   &$1.492$&  $0.340$    & $-0.607$    & $0.284$     \\
PSR J0037-4715 &\citep{Reardon:2015kba}   &  $1.44\pm 0.07$    &  $13.6\pm 0.9$   &$1.532$&  $0.299$    & $-0.179$    & $0.299$     \\
LMC X-4 &\citep{Rawls:2011jw}         &  $1.04\pm 0.09$    &  $8.301\pm 0.2$  &$1.103$&  $0.355$    & $-0.240$    & $0.330$     \\
PSR J0740+6620 &\citep{Miller:2019cac}  &  $1.44\pm 0.16$    &  $13.02\pm 1.24$ &$1.531$&  $0.312$    & $-0.193$    & $0.307$     \\
Cen X-3 &\citep{Naik:2011qc}         &  $1.49\pm 0.49$    &  $9.178\pm 0.13$ &$1.566$&  $0.464$    & $-0.369$    & $0.388$     \\
GW170817-1  &\citep{LIGOScientific:2018cki}     &  $1.45\pm 0.09$    &  $11.9\pm 1.4$   &$1.539$&  $0.345$    & $-0.230$    & $0.325$     \\
4U 1820-30 &\citep{Guver:2010td}      &  $1.46\pm 0.2$     &  $11.1\pm 1.8$   &$1.546$&  $0.374$    & $-0.261$    & $0.340$     \\
4U 1608-52  &\citep{1996IAUC.6331....1M}     &  $1.57\pm 0.3$     &  $9.8\pm 1.8$    &$1.651$&  $0.457$    & $-0.361$    & $0.385$     \\
KS 1731-260 &\citep{Ozel:2008kb}     &  $1.61\pm 0.37$    &  $10\pm 2.2$     &$1.692$&  $0.460$    & $-0.364$    & $0.386$     \\
EXO 1745-268  &\citep{Ozel:2008kb}   &  $1.65\pm 0.25$    &  $10.5\pm 1.8$   &$1.736$&  $0.448$    & $-0.35$    & $0.380$     \\
Vela X-1 &\citep{Rawls:2011jw}        &  $1.77\pm 0.08$    &  $9.56\pm 0.08$  &$1.845$&  $0.531$    & $-0.456$    & $0.424$     \\
4U 1724-207 &\citep{Ozel:2008kb}     &  $1.81\pm 0.27$    &  $12.2\pm 1.4$   &$1.909$&  $0.423$    & $-0.318$    & $0.366$     \\
SAX J1748.9-2021 &\citep{Ozel:2008kb} &  $1.81\pm 0.3$     &  $11.7\pm 1.7$   &$1.906$&  $0.441$    & $-0.341$    & $0.376$     \\
PSR J1614-2230\footnote{It should be mentioned that the estimated mass for the massive pulsars slightly exceeds the observed values, indicating the need for stricter constraints on the parameter $\alpha_1$, which may require a value of $\alpha_1=0.02$ in order to align with the observations.} &\citep{Demorest:2010bx}  &  $1.97\pm 0.04$    &  $13\pm 2$       &$2.076$&  $0.432$    & $-0.32$    & $0.371$     \\
PSR J0348+0432 &\citep{Antoniadis:2013pzd}   &  $2.01\pm 0.04$    &  $13\pm 2$       &$2.117$&  $0.441$    & $-0.341$    & $0.376$     \\
\hline
\end{tabular*}
\end{table*}
\begin{table*}
\caption{Calculation of the most interest  physical quantities.}
\label{Table2}
\begin{tabular*}{\textwidth}{@{\extracolsep{\fill}}lccccccccc@{\extracolsep{\fill}}}
\hline
%Equil. & \multicolumn{1}{c}{$x$} & \multicolumn{1}{c}{$y$} & \multicolumn{1}{c}{$z$} & \multicolumn{1}{c}{$C$} & S \\
%Points \\
{Pulsar}                              &{$\rho(0)$} &      {$\rho_\textbf{L}$} &   \multicolumn{1}{c}{$v_1^2(0)/c^2$}  &    \multicolumn{1}{c}{$v_1^2(\textbf{L})/c^2$}  & \multicolumn{1}{c}{$v_2^2(0)/c^2$} & \multicolumn{1}{c}{$v_2^2(\textbf{L})/c^2$}  &  {$\rho c^2-p_1-2p_2|_0$}&{$\rho-p_1-2p_2|_\textbf{L}$}& \multicolumn{1}{c}{$Z_\textbf{L}$}\\
                                        &{[$g/cm^3$]}&     {[$g/cm^3$]} &   {}                &    {}                & {}               & {}                &  {[$Pa$]}                &{[$Pa$]}            & {}\\
\hline
Her X-1            &8.23$\times10^{14}$     &6.54$\times10^{14}$  &  0.262   &   0.25     &  0.068 & 0.064 &8.59$\times10^{34}$ & 6.76$\times10^{34}$ & 0.204  \\
RX J185635-3754       &2.31$\times10^{15}$     &1.62$\times10^{15}$  &  0.346   &   0.296     &  0.146 & 0.122 & 2.69$\times10^{35}$ & 1.81$\times10^{35}$ & 0.340  \\
LMC X-4            &9.67$\times10^{14}$     &7.27$\times10^{14}$  &  0.3   &   0.269     &  0.998 & 0.878 & 1.05$\times10^{35}$ & 7.77$\times10^{34}$ & 0.260  \\
GW170817-2            &3.89$\times10^{14}$     &3.08$\times10^{14}$  &  0.274   &   0.252     &  0.73 & 0.665 & 4.07$\times10^{34}$ & 3.19$\times10^{34}$ & 0.208  \\
EXO 1785-248       &1.04$\times10^{15}$     &7.31$\times10^{14}$  &  0.34   &   0.293     &  0.14 & 0.117 & 1.19$\times10^{35}$ & 8.12$\times10^{34}$ & 0.329  \\
PSR J0740+6620       &3.36$\times10^{14}$     &2.67$\times10^{14}$  &  0.272   &   0.251     &  0.713 & 0.651 & 3.51$\times10^{34}$ & 2.76$\times10^{34}$ & 0.205  \\
M13       &7.63$\times10^{14}$     &5.51$\times10^{14}$  &  0.323   &   0.283     &  0.123 & 0.105 & 8.59$\times10^{34}$ & 6.04$\times10^{34}$ & 0.302  \\
LIGO       &3.38$\times10^{14}$     &2.66$\times10^{14}$  &  0.276   &   0.253     &  0.754 & 0.684 & 3.55$\times10^{34}$ & 2.76$\times10^{34}$ & 0.213  \\
X7       &2.61$\times10^{14}$     &2.14$\times10^{14}$  &  0.424   &   0.366     &  0.226 & 0.183 & 1.85$\times10^{34}$ & 1.69$\times10^{34}$ & 0.183  \\
PSR J0037-4715       &2.95$\times10^{14}$     &2.34$\times10^{14}$  &  0.273   &   0.251     &  0.719 & 0.656 & 3.08$\times10^{34}$ & 2.42$\times10^{34}$ & 0.206  \\
PSR J0740+6620       &3.39$\times10^{14}$     &2.65$\times10^{14}$  &  0.279   &   0.255     &  0.782 & 0.707 & 3.58$\times10^{34}$ & 2.77$\times10^{34}$ & 0.219  \\
GW170817-1       &4.55$\times10^{14}$     &3.45$\times10^{14}$  &  0.295   &   0.265     &  0.944 & 0.836 & 4.92$\times10^{34}$ & 3.67$\times10^{34}$ & 0.250  \\
4U 1820-30         &5.74$\times10^{14}$     &4.24$\times10^{14}$  &  0.311   &   0.275     &  0.11 & 0.095 & 6.35$\times10^{34}$ & 4.58$\times10^{34}$ & 0.279  \\
Cen X-3            &1.1$\times10^{15}$     &7.35$\times10^{14}$  &  0.376   &   0.313     &  0.176 & 0.142 & 1.32$\times10^{35}$ & 8.43$\times10^{34}$ & 0.386  \\
4U 1608-52         &9.45$\times10^{14}$     &6.38$\times10^{14}$  &  0.37   &   0.31     &  0.171 & 0.138 & 1.13$\times10^{35}$ & 7.28$\times10^{34}$ & 0.378  \\
KS 1731-260         &9.14$\times10^{14}$     &6.15$\times10^{14}$  &  0.372   &   0.311     &  0.173 & 0.139 & 1.1$\times10^{35}$ & 7.03$\times10^{34}$ & 0.381  \\
EXO 1745-268         &8.03$\times10^{14}$     &5.48$\times10^{14}$  &  0.362   &   0.306     &  0.163 & 0.133 & 9.53$\times10^{34}$ & 6.21$\times10^{34}$ & 0.366  \\
Vela X-1           &1.21$\times10^{15}$     &7.47$\times10^{14}$  &  0.448   &   0.349     &  0.25 & 0.186 & 1.59$\times10^{35}$ & 8.99$\times10^{34}$ & 0.486  \\
4U 1724-207      &5.52$\times10^{14}$     &3.87$\times10^{14}$  &  0.343   &   0.294     &  0.143 & 0.119 & 6.38$\times10^{34}$ & 4.32$\times10^{34}$ & 0.334  \\
SAX J1748.9-2021  &6.34$\times10^{14}$     &4.36$\times10^{14}$  &  0.357  &   0.302     &  0.157 & 0.129 & 7.46$\times10^{34}$ & 4.92$\times10^{34}$ & 0.357  \\
PSR J1614-2230     &5.00$\times10^{14}$     &3.47$\times10^{14}$  &  0.35   &   0.298     &  0.15 & 0.124 & 2.65$\times10^{34}$ & 5.83$\times10^{34}$ & 0.346  \\
PSR J0348+0432     &5.13$\times10^{14}$     &3.53$\times10^{14}$  &  0.357   &   0.302     &  0.157 & 0.129 & 6.04$\times10^{34}$ & 3.98$\times10^{34}$ & 0.357 \\
\hline
\end{tabular*}
\end{table*}
\begin{table*}
\caption{Calculation of the adiabatic indices at the center of the stellar.}
\label{Table2}
\begin{tabular*}{\textwidth}{@{\extracolsep{\fill}}lccc@{\extracolsep{\fill}}}
\hline
%Equil. & \multicolumn{1}{c}{$x$} & \multicolumn{1}{c}{$y$} & \multicolumn{1}{c}{$z$} & \multicolumn{1}{c}{$C$} & S \\
%Points \\
{Pulsar}                              &{$\Gamma_1(0)$} &      {$\Gamma_2(0)$} &   {$\gamma$} \\
\hline
Her X-1            &5.34    &1.39 &  1.83    \\
RX J185635-3754       &3.9     &1.65 &  1.72   \\
LMC X-4            &4.54    &1.51&  1.78  \\
GW170817-2            &5.25    &1.4 &  1.82  \\
EXO 1785-248       &3.96    &1.63 &  1.73  \\
PSR J0740+6620       &5.31     &1.4  &  1.83  \\
M13       &4.16    &1.58 & 1.75      \\
LIGO       &5.17    &1.41&  1.82   \\
X7       &5.77     &1.33  & 1.85  \\
PSR J0037-4715       &5.3    &1.4 & 1.82  \\
PSR J0740+6620       &5.1   &1.42  &1.81 \\
GW170817-1       &4.66    &1.49 & 1.79   \\
4U 1820-30         &4.35     &1.54  &  1.76\\
Cen X-3            &3.67     &1.72 &  1.69 \\
4U 1608-52         &3.7     &1.71 &  1.7    \\
KS 1731-260         &3.69    &1,72  & 1.69   \\
EXO 1745-268         &3.76     &1.69 & 1.7    \\
Vela X-1           &3.36   &1.9  & 1.63   \\
4U 1724-207      &3.93     &1.64  &  1.72 \\
SAX J1748.9-2021  &3.8    &1.68  &  1.71   \\
PSR J1614-2230     &3.87    &1.65& 1.71    \\
PSR J0348+0432     &3.81   &1.68  &  1.71   \\
\hline
\end{tabular*}
\end{table*}
\section{Mass-Radius diagram}\label{MR}
\begin{figure*}
\centering
\subfigure[~Figure of Compactness and Radius relation]{\label{fig:Comp}\includegraphics[scale=0.28]{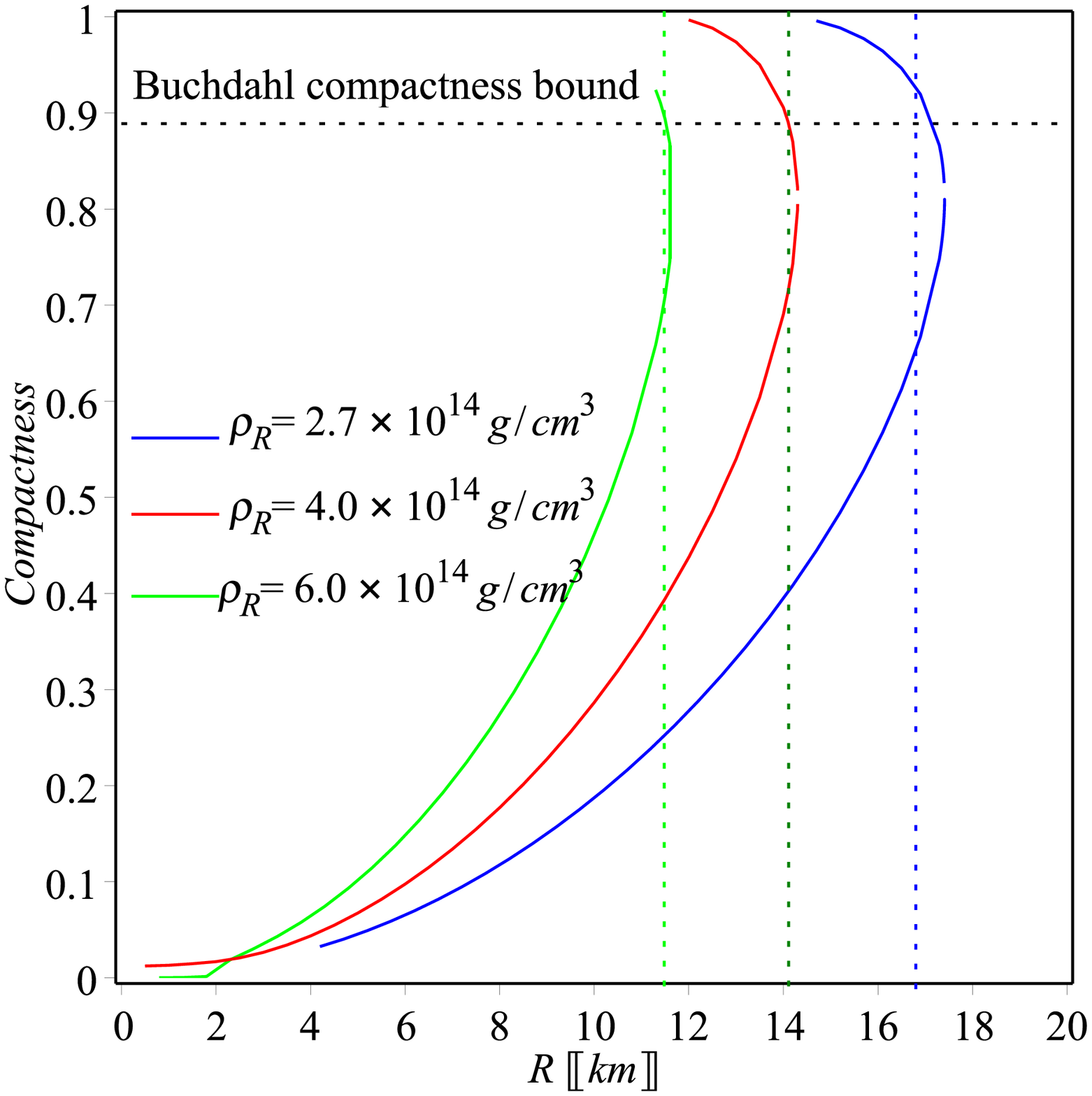}}
\subfigure[~Figure of Mass and Radius relation]{\label{fig:MR}\includegraphics[scale=.28]{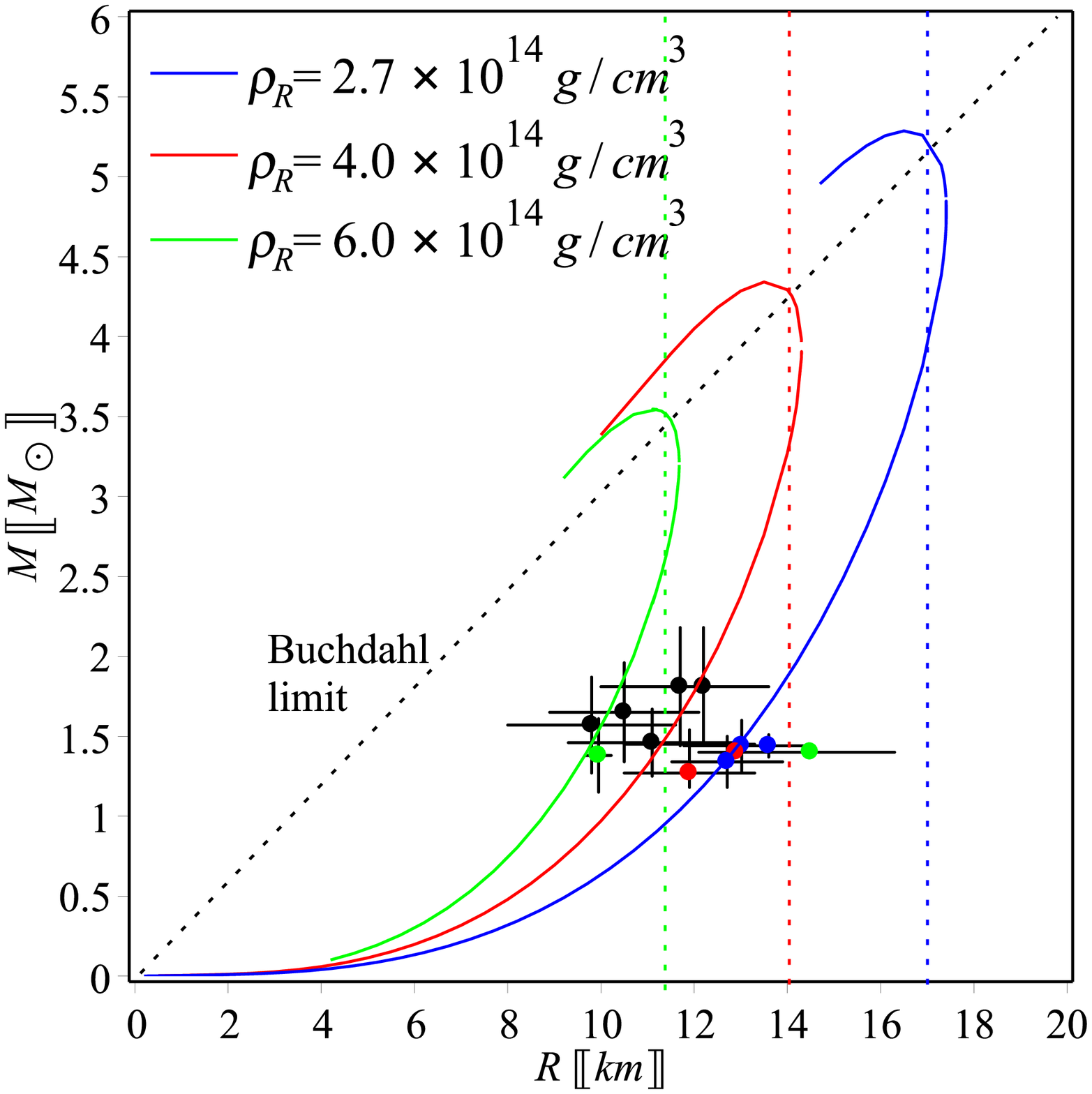}}
\subfigure[~LIGO-Virgo and NICER  restrictions]{\label{fig:NICER}\includegraphics[scale=.28]{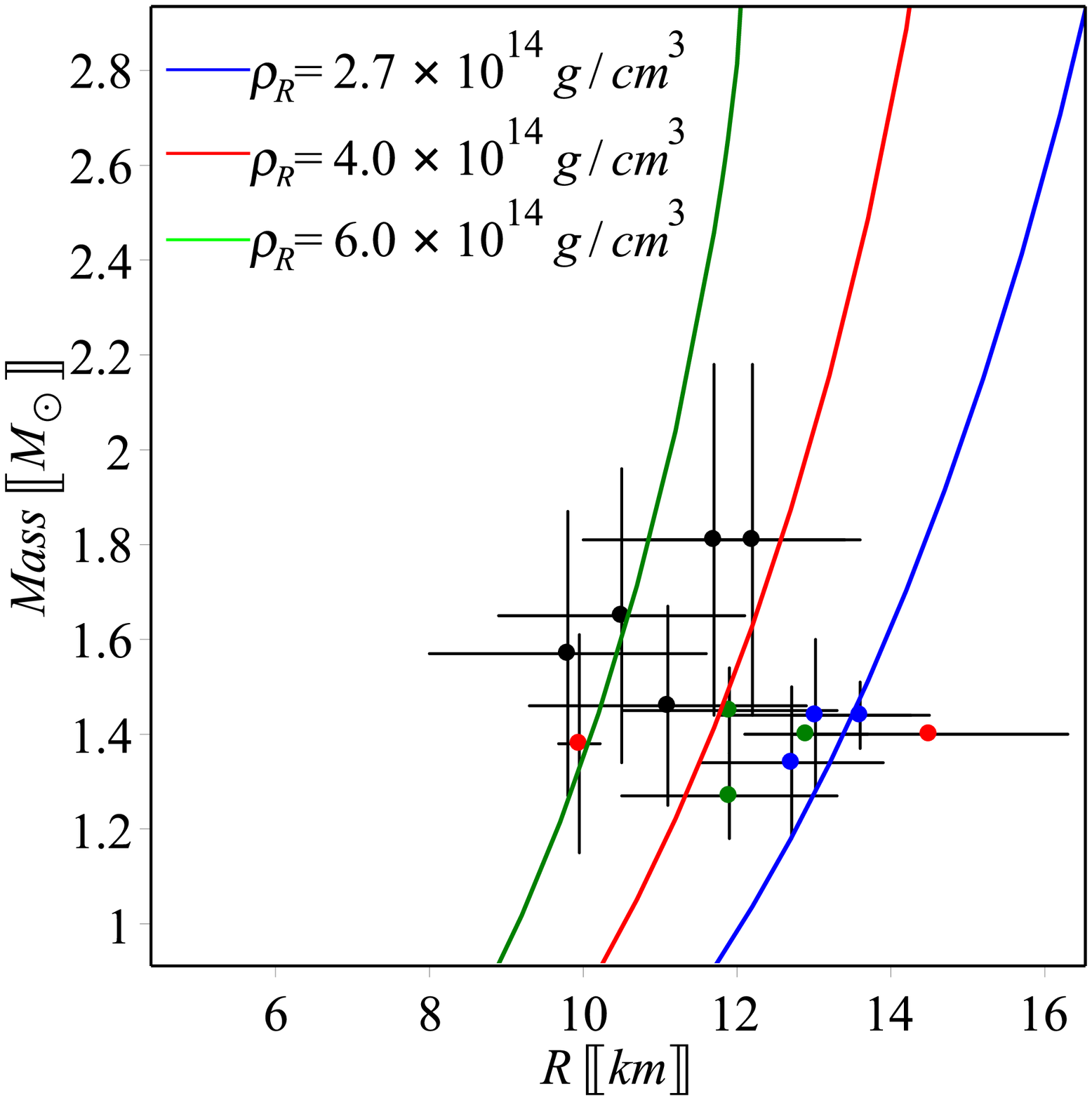}}
\caption{\subref{fig:Comp} The curves depicting the relationship between compactness and radius for different surface energy densities demonstrate that the maximal compactness is $0.92$, slightly surpassing the Buchdahl limit of $8/9$.
\subref{fig:MR} The curves representing the relationship between mass and radius reveal that the diagonal dashed line, representing the DEC (diagonal excluded curve), corresponds to an upper mass limit of $M\approx 5.25 M_\odot$ within the smaller mass, with a radius of $\textbf{L}\approx 16.6$ km. This limit is established when the surface energy density is set to the saturation nuclear density of $\rho_{{nuc}}=2.7\times 10^{14}$ g/cm$^3$. The low-mass X-ray binaries are denoted by solid blue circles, the millisecond pulsars (MSPs) are represented by solid green circles, and the gravitational wave (GW) signals are depicted by solid red circles, as indicated in Table \ref{Table1}.
\subref{fig:NICER}Upon closer examination of several intriguing pulsars, it is observed that the LIGO-Virgo restrictions on the radius, the NICER restrictions on   $PSR J0030+0451$, and the combined NICER+XMM constraints on $PSR J0740+6620$ are consistent with the $\rho_{nuc}$-curve.}
\label{Fig:CompMR}
\end{figure*}

Table \ref{Table2} provides the range of surface densities as $2.14\times 10^{14} \lesssim \rho_{\textbf{L}} \lesssim 7.47\times 10^{14}$ g/cm$^{3}$. Based on this information, we select three specific  densities at the surface as: $\rho_\textbf{L}=\rho_{nuc}=2.7\times 10^{14}$ g/cm$^3$, $\rho_\textbf{L}=4\times 10^{14}$ g/cm$^3$ and $\rho_\textbf{L}=6\times 10^{14}$ g/cm$^3$ to encompass the density at which nuclear solidification occurs. Next, for the dimensionless parameter $\alpha_1=0.02$, we establish a correlation between  $\textbf{L}$ and the compactness $C$ for each boundary condition. This relationship is derived by employing the density equation (\ref{sys}), specifically $\rho(r=\textbf{L})=\rho_\textbf{L}$. In Figure \ref{Fig:CompMR}\subref{fig:Comp}, we present the curves illustrating the relationship between compactness and radius, considering the selectedsurface density conditions. The plots demonstrate  the maximal  compactness observed is $0.92$, exceeding the limit of Buchdahl.
%Remarkably the maximal compactness approaches a saturation-like pattern which reflects the capability of the NSs to gain more sizes with small increase of masses at this limit. However, as we obtained earlier in this section the energy dominance constraint sets an upper bound on the compactness $C=0.735$ (represented by a horizontal dashed line) for stable anisotropic configuration in KB spacetime, which consequently determines the maximal allowed radius for each boundary density as represented by the vertical dotted lines in Fig. \ref{Fig:CompMR}\subref{fig:Comp}.
%

In the following, we illustrate the mass-radius curves, accompanied by the observational data obtained in Table \ref{Table1}, as shown in Figure \ref{Fig:CompMR}\subref{fig:MR}. By employing the DEC constraint, we determine that the maximal permissible mass is $M_{\text{max}}=5.25 M_\odot$, with a corresponding maximal radius of $\textbf{L}{\text{max}}=16.6$ km. This calculation is based on the surface density at the nuclear saturation   of $\rho{\text{nuc}}=2.7\times 10^{14}$ g/cm$^3$.  Significantly, within the same boundary condition, GR yields a maximal mass of $4.1 M_\odot$ with a corresponding maximum radius of $\textbf{L}_{\text{max}}=16.8$ km.  When compared to the predictions of  GR, the $f(R,T)$ theory with a positive parameter $\alpha_1=0.02$ forecasts a mass that is nearly the same, albeit with a slightly smaller size of approximately 0.2 km. Similarly, when considering surface densities of $\rho_\textbf{L}=4\times 10^{14}$ g/cm$^3$ and $\rho_\textbf{L}=6\times 10^{14}$ g/cm$^3$, we determine the corresponding maximum masses and radii as follows: ($M_{\text{max}}=4.28 M_\odot, \textbf{L}{\text{max}}=13.8$ km) and ($M{\text{max}}=3.5 M_\odot, \textbf{L}_{\text{max}}=11.4$ km). These values are consistent with the nuclear solidification density.
%Maximal radii (masses) allowed by the DEC (diagonal dashed line) are represented by vertical (horizontal) dotted lines for each boundary density on the mass-radius graph in Fig. \ref{Fig:CompMR}\subref{fig:MR}. All pulsars lie in the physical region, as allowed by the DEC.
Surprisingly, the present model has the ability to produce a NS  within the mass range of $3.5-5.25 M_\odot$, which leaves open the possibility that the companion of the binary system GW190814, with a mass of $M=2.6 M_\odot$, could be a NS. The calculated boundary density of this NS is consistent with the  saturation density and follows an EoS in a linear.

In Figure \ref{Fig:CompMR}\subref{fig:NICER}, we stress the analysis on several noteworthy stars. We  show the pulsars $PSR J0030+0451$ (observed by NICER) and  $PSR J0740+6620$ (observed by NICER+XMM) align well with the $\rho_{nuc}$. Additionally, we observe that the restrictions on the radius of a canonical neutron star imposed by LIGO-Virgo are in line with the surface density corresponding to $\rho_{nuc}$.

%%%%%%%%%%%%%%%%%%%%%%%%%%%%%%%%%%%%%%%%%%%%%% Section 6 %%%%%%%%%%%%%%%%%%%%%%%%%%%%%%%%%%%%%%%%%%%%%%%%%%%%%%%%%%%%%%%
\section{Summary of the results of the present work}\label{Sec:Conclusion}
Throughout this study, we looked at how the mass-radius  of stellar objects was affected by the non-minimal coupling between matter and geometry, as suggested by \citep{Harko:2011kv}. The Ricci scalar and the trace of the energy-momentum tensor are combined linearly in the Harko theory in its linear version, and they are connected by a dimensional constant called $\alpha$ whose vanishing reduces the theory to GR theory. Given the fact that spacetime is substantially curved, this phenomenon is expected to be carefully explored by the stellar structures of compact objects. Exact measurements of the pulsar's mass and radius, ${\textit HerX1}$, would allow for a more precise estimate of the parameter $\alpha_1$. We summarized the output of this study  as follows:
\begin{itemize}
\item We demonstrated that the anisotropy in $f(R,T)=R+\alpha_1 T$ is the same as in GR for a  spacetime has spherical symmetry with an anisotropic matter source, making deviations from GR in relation to the coupling between matter and geometry easier to distinguish. By assuming the form of the metric potential $g_{rr}$ and a specific form of the anisotropy, we were able to derive the explicate form of the components of the energy-momentum tensor, which allows us to write all physical quantities in relation to the parameters $\alpha_1$ and $C$. Because of the exact mass-radius observational constraints from observations of pulsar $\textit HerX1$, we were able to estimate the parameter $\alpha_1$ to be in the positive range at $\alpha_1=0.02$. Higher compactness values are possible within the $f(R,T)=R+\alpha_1 T$ context because this case predicts a bigger size for a given mass than GR. We demonstrated that the extra force of $f(R,T)=R+\alpha_1 T$ contributes to the hydrodynamic balance to counterbalance or offset some of the gravitational force, enabling the existence of more compact stars, relative to those predicted by  GR, for a specific mass. Furthermore, we have shown the maximum value  of $C$ is $ 0.92$, which slightly exceeds the   limit of Buchdahl $ 8/9$, but it does not approach the bound imposed by the Schwarzschild radius.

\item Surprisingly, although we did not make any specific EoS assumptions, the model exhibits excellent agreement with a linear equation of state (EoS) that includes a bag constant. Interestingly, we discovered that in the NS core, the maximum squared sound speeds are approximately $v^2_{1}\approx 0.52 c^2$ in the radial direction and $v^2_{2}\approx 0.26 c^2$ in the tangential direction. This is in contrast to the behavior observed in the case of general relativity (GR).
%    . The sound speed, on the other hand, is severely broken in hadronic EoS models and the Gaussian process non-parametric EoS approach (using NICER+XMM data from the pulsar PSR J0740+6620).

\item The model permits a maximum mass of $5.25 M \odot$ with a radius of $\textbf{L}=16.6$ km, corresponding to a surface density of $\rho_{nuc}=2.7\times 10^{14}$ g/cm$^3$. Notably, this maximum mass results in a larger compactness value compared to the prediction of general relativity (GR) for the same boundary condition of surface density.  This result leaves open the possibility that GW190814's companion could be an anisotropic NS without requiring the assumption of any unusual matter sources.

\item We conclude that Einstein's theory of gravity and $f(R,T)=R+\alpha_1 T$ modified gravity are not comparable in the astrophysical domain, where the energy-momentum tensor does not vanish, as it does in the vacuum case \citep{Nashed:2018oaf,Nashed:2018piz,Nashed:2018efg}. Contrary to GR, our extensive  analysis demonstrates that the coupling between matter and geometry in $f(R,T)=R+\alpha_1 T$ gravity resolves the discrepancy between the sound speed in compact objects with large masses and the conformal upper bound. This significant finding, together with additional relevant studies, provides evidence for the differentiation between $f(R,T)=R+\alpha_1 T$ gravity and GR .
\end{itemize}
\newpage
\appendix
\section{The explicate forms of the parameter $a_0$, $a_1$ and $a_2$ in terms of the compactness \label{A}}
\begin{eqnarray}
 \label{df1}
&a_0=4\left\{\left[ 2- 5C- {C}^{3}+ 4{C}^{2}+ 2{\mathbb C}^{1/2} {C}^{2}- 4 {\mathbb C}^{5/4}- 4\,{C}^{2}{\mathbb C}^{3/4}- 4{\mathbb C}^{3/2}C- 4C{\mathbb C}^{1/2}+ 8\,C{\mathbb C}^{3/4}+ 4{\mathbb C}^{3/2}+ 2{\mathbb C}^{1/2}- 4 {\mathbb C}^{3/4}\right.\right.\nonumber\\
&\left.\left.+ 4{\mathbb C}^{5/4}C\right]^{1/2} \left(  7828813080\alpha_1 {\mathbb C}^{1/4}- 320\alpha_1{\mathbb C}^{1/2}+ 280\,\alpha_1{\mathbb C}^{3/4}+ 23486438720\alpha_1 {\mathbb C}^{5/4}+ 31315251720\alpha_1C- C- 4{\mathbb C}^{3/4}+ 2\right.\right.\nonumber\\
&\left.\left.+ 1467902415 {\mathbb C}^{1/4}+ 6 {\mathbb C}^{1/2}- 1467902419 {\mathbb C}^{5/4 }- 31315251760\,\alpha_1 \right) \right\}\left\{ 12{\mathbb C}+ 406\,\alpha_1+ 7339512089 {\mathbb C}^{5/4}- 10275316930 {\mathbb C}^{3/2}\right.\nonumber\\
&\left.+ 1467902426C{\mathbb C}^{1/4}+ 1467902414C{\mathbb C}^{1/2}+ 4403707253 {\mathbb C}^{3/4}- 1467902416 {\mathbb C}^{1/2}- 1467902425{\mathbb C}^{1/4}+ 7828812960\alpha_1 C {\mathbb C}^{1/2}\right.\nonumber\\
&\left.- 41101267740\,\alpha_1\,C{\mathbb C}^{3/4}+ 41101267330\,\alpha_1 {\mathbb C}^{3/4}+ 25443641660\,\alpha_1 {\mathbb C}^{5/4}- 27400845040\,\alpha_1 {\mathbb C}^{3/2}- 23486438720\,\alpha_1\,{\mathbb C}^{7/4}\right.\nonumber\\
&\left.+ 7828812840\alpha_1\,C{\mathbb C}^{1/4}- 4403707256C {\mathbb C}^{3/4}+ 1467902419{\mathbb C}^{7/4}- 7828812754\alpha_1{\mathbb C}^{1/2}- 7828812880\alpha_1{\mathbb C}^{1/4}- 406\alpha_1C\right\}^{-1}\,, \nonumber \\
&a_1=\frac{1}{ 2 {\mathbb C}^{3/4} -{\mathbb C} ^{1/2}-{\mathbb C} }\left\{ - 4\left[\left( 2- 5C-{C}^{3}+ 4{C }^{2}+ 2{\mathbb C}^{1/2}{C}^{2}- 4{\mathbb C}^{5/4}- 4{C}^{2}{\mathbb C}^{3/4}- 4 {\mathbb C}^{3/2}C- 4C{\mathbb C}^{1/2}+ 8C{\mathbb C}^{3/4}+ 4{\mathbb C}^{3/2}\right.\right.\right.\nonumber \\
&\left.\left.\left.+ 2{\mathbb C}^{1/2}- 4{\mathbb C}^{3/4}+ 4{\mathbb C}^{5/4}C\right)^{1/2} \left( 7828813080\alpha_1{\mathbb C}^{1/4}- 320\alpha_1{\mathbb C}^{1/2}+ 280\alpha_1{\mathbb C}^{3/4}+ 23486438720\alpha_1{\mathbb C}^{5/4}+ 31315251720\alpha_1C\right.\right.\right.\nonumber \\
&\left.\left.\left.- C- 4{\mathbb C}^{3/4}+ 2+ 1467902415\sqrt [4]{{\mathbb C} }+ 6{\mathbb C}^{1/2}- 1467902419{\mathbb C}^{5/4}- 31315251760\alpha_1 \right) {\mathbb C}^{1/4}\right]\left\{ 12{\mathbb C}+ 406\alpha_1+ 7339512089{\mathbb C}^{5/4}\right.\right.\nonumber \\
&\left.\left.- 10275316930{\mathbb C}^{3/2}+ 1467902426C{\mathbb C}^{1/4}+ 1467902414C{\mathbb C}^{1/2}+ 4403707253{\mathbb C}^{3/4}- 1467902416{\mathbb C}^{1/2}- 1467902425{\mathbb C}^{1/4}\right.\right.\nonumber \\
&\left.\left.+ 7828812960\alpha_1C{\mathbb C}^{1/2}- 41101267740 \alpha_1C{\mathbb C}^{3/4}+ 41101267330\alpha_1{\mathbb C}^{3/4}+ 25443641660\alpha_1{\mathbb C}^{5/4}- 27400845040\alpha_1{\mathbb C}^{3/2}\right.\right.\nonumber \\
&\left.\left.- 23486438720\alpha_1{\mathbb C}^{7/4}+ 7828812840\alpha_1C{\mathbb C}^{1/4}- 4403707256C{\mathbb C}^{3/4}+ 1467902419{\mathbb C}^{7/4}- 7828812754\alpha_1{\mathbb C}^{1/2}\right.\right.\nonumber \\
&\left.- 7828812880\alpha_1{\mathbb C}^{1/4}- 406\alpha_1C\right\}^{-1}+ 4 \left\{\left[ 2- 5C- 1{C}^{3}+ 4{C}^{2}+ 2{\mathbb C}^{1/2}{C}^{2}- 4{\mathbb C}^{5/4}- 4{ C}^{2}{\mathbb C}^{3/4}- 4{\mathbb C}^{3/2}C- 4C{\mathbb C}^{1/2}\right.\right.\nonumber \\
&\left.\left.\left.+ 8C {\mathbb C}^{3/4}+ 4{\mathbb C}^{3/2}+ 2 {\mathbb C}^{1/2}- 4{\mathbb C}^{3/4}+ 4{\mathbb C}^{5/4}C\right]^{1/2} \left(  7828813080\alpha_1{\mathbb C}^{1/4}- 320\alpha_1{\mathbb C}^{1/2}+ 280\alpha_1{\mathbb C}^{3/4}+ 23486438720\alpha_1{\mathbb C}^{5/4}\right.\right.\right.\nonumber \\
&\left.\left.\left.+ 31315251720\alpha_1C- C- 4{\mathbb C}^{3/4}+ 2+ 1467902415{\mathbb C}^{1/4}+ 6 {\mathbb C}^{1/2}- 1467902419{\mathbb C}^{5/4 }- 31315251760\alpha_1 \right){\mathbb C}^{1/2}\right\}\left\{12{\mathbb C}+ 406\alpha_1\right.\right.\nonumber \\
&\left.\left.+ 7339512089{\mathbb C}^{5/4}- 10275316930{\mathbb C}^{3/2}+ 1467902426C {\mathbb C}^{1/4}+ 1467902414C{\mathbb C}^{1/2}+ 4403707253{\mathbb C}^{3/4}- 1467902416 {\mathbb C}^{1/2}\right.\right.\nonumber \\
&\left.\left.- 1467902425{\mathbb C}^{1/4}+ 7828812960\alpha_1C{\mathbb C}^{1/2}- 41101267740\alpha_1C{\mathbb C}^{3/4}+ 41101267330\alpha_1{\mathbb C} ^{3/4}+ 25443641660\alpha_1{\mathbb C}^{5/4}\right.\right.\nonumber \\
&\left.\left.- 27400845040\alpha_1{\mathbb C}^{3/2}- 23486438720\alpha_1{\mathbb C}^{7/4}+ 7828812840\alpha_1C{\mathbb C}^{1/4}- 4403707256C{\mathbb C}^{3/4}+ 1467902419{\mathbb C}^{7/4}- 406\alpha_1C\right.\right.\nonumber \\
&\left.\left.- 7828812754\alpha_1{\mathbb C}^{1/2}- 7828812880\alpha_1{\mathbb C}^{1/4}\right\}^{-1}+4\left[2 -5C-{C}^{3}+4{C}^{2}+2{\mathbb C}^{1/2}{C}^{2}-4 {\mathbb C}^{5/4}-4{C}^{2}{\mathbb C}^{3/4}-4{\mathbb C}^{3 /2}C\right.\right.\nonumber \\
&\left.\left.-4C{\mathbb C}^{1/2}+8C {\mathbb C}^{3/4}+4{\mathbb C}^{3/2}+2{\mathbb C}^{1/2}-4{\mathbb C}^{3/4}+4{\mathbb C}^{5/4}C\right]^{1/2} \right\} \,, \qquad a_2=\sqrt{1-{\mathbb C}^{1/4}},\qquad \mbox{where} \qquad  {\mathbb C}=(1-C)\,.
\end{eqnarray}
Equation (\ref{df1})  yields the following form upon  the limit $\alpha_1\to 0$, \citep{Roupas:2020mvs}
\begin{eqnarray}
&a_0=8\left\{\left[ 2- 5C-{C}^{3}+ 4{C}^{2}+ 2{\mathbb C}^{1/2}{C}^{2}- 4 {\mathbb C}^{5/4}- 4{C}^{2} {\mathbb C}^{3/4}- 4 {\mathbb C}^{3/2}C- 4C
{\mathbb C}^{1/2}+ 8C{\mathbb C}^{3/4}+ 4 {\mathbb C}^{3/2}+ 2{\mathbb C}^{1/2}- 4 {\mathbb C}^{3/4}\right.\right.\nonumber \\
&\left.\left.+ 4 {\mathbb C}^{5/4}C\right]^{1/2} \left( - 576443933- 576443932{\mathbb C}^{1/4}- 576443932{\mathbb C}^{1/2}- 576443932 {\mathbb C}^{3/4}+ 576443932C \right)\right\}\left\{ 8{\mathbb C}- 2305775729{\mathbb C}^{1/4}\right.\nonumber \\
&\left.
 -3458663592{\mathbb C}^{1/2}- 7 {\mathbb C}^{3/4}+ 6917327184 {\mathbb C}^{5/4}- 1152887864 {\mathbb C}^{3/2}+ 2305775731C {\mathbb C}^{1/4}+
  3458663592C {\mathbb C}^{1/2}\right\}^{-1}\,,\nonumber\\
&a_1=\frac{1}{\left(2 {\mathbb C}^{3/4} - {\mathbb C}^{1/2}-{\mathbb C}\right)} \left(  8\left\{\left[ 2- 5C- {C}^{3}+ 4{C} ^{2}+ 2{\mathbb C}^{1/2}{C}^{2}- 4 {\mathbb C}^{5/4}- 4{C}^{2} {\mathbb C}^{3/4}- 4  {\mathbb C}^{3/2}C- 4C{\mathbb C}^{1/2}+ 8C {\mathbb C}^{3/4}+ 4 {\mathbb C}^{3/2}\right.\right.\right.\nonumber\\
&\left.\left.\left.+ 2{\mathbb C}^{1/2}- 4 {\mathbb C}^{3/4}+ 4 {\mathbb C}^{5/4}C\right]^{1/2} \left( - 576443933- 576443932\sqrt [4]{{\mathbb C}}- 576443932 {\mathbb C}^{1/2}- 576443932 {\mathbb C}^{3/4} + 576443932C \right){\mathbb C}^{1/4}\right\}\right.\nonumber\\
&\left.\left\{ 8{\mathbb C}- 2305775729 {\mathbb C}^{1/4}- 3458663592{\mathbb C}^{1/2}- 7 {\mathbb C} ^{3/4}+ 6917327184{\mathbb C}^{5/4}- 1152887864 {\mathbb C}^{3/2}+ 2305775731C{\mathbb C}^{1/4}\right.\right.\nonumber\\
&\left.\left.+ 3458663592C{\mathbb C}^{1/2}\right\}^{-1}- 8\left\{\left[2- 5C- 1{C}^{3}+ 4{C}^{ 2}+ 2{\mathbb C}^{1/2}{C}^{2}- 4 {\mathbb C}^{5/4}- 4{C}^{2} {\mathbb C} ^{3/4}- 4  {\mathbb C}^{3/2}C- 4C{\mathbb C}^{1/2}+ 8C {\mathbb C}^{3/4}+ 4 {\mathbb C}^{3/2}\right.\right.\right.\nonumber\\
&\left.\left.\left.+ 2{\mathbb C}^{1/2}- 4 {\mathbb C}^{3/4}+ 4 {\mathbb C}^{5/4}C\right]^{1/2} \left( - 576443933- 576443932{\mathbb C}^{1/4}- 576443932 {\mathbb C}^{1/2}- 576443932 {\mathbb C}^{3/4} + 576443932C \right) {\mathbb C}^{1/2}\right\}\right.\nonumber\\
&\left.+\left\{ 8{\mathbb C}- 2305775729{\mathbb C}^{1/4}- 3458663592{\mathbb C}^{1/2}- 7 {\mathbb C}^{3/4}+ 6917327184{\mathbb C}^{5/4}- 1152887864 {\mathbb C}^{3/2}+ 2305775731C{\mathbb C}^{1/2}\right.\right.\nonumber\\
&\left.\left.+ 3458663592C{\mathbb C}^{1/2}\right\}^{-1}+4\left[2-5C-{C}^{3}+4{C}^{2}+22{\mathbb C}{C}^{2}-4 2{\mathbb C}^{5/4}-4{C}^{2} 2{\mathbb C}^{3/4}-42{\mathbb C}^{3/2}C-4C2{\mathbb C}^{1/2}+8C{\mathbb C}^{ 3/4}\right.\right.\nonumber\\
&\left.\left.+4{\mathbb C}^{3/2}+2{\mathbb C}^{1/2}-4{\mathbb C}^{3/4}+4 {\mathbb C}^{5/4}C\right]^{1/2} \right)\,, \qquad \qquad  {\mathbb C}=(1-C)\,.
\end{eqnarray}
%%%%%%%%%%%%%%%%%%%%%%%%%%
\section{Radial gradients\label{B}}
%\subsection{Radial gradients}
We derive in this appendix the explicate forms of the radial derivative of the fluid density and pressures and get:
\begin{eqnarray}\label{eq:dens_grad}
&\rho'=a_2^{4} \left(1392{\textbf{L}}^{4}a_1 a_2^{4}{r}^{3}-504{\textbf{L}}^{6}a_1a_2^{2}r-2688{\textbf{L} }^{6}a_1a_2^{2}\alpha_1r+7424{\textbf{L}}^{4}a_1a_2^{4}\alpha_1{r}^{3}-1320{\textbf{L}}^{6}\alpha_1a_0r- 180{\textbf{L}}^{6}a_0r-1332{\textbf{L}}^{2}a_1a_2^{6}{r}^{5}\right.\nonumber\\
&\left.-7104{\textbf{L}}^{2}a_1a_2^{5}{r}^{5}\alpha_1+2512{\textbf{L}}^{4}a_2^{2}\alpha_1a_0r^{3}+336 {\textbf{L}}^{4}a_2^{2}{r}^{3}a_0+2304a_1a_2^{ 8}{r}^{7}\alpha_1+432a_1a_2^{8}{r}^{7}-1224{a_2}^{4}\alpha_1a_0{\textbf{L}}^{2}{r}^{5}-162a_2^{4 }{r}^{5}a_0{\textbf{L}}^{2} \right)\nonumber\\
&\times \left\{3{\kappa}^{2} \left( 1+2\alpha_1 \right)  \left( 8\alpha_1+1 \right) {\textbf{L}}^{8} \left( a_0{\textbf{L}}^{2}+2a_1a_2^{2}{\textbf{L}}^{2}-2a_1a_2^{4}{r}^{2} \right) {c}^{2}\right\}^{-1}\,,\nonumber\\
\end{eqnarray}
\begin{eqnarray}\label{eq:pr_grad}
  & p'_1=-a_2^{2}\left( -120\,{\textbf{L}}^{6}a_1a_2^{4}r+384{\textbf{L}}^{6}a_1\,a_2^{4}\alpha_1r +240\,{\textbf{L}}^{4}a_1a_2^{6}{r}^{3}-1792{\textbf{L}}^{ 4}a_1a_2^{6}\alpha_1{r}^{3}+408{\textbf{L}} ^{6}a_2^{2}\alpha_1a_0r+36{\textbf{L}}^{6}a_2^{2}a_0r+48a_1a_2^{10}{r}^{7}\right.\nonumber\\
&\left.-180{\textbf{L}}^{2}a_1a_2 ^{8}{r}^{5}+2112{\textbf{L}}^{2}a_1a_2^{8}{r}^{5} \alpha_1-944{\textbf{L}}^{4}a_2^{4}\alpha_1a_0{r}^{3}-96{\textbf{L}}^{4}a_2^{4}{r}^{3}{\kappa}^{2 }a_0-768a_1a_2^{10}{r}^{7}\alpha_1+504a_2^{6}\alpha_1a_0{\textbf{L}}^{2}{r}^{5}+54a_2^{6}{r}^{5}a_0{\textbf{L}}^{2} \right)\nonumber\\
&\times\left\{3{\textbf{L}}^{8} {\kappa}^{2}\left(1+ 10{\kappa}^{2}\alpha_1+16{\kappa}^{2}{\alpha_1}^{2} \right)  \left( a_0{\textbf{L}}^{2}+2a_1 a_2^{2}{\textbf{L}}^{2}-2a_1a_2^{4}{r}^{2} \right) \right\}^{-1}\,,
\end{eqnarray}
\begin{eqnarray}\label{eq:pt_grad}
&p'_t=-4a_2^{2} \left[144{\textbf{L}}^{4}a_1a_2^{6}{r}^{3} -48{\textbf{L}}^{6}a_1a_2^{4}r-48{\textbf{L}}^{6}a_1a_2^{4}\alpha_1r+224{\textbf{L}}^{4} a_1a_2^{6}\alpha_1{r}^{3}+30{\textbf{L}}^{6 }a_2^{2}\alpha_1a_0r-264{\textbf{L}}^{2}a_1a_2^{8}{r}^{5}\alpha_1-144{\textbf{L}}^{2}a_1a_2^{8}{r}^{5}\right.\nonumber\\
&\left.-44{\textbf{L}}^{4}a_2^{4}{ \kappa}^{2}\alpha_1a_0{r}^{3}+48a_1a_2^ {10}{r}^{7}+96a_1a_2^{10}{r}^{7}{\kappa} ^{2}\alpha_1+18a_2^{6}\alpha_1a_0 {\textbf{L}}^{2}{r}^{5} \right] \left\{3{\textbf{L}}^{8} {\kappa}^{2}\left(1+ 10{\kappa}^{2}\alpha_1+16{\kappa}^{2}{\alpha_1}^{2} \right) \right.\nonumber\\
&\times\left. \left( a_0{\textbf{L}}^{2}+2a_1a_2^{2}{\textbf{L}}^{2 }-2a_1a_2^{4}{r}^{2} \right)\right\}^{-1}\,,\nonumber\\
\end{eqnarray}
where $'\equiv \frac{d}{dr}$. Equations (\ref{eq:dens_grad}), (\ref{eq:pr_grad}) and (\ref{eq:pt_grad}) which represent the derivatives of the components of the energy-momentum tensor are important since they investigate  if  the physical stability of the compact object is satisfied or not as we will  show in the coming section.
\bibliography{Refs}{}
\bibliographystyle{aasjournal}

\end{document}